%% file: buder_galah_accretion_chem.tex
\documentclass[fleqn,usenatbib]{mnras}
\usepackage[T1]{fontenc}
\DeclareRobustCommand{\VAN}[3]{#2}
\let\VANthebibliography\thebibliography
\def\thebibliography{\DeclareRobustCommand{\VAN}[3]{##3}\VANthebibliography}

\usepackage{graphicx}	
\usepackage{amsmath}	
\usepackage{amssymb}	
\usepackage{xspace} 
\usepackage{xcolor}
\usepackage{CJK}
\usepackage{fontawesome}
\usepackage{gensymb}
\usepackage{multirow}

\newcommand{\added}[1]{#1}

\newcommand{\dex}{\,\mathrm{dex}}	
\newcommand{\Msol}{\,\mathrm{M_\odot}} 
\newcommand{\kpc}{\,\mathrm{kpc}}	
\newcommand{\Gyr}{\,\mathrm{Gyr}}	
\newcommand{\kms}{\,\mathrm{km\,s^{-1}}}	
\newcommand{\kpckms}{\,\mathrm{kpc\,km\,s^{-1}}}	
\newcommand{\kmkmss}{\,\mathrm{km^2\,s^{-2}}}	

\newcommand{\Gaia}{\textit{Gaia}\xspace} 

\definecolor{linkcolor}{rgb}{0.1216,0.4667,0.7059}
\newcommand{\codeicon}{{\faCloudDownload}}
\newcommand{\codelink}[1]{\href{https://github.com/svenbuder/Accreted-stars-in-GALAH-DR3/tree/main/figures/#1.ipynb}{\codeicon}\,\,}
\newcommand{\oscaption}[2]{\caption{#2 \codelink{#1}}}

\newcommand{\figuretextwidth}[4]{\begin{figure*} \centering \includegraphics[width=#1]{figures/#2.png}\oscaption{#3}{#4}\label{fig:#2} \end{figure*}}
\newcommand{\figurecolumnwidth}[3]{\begin{figure} \centering \includegraphics[width=\columnwidth]{figures/#1.png}\oscaption{#2}{#3}\label{fig:#1} \end{figure}}

\usepackage{newtxtext,newtxmath}

\title[Accreted stars in GALAH+ DR3]{The GALAH Survey: Chemical tagging and chrono-chemodynamics of accreted halo stars with GALAH+ DR3 and \Gaia eDR3\thanks{\added{GALAH+ combines observations obtained through the main GALAH proposals as well as K2-HERMES, TESS-HERMES, and partner proposals.}}\thanks{All code, data, figures, and tables available at \url{https://github.com/svenbuder/Accreted-stars-in-GALAH-DR3}.}}

\author[S. Buder et al.]{Sven Buder,$^{1,2}$\thanks{E-mail: sven.buder@anu.edu.au}
Karin~Lind$^{3}$, 
Melissa~K.~Ness$^{4,5}$, 
Diane~K.~Feuillet$^{6}$,
Danny~Horta$^{7}$,
Stephanie~Monty$^{1,2}$, \newauthor
Tobias~Buck$^{8}$, 
Thomas~Nordlander$^{1,2}$,
Joss~Bland-Hawthorn$^{9,2}$,
Andrew~R.~Casey$^{10,11}$,\newauthor
Gayandhi~M.~De~Silva$^{12}$,
Valentina~{D'Orazi}$^{13}$,
Ken~C.~Freeman$^{1,2}$, 
Michael~R.~Hayden$^{9,2}$,
Janez~Kos$^{14}$, \newauthor
Sarah~L.~Martell$^{15,2}$, 
Geraint~F.~Lewis$^{9}$,
Jane~Lin$^{1,2}$, 
Katharine.~J.~Schlesinger$^{1}$, 
Sanjib~Sharma$^{9,2}$, \newauthor
Jeffrey~D.~Simpson$^{15,2}$, 
Dennis~Stello$^{15,9,16,2}$,
Daniel~B.~Zucker$^{12,17,2}$, 
Toma\v{z}~Zwitter$^{14}$,
Ioana~Ciuc\u{a}$^{1,2,18}$,\newauthor
Jonathan~Horner$^{19}$,
Chiaki~Kobayashi$^{20,2}$,
Yuan-Sen~Ting (丁源森)$^{1,21}$,
Rosemary~F.~G.~Wyse$^{22}$, \newauthor
and the GALAH collaboration
\\
\\
(Affiliations listed after the references)}

\date{Accepted 2021 November 26. Received 2021 November 3; in original form 2021 September 9}
\pubyear{2021}

\begin{document}
\begin{CJK*}{UTF8}{gbsn}
\label{firstpage}
\pagerange{\pageref{firstpage}--\pageref{lastpage}}
\maketitle
\end{CJK*}

\begin{abstract}
Since the advent of \Gaia astrometry, it is possible to identify massive accreted systems within the Galaxy through their unique dynamical signatures. One such system, \Gaia-Sausage-Enceladus (GSE), appears to be an early ``building block'' given its virial mass $> 10^{10}\Msol$ at infall ($z\sim 1-3$). In order to separate the progenitor population from the background stars, we investigate its chemical properties with up to 30 element abundances from the GALAH+ Survey Data Release 3 (DR3). To inform our choice of elements for purely chemically selecting accreted stars, we analyse 4164 stars with low-$\alpha$ abundances and halo kinematics. These are most different to the Milky Way stars for abundances of Mg, Si, Na, Al, Mn, Fe, Ni, and Cu. Based on the significance of abundance differences and detection rates, we apply Gaussian mixture models to various element abundance combinations. We find the most populated and least contaminated component, which we confirm to represent GSE, contains 1049 stars selected via [Na/Fe] vs. [Mg/Mn] in GALAH+ DR3. We provide tables of our selections and report the chrono-chemodynamical properties (age, chemistry, and dynamics). Through a previously reported clean dynamical selection of GSE stars, including $30 < \sqrt{J_R / \kpckms} < 55$, we can characterise an unprecedented 24 abundances of this structure with GALAH+ DR3. \added{With our chemical selection we characterise the dynamical properties of the GSE, for example mean $\sqrt{J_R / \kpckms} = $ \input{depending_text/chem_percentiles_sqrt_J_R}. We find only $(29\pm1)\%$ of the GSE stars within the clean dynamical selection region. Our methodology will improve future studies of accreted structures and their importance for the formation of the Milky Way.} \href{https://github.com/svenbuder/Accreted-stars-in-GALAH-DR3}{\faGithub}
\end{abstract}


\section{Introduction} \label{sec:introduction}

Significant investment has been made in the pursuit of understanding how the Milky Way, as a benchmark spiral galaxy, has formed. To unravel our Galactic history we need large inventories of stellar spatial/dynamical information \citep[e.g.][]{Brown2021}, as well as chemical abundances \citep{Jofre2019}. Stellar ages \citep{Soderblom2010}, even at low precision, in concert with this information are key in connecting the Milky Way today to its past.

Holistically, the Milky Way has been described as comprised of an ensemble of populations, identified as major overdensities. These include a thin and a thick disk component, the bulge, and the halo \citep[see e.g.][for a review]{BlandHawthorn_Gerhard2016}. With improvements in both the quantity and diversity of the data, it has become clear that the two disk components overlap not only spatially but also dynamically \citep[e.g.][]{Bovy2012b}. Recent studies argue that the disk populations are better disentangled using their (fixed) chemical abundances rather than their (evolving) orbital properties; as young (thin) low-$\alpha$ and old (thick) high-$\alpha$ \citep[e.g.][]{Bensby2014, Buder2019, BlandHawthorn2019}. As the kinematic and dynamic properties of stars change with time as the Galaxy evolves, we see that structures identified chemically that have likely been born with discrete and separate orbital properties now overlap. This includes populations of stars with disk-like chemistry on halo-like orbits and vice-versa \citep[e.g.][]{Belokurov2020}. Coarse kinematic/dynamic selections are therefore likely to be significantly contaminated. A possible way forward is to concentrate on the chemical abundances of stars and select (or ``tag'') stars chemically \citep[see e.g.][for a review on chemical tagging]{FreemanBlandHawthorn2002} as a way to identify signatures of the Milky Way's formation. The basic assumption here is that element abundances of stars are similar if they are born together, do not change significantly over time, and are significantly distinct from other populations/birth sites. In the disk it appears that the chemical abundance variance is low \citep{Bovy2016b, Ness2018, Ness2019b, Weinberg2021, Ting2021}. However, the stellar halo has a much more diverse and composite origin \citep[e.g.][]{Helmi2020, Naidu2020}. 

The stellar halo captures the story of the earliest moments in the assembly of the Milky Way, as well as its cosmological encounters, via accreted populations over time. One big and outstanding question in the realm of the halo is: to what level did accreted stars and mergers play an important role in the Milky Way's formation? The importance of accretion in the build-up of the halo - and its connection with the disk due to their co-existence and thus likely interaction - is still enigmatic. This also includes the linked question of what fraction of the halo formed in-situ \citep{BlandHawthorn_Gerhard2016}. Mergers lead to complex phase-space structure and a wide range of both orbital properties and chemical abundances \citep[e.g.][]{Amorisco2017, JeanBptiste2017, Monachesi2019, Koppelman2020b}. As we gather more data, we hope to be able to decipher this puzzle. Ultimately we will need to link our observations with theoretical predictions to find the most likely formation scenarios.

Opening a new chapter in the understanding of the Galactic halo, \citet{Nissen1997b} found differences between the chemical abundances\footnote{Chemical abundances of an arbitrary element X are reported either with an absolute logarithmic ratio of the number densities with respect to H, that is $\mathrm{A(X)} = \log \left(N_\mathrm{X}/N_\mathrm{H} \right) + 12$, or as a ratio of elements X and Y relative to the Solar values ($\odot$), that is $\mathrm{[X/Y]} = \left( \mathrm{A(X)} -\mathrm{ A(Y)} \right) - \left( \mathrm{A(X)}_\odot -\mathrm{A(Y)}_\odot \right)$.} of halo stars even though their metallicities and iron abundances exhibit significant overlap. When expanding the sample from 13 halo and 16 disk stars to a total of 94 stars, two clear sequences of low- and high-$\alpha$ halo stars became evident in the [Fe/H] vs. [Mg/Fe], that is, the Tinsley-Wallerstein diagram \citep{Nissen2010}. Such differences were also found to be clearly visible for other nucleosynthesis channels \citep{Nissen2010, Nissen2011, Ting2012, Hawkins2015}, among them light odd-Z elements like Na and Al or iron-peak elements like Ni, Mn, and Cu. Guided by our theoretical understanding of the metallicity-dependent nucleosynthesis of Na, Al, and Cu through massive stars and in particular supernovae (SNe) II as well as Mn via SNIa \citep[e.g.][]{Kobayashi2006,Kobayashi2020}, the low enrichment in these elements suggested that these stars were born outside of the Milky Way. This picture was further supported by the very different overall kinematic and dynamic properties of these stars compared to the Milky Way \citep{Nissen2010, Schuster2012}, but was limited to few stars.

Astrometric data provided by the \Gaia satellite \citep{Brown2016} have been revolutionary. These data have enabled the discovery of accreted structures in dynamical space, most notably the \Gaia-Sausage-Enceladus \citep[GSE, see e.g.][]{Belokurov2018, Helmi2018, Helmi2020}. The stellar and virial masses of the \Gaia-Enceladus-Sausage progenitor satellite has been estimated in the range of $M_\star \sim 10^{8.7-9.85}\Msol$ \citep{Feuillet2020,Naidu2021} and $M_\text{vir} > 10^{10}\Msol$ \citep{Belokurov2018}, or a mass ratio at infall (with respect to the early Milky Way) between 1:4 and 1:2.5 \citep{Helmi2018,Naidu2021}. According to preliminary chemical studies of this overdensity \citep{Das2020}, it seems likely that it could contribute between 20--30\,\% to the metal-poor stars below iron abundances $\mathrm{[Fe/H]} < -1$. According to estimates by \citet{Naidu2020}, the GSE contributes significantly to the inner halo ($R < 15\kpc$), but even dominates the halo within $z \approx 10-20 \kpc$ and $R \approx 15-25 \kpc$. The observational evidence \citep[for reviews see][]{Nissen2018, Helmi2020} seems to support the picture suggested by \citet{Searle1978}, where accretion processes contribute massively to the build-up of the halo, in addition to an in-situ inner halo population that formed during a dissipative collapse. 

In recent years, with the advent of revolutionary data from massive stellar surveys. Especially the combination of data from \Gaia with spectroscopic surveys like the SDSS SEGUE and APOGEE Surveys \citep[e.g.][]{Belokurov2018, Helmi2018, Mackereth2018, Hayes2018, Myeong2018a, Das2020} and the H3 Survey \citep{Conroy2019, Naidu2020, Bonaca2020} has helped to identify a wealth of substructure in the stellar halo of the Galaxy. The excitement of such discoveries in such a short period of time has lead to a plethora of different conjectured accretion events (along with their nomenclature), whose reality and distinction still needs to be fully established. Certainly reviewed, it would be useful to have more consistency in the different structures reported in the literature \citep{Helmi2020} or consensus in adopted nomenclature \citep[see e.g][]{Naidu2020}. 

In this paper, we will therefore assume that the low-$\upalpha$ halo stars \citep{Nissen2010, Hayes2018}, blob \citep{Koppelman2018, Das2020}, Sausage \citep{Belokurov2018}, and \Gaia-Enceladus \citep{Helmi2018} are more or less contaminated selections of the same substructure, which we will refer to as GSE. Several of these assumptions have already been convincingly demonstrated to be true, e.g. for low-$\alpha$ halo and GSE to first order \citep{Haywood2018, Mackereth2019}. We emphasise, however, that different techniques might actually select not only the GSE, but also from other separate substructures. Several substructures, like Sequoia \citep{Barba2019, Myeong2019, Monty2020} on significantly retrograde orbits, have been found, which might be ``contaminating'' the GSE selection. We revisit this problem especially in the discussion at the end of our study.

We are just at the beginning of understanding how we can use our ``tools'' \citep{Helmi2020}, that is astrophysical ones, like chemical composition and age, as well as kinematic/dynamical ones, to identify accreted stars. We provide a list of previously used tools to identify accreted stars in App.~\ref{sec:selection_techniques}, sorted by the categories of information they use from purely kinematic over chemodynamic to purely chemical. Future work should study how (dis-)similar the selection of stars using these different techniques can be.

In this paper, we aim to identify, or ``tag'', accreted stars to first order via their chemical composition, a technique proposed by \citet{FreemanBlandHawthorn2002} to identify the signatures of galaxy formation. We use estimates of the chemical composition from the stellar spectroscopic survey GALactic Archaeology with HERMES \citep[GALAH, ][]{DeSilva2015, Buder2021} aided by the astrometric data from the \Gaia satellite \citep{Brown2021}. The combination of these data sets together with age estimates from isochrone fitting allows us to study the ages, chemistry, and dynamics (chrono-chemodynamics) of the selected stars, that is, their stellar ages as well as their chemical and dynamical properties.

The data of the GALAH survey, exceeds the data by \citet{Nissen2010} and APOGEE both in the number of stars and the number of element abundances. GALAH+ DR3 delivers up to 30 element abundances. 2\% of its 588 571 stars are metal poor with $\mathrm{[Fe/H]} < -1$ and 4\% exhibit halo kinematics \citep{Buder2021}. In this observational paper we therefore aim to address the following questions:
\begin{enumerate}
\item How can we best select accreted stars chemically within GALAH+ DR3 data?
\item Avoiding circular arguments, what dynamical space do the chemically selected GSE stars occupy and what are the chemical properties of the dynamically selected GSE stars?
\item Are the dynamically and chemically selected substructures truly the same, that is what is the quantitative overlap?
\item What can we learn from the stars of the chemical and dynamical selection that do and do not overlap?
\end{enumerate}

In our initial search for chemical differences between accreted halo stars and in-situ Milky Way stars we are guided by the sample from \citet{Nissen2010}, which comprises the largest number of abundances studied for accreted halo stars and compare with the more recent literature achieved with data from APOGEE and H3 when putting our results into context.

We present the data used for this study in Sec.~\ref{sec:data}, together with a description of different quality cuts that we perform, before trying to find the best chemical and dynamical selection of accreted stars in Sec.~\ref{sec:our_selection_techniques}. We compare the samples of these techniques, and in particular their chrono-chemodynamic properties in Sec.~\ref{sec:chronochemodynamics}. In that section, we will also include the current literature for each of the properties. This allows us to then put our results into context during our discussion Sec.~\ref{sec:discussion}. Here we put the purely observational constraints from Sec.~\ref{sec:our_selection_techniques} in the context of the theoretical framework of Galactic chemical evolution and nucleosynthesis pathways to discuss the prospects of chemically tagging the accreted halo (Sec.~\ref{sec:prospects_chem_tagg}), discuss the (dis-)similarities of different selections (Sec.~\ref{sec:dissimilarity}), how we can combine selection criteria for a chemodynamical selection of the GSE (Sec.~\ref{sec:towards_chemodyn}), and the implication of the stellar age distribution of the GSE on different formation and accretion scenarios (Sec.~\ref{sec:age_timescale}). We conclude our study in Sec.~\ref{sec:conclusions} and give an outlook in Sec.~\ref{sec:Outlook}, including remarks on the way forward by combining chemistry and dynamics to identify and analyse chemodynamic substructure, for example in abundance-action space.

\section{Data: GALAH+ DR3 and its Value-Added-catalogues (VACs)} \label{sec:data}

For this study, we use the chemical abundance data from GALAH+ DR3 \citep{Buder2021} together with the spatial and astrometric information from the \Gaia mission \citep{Gaia-Collaboration2016}, namely \Gaia eDR3 \citep{Brown2021}, and include  corrections of parallax zero points \citep{Lindegren2021a, Lindegren2021b}.

GALAH+ DR3 provides elemental abundances based on high-resolution ($R \sim 28\,000$) spectra from the four optical bands of the HERMES spectrograph \citep{Sheinis2015} at the Anglo-Australian Telescope. In brief, stellar parameters ($T_\text{eff}$, $\log g$, [Fe/H], $v_\text{mic}$, $v_\text{broad}$, and $v_\text{rad}$) and abundances for up to 30 different elements are estimated using our modified version of the spectrum synthesis code Spectroscopy Made Easy \citep[\textsc{sme}][]{Valenti1996, Piskunov2017} and 1D \textsc{marcs} model atmospheres \citep{Gustafsson2008}. Eleven elements are computed in non-LTE \citep{Amarsi2020}, the others in local thermodynamic equilibrium (LTE). Combining GALAH+ DR3 with \Gaia eDR3 provides a dataset with chemical abundances for up to 30 different elements and kinematic as well as dynamic properties and isochrone interpolated stellar ages for 678\,324 spectra of 588\,571 stars. Here, we use the value-added-catalogues (VACs) of stellar ages and dynamics provided as part of GALAH+ DR3 \citep{Buder2021}. 

We apply some basic cuts to each selection that will be used throughout this study. We expect the stars to have passed the spectroscopic quality check \texttt{flag\_sp}, be part of the GALAH main survey or the K2/TESS-HERMES follow-up (to exclude observations of the bulge and open/globular clusters, such as $\omega$\,Cen), be within $D_\varpi < 10 \kpc$ (to exclude LMC and SMC), and have available dynamic/age data and unflagged, that is, reliably measured abundances for each of the particular set of elements X used:\begin{equation} \label{eq:basic_cuts}
\text{Basic cuts} = 
\begin{cases}
\texttt{flag\_sp} = 0, \texttt{flag\_fe\_h} = 0, \\
\texttt{survey} \neq \text{``other''}, D_\varpi < 10\kpc, \\
\texttt{L\_Z}\text{, }\texttt{J\_R}\text{, }\texttt{ecc}\text{ \& }\texttt{age\_bstep} \text{ finite and} \\
\texttt{flag\_X\_fe} = 0 \text{ for each used element X} \\
\end{cases}
\end{equation}

We focus on field stars, as we know that globular clusters exhibit significant abundance trends due to multiple stellar populations \citep[e.g.][]{Carretta2009}. We stress, however, that these clusters also hold valuable information and, according to current studies \citep[e.g.][]{Massari2019, KochHansen2021}, $\sim$35\% of them appear to be linked to merger events.

\added{The requirement of finite age estimates from BSTEP \citep[a Bayesian isochrone interpolation tool used as part of GALAH+ DR3][]{Sharma2018} ensures that the fitting of both ages and distances via BSTEP was successful. As a result, we can use the distances $D_\varpi$ for all stars of our base sample that are informed by spectroscopic, photometric, and astrometric information, rather than the photogeometric or geometric distances by \citet{BailerJones2021} that are also provided in the GALAH+ DR3 VACs.}

For the radial velocities, we prefer to use the template matched values \texttt{rv\_obst} provided by the GALAH DR3 RV VAC, otherwise those from the \textsc{sme} pipeline (\texttt{rv\_sme\_v2}) and else those from \Gaia eDR3 \citep{Katz2019}, which originate in \Gaia DR2. We always the radial velocity $v_\text{rad}$ with the smallest uncertainty:
\begin{equation} \label{eq:best_rv}
\centering
v_\text{rad} =
\begin{cases}
\texttt{rv\_obst} \text{ if available w/ smallest unc., else} \\
\texttt{rv\_sme\_v2} \text{ if avail. w/ smallest unc., else} \\
\texttt{dr2\_radial\_velocity} \text{ if avail. w/ smallest unc.}
\end{cases}
\end{equation}

\added{In practice, we use 97\% radial velocities from template matching and 3\% GALAH radial velocities where no template was available. We only use \Gaia DR2 radial velocities for 185 stars of the base sample ($\ll 1\%$) and 4 stars of our final chemical and dynamical selections ($\ll1\%$), none of them being main-sequence turn-off (MSTO) stars.}

We use kinematic and dynamic properties like orbit actions and eccentricities as reported in the GALAH DR3 VAC for dynamics. These calculations were performed by using {\sc galpy} \citep{Bovy2015} and its {\sc orbit} module as well as {\sc galpy}'s {\sc actionAngleStaeckel} approximations via the Staeckel fudge  \citep{Binney2012, Mackereth2018}. Calculations assumed the axisymmetric potential by \citet{McMillan2017} and a circular velocity of $233.1\kms$. The Sun is positioned at $R = 8.21\kpc$ and $z=0.025\kpc$ \citep{Juric2008} with space motions $U_\odot = -11.1\kms$, $V_\odot = 15.17\kms$, $W_\odot = 7.25\kms$ \citep{Schoenrich2010,Reid2004} relative to the Local Standard of Rest (LRS).  For more details see \citet{Buder2021}. We use isochrone-interpolated stellar ages from BSTEP \citep{Sharma2018}. The most reliable isochrone-interpolated stellar ages of our data set are determined for
\begin{equation} \label{eq:msto}
\text{MSTO stars} = 
\begin{cases}
T_\text{eff} \geq 5350\,\mathrm{K} \text{ and} 
\log g \geq 3.5\,\mathrm{\log \left(cm\,s^{-2} \right)} 
\end{cases}
\end{equation}

\figuretextwidth{17cm}{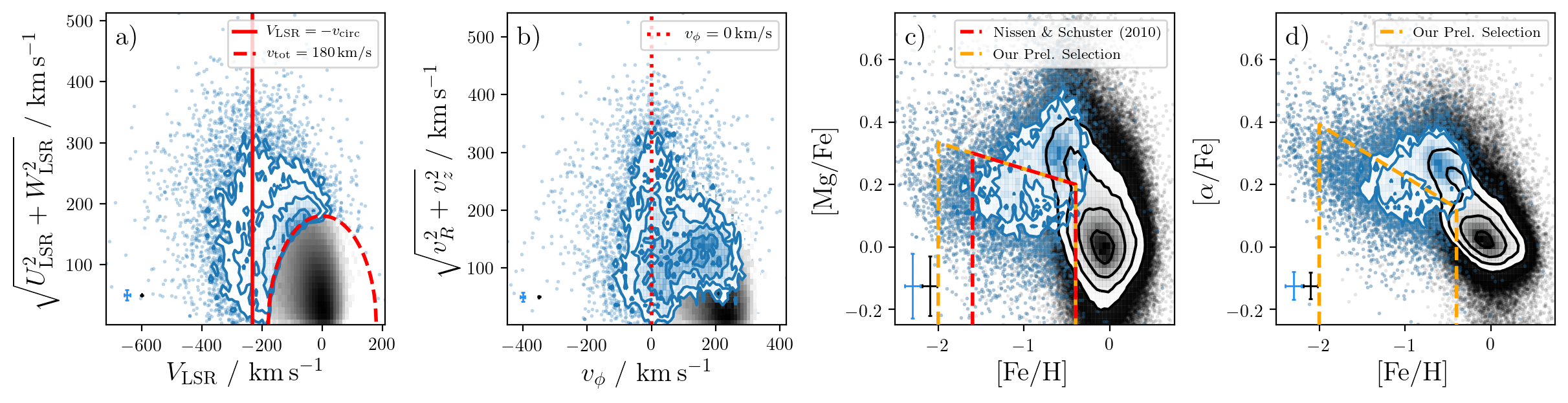}{chemical_differences}{
\textbf{Visualisation of the preliminary selection of low-$\alpha$ stars (see Eq.~\ref{eq:prelim_low_alpha_halo}) from GALAH+ DR3 based on the selection by \citet{Nissen2010}.}
\textbf{Panel a)} Initial selection (shown with red dashed line) of stars via a cut in total velocity $v_\text{tot} > 180\,\mathrm{km\,s^{-1}}$, here shown in the classical Toomre diagram $V$ vs. $\sqrt{U^2 + W^2}$, relative to the local standard of rest (LSR). Stars on retrograde orbits are left of the red line of $V = -233.1\kms$.
\textbf{Panel b)} Same stars, but in the Galactocentric reference frame.
\textbf{Panel c)} [Fe/H] vs. [Mg/Fe] diagram with the chemical selection of low-$\alpha$ halo stars by \citet{Nissen2010} shown as red dashed box. Our selection (orange dashed box) is extended towards lower $\mathrm{[Fe/H]}$ to built a larger sample.
\textbf{Panel d)} [Fe/H] vs. global [$\alpha$/Fe] diagram showing an additional cut (orange dashed box) to clean our selection from contamination due to the lower precision of our sample relative to \citet{Nissen2010}.
Error bars in the bottom left of each panel show the median uncertainties for our base sample (black) and high $v_\text{tot}$ samples (blue).
}

\section{Chemical/dynamical selections} \label{sec:our_selection_techniques}

As we describe in Sec.~\ref{sec:introduction}, a plethora of different techniques exist to enable the selection of accreted stars (see also again Table~\ref{tab:selection_techniques}). In this section, we seek the best way to chemically tag \citep{FreemanBlandHawthorn2002, Ting2015} accreted stars. This refers to tracing a common origin through similarities in chemical composition, under the assumption that each origin is chemically distinct. In a similar context, \citet{Rix2013} advocated strongly for the use of mono-abundance populations as a productive way forward, both when applied via the selection of chemical cells \citep{Ting2015, Lu2021} for observational data \citep[e.g.][]{Bovy2012, Bovy2012b, Bovy2016} as well as models \citep[e.g.][]{Ting2013,Bird2013}, when keeping in mind that these are not necessarily mono-age populations \citep{Minchev2017}. This approach is effectively one application of strong chemical tagging, because it does not at all rely on non-chemical data for the selection of accreted substructures.

In order to find the best chemical selection of accreted stars in GALAH+ DR3, we are, however, limited by the data. We therefore first have to assess the enrichment differences between the halo and the disk among elements reported in GALAH+ DR3. While initial applications of mono-abundance populations have been performed in 2-dimensional space \citep[e.g.][]{Navarro2011,DiMatteo2019,Carollo2021} for [Fe/H] and [Mg/Fe] or [$\alpha$/Fe], the use of more abundances and especially nucleosynthesis dimensions, seems advisable and can be based on already existing literature \citep{Nissen2010,Ting2012, Hawkins2015, Hayes2018, Das2020}. We therefore first study the quantitative enrichment differences as found in GALAH+ DR3 in Sec.~\ref{sec:enrichment_differences} and then assess the most promising combination of abundances in Sec.~\ref{sec:choosing_chemical_selection}, before finding our final chemical selection of accreted stars (Sec.~\ref{sec:gaussian_mixture_models}), which we aim to compare to the dynamical selection introduced in Sec.~\ref{sec:dynamical_selection}. We put our findings of chemical differences into the theoretical context of nucleosynthesis processes like SNIa and SNII in Sec.~\ref{sec:prospects_chem_tagg}.

\subsection{Chemical differences of kinematic low-/high-\texorpdfstring{$\alpha$}{alpha} halo stars} \label{sec:enrichment_differences}

The studies by \citet{Nissen2010,Nissen2011,Nissen2012} and \citet{Nissen2014} have found significant differences between high-velocity stars of the disk and accreted stars (low-$\alpha$ halo in their study), when using the differences in [Mg/Fe] and [Na/Fe] as a baseline. We use the selection of accreted (low-$\alpha$ halo) stars and abundances reported in these studies, which are among the most precise measurements across nucleosynthesis channels of halo stars to date, as a starting point to learn about the enrichment differences between the halo and disk for different elements within GALAH+ DR3. These also serve as an additional reliability check of GALAH+ DR3 data in this parameter space. We find three stars (2MASS IDs 07434398-0004006, 08584388-1607583, 13535810-4632194) overlapping between GALAH+ DR3 and the sample from \citet{Nissen2010}\added{, with [Fe/H] values of -1.27, -0.86, and -0.73, respectively.}. Their stellar parameters agree within the uncertainties for all stellar parameters (we note parallax uncertainties of less than 1\%) and the abundances typically differ by less than $0.05\dex$, with Cr being the only exception with a difference of $0.1\dex$.
\added{We note only 195 (5\%) of our low-$\alpha$ halo sample are MSTO stars, similar to the selection by \citet{Nissen2010}. The majority of our sample are giant stars. Found differences and scatters between the literature and our sample could thus be real or influenced by non-LTE effects as well as analysis effects like the choice of analysed lines and our prescription of microturbulence \citep[see discussion in Sec.~6.4 of][]{Buder2021}.}

\subsubsection{Separating kinematic low- and high-\texorpdfstring{$\alpha$}{alpha} halo stars}

For the comparison with the literature data of the low-$\alpha$ halo, we perform very similar cuts to the GALAH+ DR3 data as \cite{Nissen2010}. We apply an initial cut in the total velocity of $v_\text{tot} > 180\kms$ with respect to the LSR. We plot the velocity distribution (grey-scaled density) in Fig.~\ref{fig:nissen_selection_corner}a in a classic Toomre diagram of space velocities with respect to the LSR for the GALAH+ DR3 data and with Galactocentric space velocities in Fig.~\ref{fig:nissen_selection_corner}b. We only show data with reliable (unflagged) [Mg/Fe] and [$\alpha$/Fe] (an error-weighted average of Mg, Si, Ca, and \added{\ion{Ti}{i} lines}) in addition to the basic quality cuts of Eq.~\ref{eq:basic_cuts}. \added{We adopt this definition of [$\alpha$/Fe] from GALAH+ DR3 and explicitly treat O separately, because of its significantly different trend compared to the other $\alpha$-process elements \citep[see][for a detailed discussion]{Buder2018}}. In this projection, stars that move similar to the LSR are located close to the origin of coordinates, like the Sun. Almost all stars of GALAH+ DR3 have small total motions compared to the LSR, with only \input{depending_text/percentage_vtot.tex} and \input{depending_text/percentage_vtan.tex} above a total or tangential velocity ($v_\text{tot}$ or $v_T$) above $180\kms$, respectively. These stars, shown in a blue density distribution in Fig.~\ref{fig:nissen_selection_corner}, are typically assigned to the kinematic halo \citep[e.g.][]{Venn2004} and are thought to cover both accreted stars as well as in-situ halo and/or disk stars on dynamically hot and heated orbits.
In addition to this kinematic cut, we apply a cut in both $\alpha$-enhancement and iron abundance, to get a preliminary selection of the low-$\alpha$ halo as reported by \citet{Nissen2010}. However, we expand the selection by \citet{Nissen2010}, shown as the red dashed lines in Fig.~\ref{fig:nissen_selection_corner}c, which is limited to $-1.6 < \mathrm{[Fe/H]} < -0.4$ down to an iron abundance of $\mathrm{[Fe/H]} \sim -2.0$, where the onset of SNIa contributions for the GSE was found by \citet{Matsuno2019}. This includes more stars in our preliminary selection (see the difference between red and orange dashed lines in Fig.~\ref{fig:nissen_selection_corner}c), as the low-$\alpha$ halo stars clearly extend past the original selection by \citet{Nissen2010}. We acknowledge that this preliminary selection excludes the most metal-poor stars of the GSE \citep{Cordoni2021}. Our precision for kinematic halo stars is on average lower by a factor of 2-3 compared to \citet{Nissen2010}, for example \protect\input{depending_text/median_uncertainties_fehmgfealphafe.tex} compared to 0.03, 0.03, and 0.02 for [Fe/H], [Mg/Fe], and [$\alpha$/Fe]. We thus see a significant contamination of our [Mg/Fe] measurements by the high-$\alpha$ halo, located at \input{depending_text/high_alpha_halo_fehmgfealphafe.tex}. 
We therefore apply a second chemical cut, estimated from the data by \citet{Nissen2010}, on the combined [$\alpha$/Fe] (see the orange dashed line in Fig.~\ref{fig:nissen_selection_corner}d). The applied cuts for the preliminary selection of low-$\alpha$ halo stars in GALAH+ DR3 data, leading to a sample of \input{depending_text/nr_preliminary_low_alpha_halo.tex} spectra (\input{depending_text/nr_preliminary_low_alpha_halo_na.tex} of them with unflagged [Na/Fe] measurements), can be summarised as
\begin{equation} \label{eq:prelim_low_alpha_halo}
\mathrm{Prel.~low-\alpha~halo} =
\begin{cases}
Eq.~\ref{eq:basic_cuts}\text{, }v_\text{tot} > 180\,\mathrm{km\,s^{-1}} \text{, } \\
\text{flags = 0 for Fe, }\alpha\text{, Mg, \& Na, } \\
-2.0 \leq \mathrm{[Fe/H]} \leq -0.4 \text{, }\\
\mathrm{[Mg/Fe]} < - \frac{1}{12} \times \mathrm{[Fe/H]} + \frac{1}{6} \text{, and}\\
\mathrm{[\alpha/Fe]} < - \frac{1}{6} \times \mathrm{[Fe/H]} + \frac{0.7}{12}.\\
\end{cases}
\end{equation}

Conversely, we describe the preliminary high-$\alpha$ halo via
\begin{equation} \label{eq:prelim_high_alpha_halo}
\mathrm{Prel.~high-\alpha~halo} =
\begin{cases}
Eq.~\ref{eq:basic_cuts}\text{, } v_\text{tot} > 180\,\mathrm{km\,s^{-1}} \text{, } \\
\text{flags = 0 for Fe, }\alpha\text{, Mg, \& Na, } \\
-2.0 \leq \mathrm{[Fe/H]} \leq -0.4 \text{, }\\
\mathrm{[Mg/Fe]} \geq - \frac{1}{12} \times \mathrm{[Fe/H]} + \frac{1}{6} \text{, and}\\
\mathrm{[\alpha/Fe]} \geq - \frac{1}{6} \times \mathrm{[Fe/H]} + \frac{0.7}{12}.\\
\end{cases}
\end{equation}

\figuretextwidth{17cm}{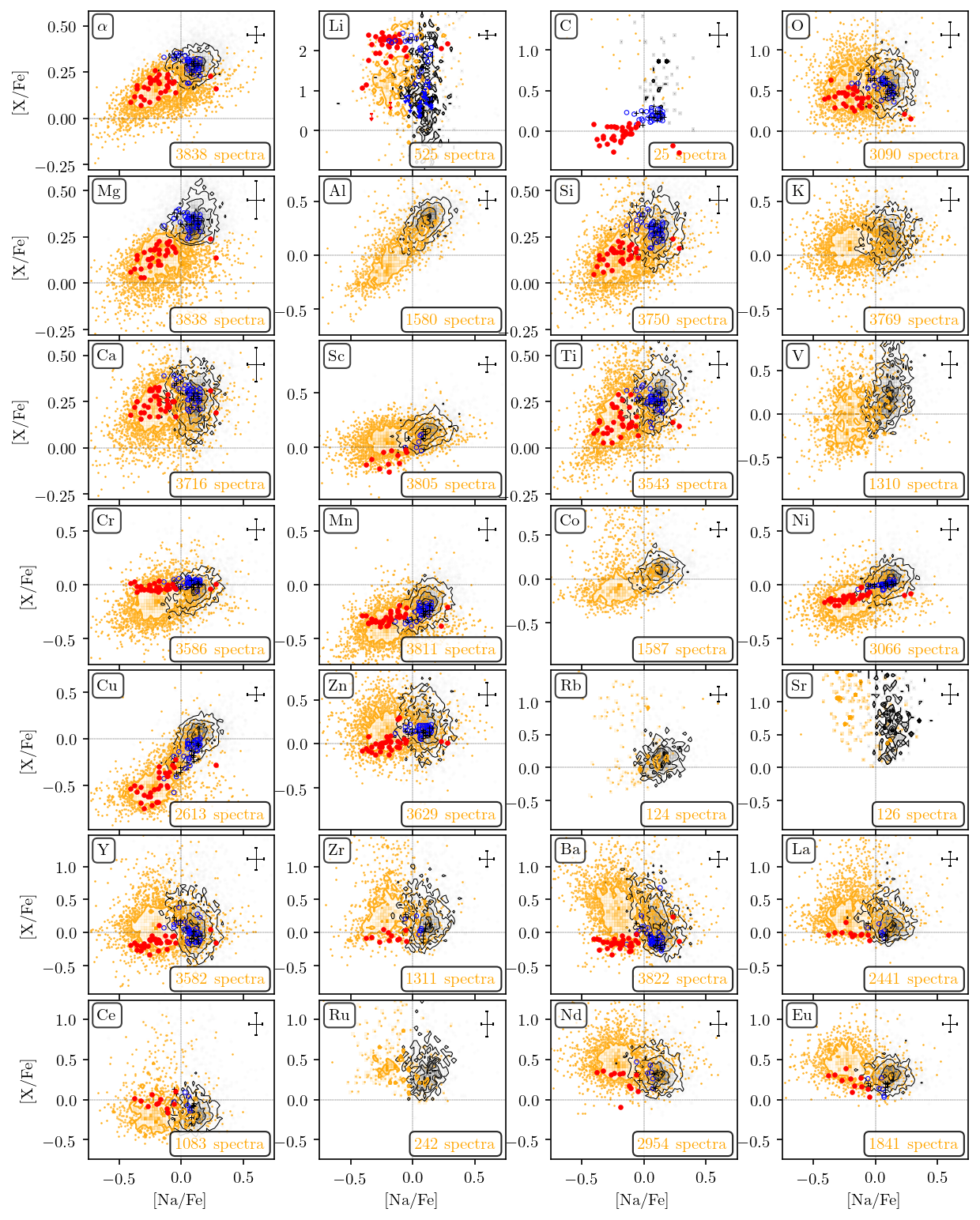}{chemical_differences}{
\textbf{Abundances [X/Fe] for the the 28 elements measured by GALAH in addition to Na and Fe, whose abundance ratio [Na/Fe] is used on the ordinate.} GALAH+ DR3 stars which are preliminary tagged to the low-$\alpha$ halo (via Eq.~\ref{eq:prelim_low_alpha_halo}) are shown in orange with numbers indicate in the bottom right. stars which are preliminary tagged to the high-$\alpha$ halo (via Eq.~\ref{eq:prelim_high_alpha_halo}) are shown in black contours.
We also show the data by \citet{Nissen2010} for $\alpha$, Na, Mg, Si, Ca, Ti, Cr, and Ni with red circles for their low-$\alpha$ halo stars, blue open circles for their high-$\alpha$ halo stars and black crossed for their thick disk stars. For the same stars of this study, we plot the data by \citet{Nissen2011} for Mn, Cu, Zn, Y, and Ba, \citet{Nissen2012} for Li (their non-LTE values with arrows for upper limits), \citet{Nissen2014} for O (their non-LTE values based on the $\lambda 7774$ \ion{O}{i} triplet), and \citet{Fishlock2017} for Sc, Zr, La, Ce, Nd, and Eu.} 

\subsubsection{Chemical differences for element groups}

To get a first impression of how significant the differences for the low- and high-$\alpha$ halo are, we follow a similar approach to \citet[][see their Fig.~5]{Nissen2011} by plotting the abundances for our preliminary low-$\alpha$ halo selection (orange) and all GALAH+ DR3 stars (greyscale) as a function of the light odd-Z element Na in Fig.~\ref{fig:nafe_xfe_nissen_all_hah}. We overplot the measurements by \citet{Nissen2010,Nissen2011,Nissen2014,Fishlock2017} for low-$\alpha$ halo (red) and high-$\alpha$ halo (blue) as well as thick disk stars (black). Although the individual figures with [Na/Fe] as their x-axis are sorted by their atomic numbers, we subsequently discuss them based on their major element group, that is 1) light elements Li and O, 2) the $\alpha$-process elements Mg, Si, Ca, and Ti, as well as their error-weighted combination noted as $\alpha$, 3) the light odd-Z elements Al and K, 4) the iron-peak elements Sc, V, Cr, Mn, Co, Ni, Cu, and Zn, 5) the s-process dominated elements Y, Zr, Ba, La, and Ce, and 6) the r-process dominated elements Nd and Eu.
\added{Both our measurements and those from the literature are assuming 1D non-LTE for Li \citet{Nissen2012}, C and O \citep{Nissen2014}. For Na, Mg, Al, Si, K, Ca, Mn, and Ba we compare our 1D non-LTE measurements with 1D LTE ones from the literature. For the other elements (Sc, Ti, V, Cr, Co, Ni, Cu, Zn, Rb, Sr, Y, Zr, La, Ce, Ru, Nd, and Eu) we compare 1D LTE measurements.}

We are looking for a way to isolate the accreted structure via its chemical signature. We aim to find those elements in Fig.~\ref{fig:nafe_xfe_nissen_all_hah}, which show both a dense concentration of accreted stars in abundance space (suggesting either a high measurement precision or a low intrinsic dispersion of the particular element in accreted stars) as well as a significant separation from the preliminary high-$\alpha$ halo as well as thick disk. In addition to this figure, we have calculated the $16^\text{th}$, $50^\text{th}$, and $84^\text{th}$ percentile for each abundance for the preliminary low-$\alpha$ and high-$\alpha$ selection and computed means $\mu_l$ and $\mu_h$, standard deviations $\sigma_l$ and $\sigma_h$ as well as skewness values $\tilde{\mu}_{l,3}$ and $\tilde{\mu}_{h,3}$ for both selections after performing 2-$\sigma$ clipping\footnote{For this, we clip the lowest and highest $2.275\%$ of the sample. Using 3-$\sigma$ clipping or no $\sigma$ clipping lead to on average 10-17\% and 12-24\% smaller significances, respectively.}. We list all values in Tab.~\ref{tab:xfe_percentiles} and include them subsequently for the assessment of the abundance differences. While these calculations allow us to quantify the distributions, we note that we do not necessarily expect the chemical enrichment in a low-mass galaxy to produce normally distributed abundances. To allow better judgement of Gaussianity beyond the calculated numbers, we append histograms for the selections in the supplementary material, again sorted by the major element groups. Readers who are not concerned with the reliability and Gaussianity of the GALAH abundances, can move on to Sec.~\ref{sec:choosing_chemical_selection}, where we choose the most promising abundances.

\input{tables/xfe_percentiles.tex} 

\paragraph*{Light elements: Li, O:}

Looking at Li in Fig.~\ref{fig:nafe_xfe_nissen_all_hah}, we do not see a significant separation of the structures, but a distribution of stars from all structures across a significant range of [Li/Fe] (with 68\% of the values between $\input{depending_text/lah_p16_p84_Li}$), which can be explained by the change of [Li/Fe] across different populations due to stellar evolutionary effects like depletion \citep[e.g.][]{Gao2020}. For O, we see that the GALAH+ DR3 data is overlapping with the data by \citet{Nissen2014}, but exhibits a larger scatter and extends to much higher [O/Fe] (with 68\% of the values between $\input{depending_text/lah_p16_p84_O}$), whereas the low-$\alpha$ halo stars by \citet{Nissen2014} only extend up to $\mathrm{[O/Fe]} \leq 0.61$. For O, especially when measured from the $\lambda 7774$ \ion{O}{i} triplet as for both our and the \citet{Nissen2014} data, 3D and non-LTE effects are known to be significant \citep{Amarsi2015, Amarsi2016b, Amarsi2019b}. Our abundance data takes into account non-LTE corrections from \citet{Amarsi2020}, and we note that \citet{Nissen2014} likewise used non-LTE corrections from \citet{Fabbian2009}. There is an extended tail towards higher [O/Fe] values (causing a slightly positive skewness of $\tilde{\mu}_{l,3} = \input{depending_text/skewness_O_lah.tex}$). This suggests an unknown error source causing spurious high abundances \citep[see discussions in][]{Buder2021}.

\paragraph*{$\alpha$-process elements: Mg, Si, S, Ca, Ti:}

For the individual $\alpha$-process elements (Fig.~\ref{fig:nafe_xfe_nissen_all_hah}), but especially for their error-weighted combination (reported as [$\alpha$/Fe] by GALAH+ DR3), we see a significantly smaller scatter than for O, that is \input{depending_text/alpha_scatter.tex}. Each of their distributions is symmetrical and agrees extremely well with the distribution of stars from \citet{Nissen2010} for $\mathrm{[Na/Fe]} < 0$. For [Mg/Fe] we see a moderately negative skewness of $\tilde{\mu}_{l,3} = \input{depending_text/skewness_Mg_lah.tex}$, which is caused by our strict linear cuts on both elements. The distribution for Ti is, contrary to those of other $\alpha$-elements, skewed towards higher values with $\tilde{\mu}_{l,3} = \input{depending_text/skewness_Ti_lah.tex}$, indicating possible issues with high [Ti/Fe] measurements (because Ti is detected in more than \input{depending_text/lah_ti_detection_percentage.tex} of the low-$\alpha$ halo).
In these panels, which have all measurements for [Mg/Fe], we notice a significant number (\input{depending_text/low_alpha_halo_nafe_0.tex} of stars preliminary selected as part of the low-$\alpha$ halo, but with $\mathrm{[Na/Fe]} > 0$. \citet{Nissen2010} found only 2 of the 38 (5\%) low-$\alpha$ halo stars in their study (G53-41 and G150-40) in this abundance space. Due to our lower precision, our sample also reaches into the super-Solar [Na/Fe] regime. \citet{Nissen2011} suggested that their two Na-enhanced stars could be halo field counterparts of the Na-enhanced globular cluster stars. While we have excluded the dedicated globular cluster observations like those of $\omega$\,Cen in our initial selection (see Eq.~\ref{eq:basic_cuts}), a follow-up of these Na-rich stars should be done in a dedicated study.

\paragraph*{Light odd-Z elements: Na, Al, K:}

Similar to Na and based on the studies by \citet{Hawkins2015} and \citet{Das2020}, we would expect Al to show a significant difference between the preliminary low- and high-$\alpha$ halo. Indeed, we see a very similar (almost 1:1 relation) between the [Na/Fe] and [Al/Fe] measurements of the low-$\alpha$ halo in Fig.~\ref{fig:nafe_xfe_nissen_all_hah}. In our sample of GALAH+ DR3, we are, however only able to estimate $\input{depending_text/lah_Al_detection_percentage.tex}$ of the Al abundances for the low-$\alpha$ halo. This is caused by the challenges involved in detecting Al lines in our spectral at the lowest [Al/Fe] in our sample, as is also indicated by the positive skewness of $\tilde{\mu}_{l,3} = \input{depending_text/skewness_Al_lah.tex}$. Contrary to this, we can measure [K/Fe] from the \ion{K}{i} resonance line for almost all stars ($\input{depending_text/lah_K_detection_percentage.tex}$). This element, however, shows only small differences in [K/Fe] between the low- and high-$\alpha$ halo.

\paragraph*{Iron-peak elements: Sc to Zn:}

For the iron-peak elements, we are able to detect Sc, Cr, Mn and Zn in more than 90\% of the sample. For Ni and Cu, the corresponding detection frequencies are $\input{depending_text/lah_Ni_detection_percentage.tex}$ and $\input{depending_text/lah_Cu_detection_percentage.tex}$. Less than half of the measurements are available for Co ($\input{depending_text/lah_Co_detection_percentage.tex}$) and V ($\input{depending_text/lah_V_detection_percentage.tex}$). Especially for the last three elements, we see that the distribution is positively skewed with $\tilde{\mu}_{l,3} = \input{depending_text/skewness_Cu_lah.tex}$, $\input{depending_text/skewness_Co_lah.tex}$, and $\input{depending_text/skewness_V_lah.tex}$ for Cu, Co, and V, respectively. For Co and V, we can explain these issues with existing measurement issues in GALAH+ DR3 \citep{Buder2021}, with large scatter for V (68\% of the values between $\input{depending_text/lah_p16_p84_V}$) and extended tails of high abundances for both V and Co. For Cu, the most likely explanation are detection limitations, as this element shows the largest difference of $\vert \mu_l - \mu_h \vert = \input{depending_text/lah_mul_muh_diff_Cu.tex}$ of all elements (except Fe) in our sample (see again Tab.~\ref{tab:xfe_percentiles}). For this element, we further see the best agreement between the distribution of abundances [X/Fe] compared to those from \citet{Nissen2010} and \citet{Nissen2011}. Their values of [X/Fe] of the low-$\alpha$ halo are typically higher and less scattered for Cr ($\mu_l \pm \sigma_l = -0.02 \pm 0.03$ compared to our $\mu_l \pm \sigma_l = \input{depending_text/lah_mul_sigmal_Cr.tex}$). For both Mn ($\mu_l \pm \sigma_l = -0.31 \pm 0.05$ compared to our $\mu_l \pm \sigma_l = \input{depending_text/lah_mul_sigmal_Mn.tex}$) and Ni ($\mu_l \pm \sigma_l = -0.10 \pm 0.05$ compared to our $\mu_l \pm \sigma_l = \input{depending_text/lah_mul_sigmal_Ni.tex}$) we find good agreement. This is especially noteworthy in the case of Mn, because the element was treated in LTE by them, but non-LTE by us with calculations based on departure coefficients by \citet{Amarsi2020}. For Zn, their values are significantly lower and less scattered than ours ($\mu_l \pm \sigma_l = 0.02 \pm 0.09$ compared to our $\mu_l \pm \sigma_l = \input{depending_text/lah_mul_sigmal_Zn.tex}$). Given that we use the same two lines with the same excitation potential, possible reasons are either the differences in our analysis ($\log(gf)$ values of $\lambda\lambda 4722, 4811$ \ion{Zn}{i}) or that the underlying selection is different.

\paragraph*{Neutron-capture elements: Rb to Eu:}

We estimate higher values than \citet{Nissen2011} for Y ($\mu_l \pm \sigma_l = -0.14 \pm 0.09$ compared to our $\mu_l \pm \sigma_l = \input{depending_text/lah_mul_sigmal_Y.tex}$). Similar to Zn, Y is estimated from two lines ($\lambda\lambda 4855,4884$ \ion{Y}{ii}) of the blue HERMES detector, the latter overlapping (but again with different $\log (gf)$ values) with $\lambda\lambda 4884,5087$ \ion{Y}{ii}, that is, the two lines used by \citet{Nissen2011}. As an effect of the high [Y/Fe] of the low-$\alpha$ halo, the difference between the means of low-$\alpha$ and high-$\alpha$ halo is only $\input{depending_text/lah_mul_muh_diff_Y.tex}$. Also for Ba, we see a significant difference between the values from \citet{Nissen2011} and our distributions ($\mu_l \pm \sigma_l = -0.16 \pm 0.09$ compared to our $\mu_l \pm \sigma_l = \input{depending_text/lah_mul_sigmal_Ba.tex}$). The scatter of our [Ba/Fe] is large, and the lack of lower [Ba/Fe] values for low $\mathrm{[Na/Fe]} \sim -0.5$ stars suggests that our values are possibly too high for the most Na-poor low-$\alpha$ halo stars. For the other three s-process elements Zr, La, and Ce, we are limited again by detectability, allowing only measurements of $\input{depending_text/lah_Zr_detection_percentage.tex}$, $\input{depending_text/lah_La_detection_percentage.tex}$, $\input{depending_text/lah_Ce_detection_percentage.tex}$ of the low-$\alpha$ halo. In general, all of the s-process elements show significant tails of high [X/Fe] values and positive skewness of $\tilde{\mu}_{l,3} = \input{depending_text/skewness_Y_lah.tex}$, $\input{depending_text/skewness_Zr_lah.tex}$, $\input{depending_text/skewness_Ba_lah.tex}$, $\input{depending_text/skewness_La_lah.tex}$, and $\input{depending_text/skewness_Ce_lah.tex}$ for Y, Zr, Ba, La, and Ce. It is noteworthy that the position of the [Ce/Fe] distribution ($\mu_l \pm \sigma_l = \input{depending_text/lah_mul_sigmal_Ce.tex}$) coincides with those of Y and Ba by \citet{Nissen2011}.

For the r-process elements we find typically positive values of [X/Fe] with $\mu_l \pm \sigma_l = \input{depending_text/lah_mul_sigmal_Nd.tex}$ and $\mu_l \pm \sigma_l = \input{depending_text/lah_mul_sigmal_Eu.tex}$ for Nd and Eu, respectively. Both are above the average values for the high-$\alpha$ halo, with mean differences $\mu_l - \mu_h$ of $\input{depending_text/lah_mul_muh_diff_Nd.tex}$ and $\input{depending_text/lah_mul_muh_diff_Eu.tex}$, respectively. This could be an effect of our measurements being close to the detection limit and possibly overestimated. The few estimates by \citet{Fishlock2017} for the low-$\alpha$ halo sample by \citet{Nissen2010} are at least always at the lower edge of our measurements.

\subsection{Choosing the most promising abundances} \label{sec:choosing_chemical_selection}

Having looked at the various elements covered by the GALAH+ DR3 data, we now consider which combination of abundances is most promising as a tool for the selection of GSE stars using GALAH+ DR3 and \Gaia eDR3. The previous research \citep{Nissen2010, Nissen2011, Ting2012, Hawkins2015, Hayes2018} provided several promising indicators for elements with significantly different nucleosynthetic sites and ejection timescales, including 2-dimensional maps of [Na/Fe] vs. [Ni/Fe] or [Al/Fe] vs. [Mg/Mn].

Based on the available abundances and their separation between the preliminary low-$\alpha$ halo from the high-$\alpha$ halo, we are now looking for the combination within GALAH+ DR3 and \Gaia eDR3 that is most promising to select as many accreted stars chemically, while avoiding significant contamination. As a guideline, we use correlation, precision, and the number of measurements to select the most promising combination from the individual elements.

Among the major element groups, we identify the $\alpha$-process elements, odd-Z elements, and iron-peak elements to have both the largest absolute distances $\mu_l - \mu_h$ between the low-$\alpha$ and high-$\alpha$ halo. Furthermore, following the arguments of \citet{Lindegren2013}, we can quantify how significant the separation between the two populations is, by taking into account their scatter within GALAH+ DR3 - caused by either their intrinsic scatter or our measurement uncertainties: 
\begin{equation}
\mu_{1,2} = \pm \frac{r \sigma}{2} \quad \rightarrow \quad r = \frac{\vert \mu_1 - \mu_2 \vert}{\sigma}.
\end{equation}
This separation significance $r$ is listed in Table~\ref{tab:xfe_percentiles}. We find the largest values ($r > 1$) for [Fe/H], [$\alpha$/Fe], [Mg/Fe], [Si/Fe], [Na/Fe], [Al/Fe], [Mn/Fe], [Ni/Fe], and [Cu/Fe]. 

To get a sense of the correlation between the individual elements, we calculate the Pearson correlation coefficients $r_P$ for the low-$\alpha$ halo stars, indicating higher correlations between \input{depending_text/pearsonr_alpha_Mg}, \input{depending_text/pearsonr_alpha_Si}, \input{depending_text/pearsonr_Na_Al}, \input{depending_text/pearsonr_Mn_Cu}, and \input{depending_text/pearsonr_Ni_Cu} but lesser correlations for \input{depending_text/pearsonr_Mg_Si} and \input{depending_text/pearsonr_Mn_Ni}. Comparing these coefficients to all other element combinations, coefficients above 0.6 appear infrequently among the preliminary low-$\alpha$ stars, that is, only for combinations of $\alpha$ with Mg or Si ($\alpha$ is computed based on Mg, Si, Ca, and Ti), Na with Al (odd-Z), Mn or Ni with Cu (iron-peak), Y with Ba as well as Zr with La with Ce (all s-process), and Nd with La/Ce/Eu (s/r-process).

We note that at higher precision, we would expect these correlation coefficients to be even larger, but at the same time would expect to see clear intrinsic differences between elements \citep[e.g.][]{Ting2021, Weinberg2021}.

Among the elements with $r > 1$, Al has the fewest measured abundances for the low-$\alpha$ halo (\input{depending_text/lah_Al_detection_percentage}), followed by Cu (\input{depending_text/lah_Cu_detection_percentage}) and Ni (\input{depending_text/lah_Ni_detection_percentage}), all other elements have close to 100\% detection rate within our selection. 

\figurecolumnwidth{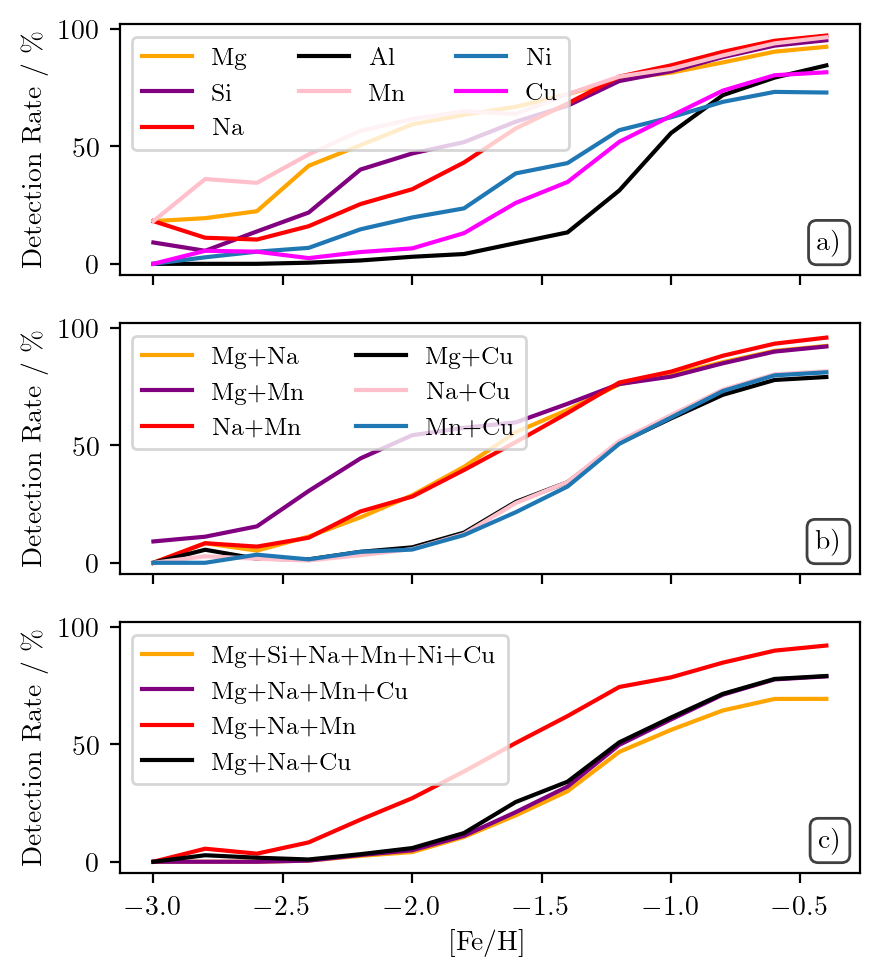}{gaussian_mixture_models}{
\textbf{Overview of completeness of the most promising elements and their combinations as a function of [Fe/H]}.
\textbf{Panel a)} for single elements.
\textbf{Panel b)} for combinations of 2 elements.
\textbf{Panel c)} for 3-6 elements.
}

We have repeated this exercise also with higher values of $v_\text{tot}$ up to $240\kms$ and a more strict limit on [Fe/H], for example only down to $-1.6\dex$. In all cases with similar results on the separation significances $r$ within 0.15, with the exception of the rarely measured element C, which increase by up to a factor of 2.

To get a different angle on the detection rate, we plot the completeness (as a function of the numbers of stars with unflagged [Fe/H] measurements) in bins of $-3.0..(0.2)..-0.4\,\mathrm{dex}$ in Fig.~\ref{fig:Completeness_Combinations} - this time for all stars and not only the preliminary low-$\alpha$ halo ones. Based on Fig.~\ref{fig:Completeness_Combinations}a, we can conclude that the detection rate for all elements decreases towards lower [Fe/H], with a significant drop below $\mathrm{[Fe/H]} \leq -1.5\dex$. We further include Al, which was previously used by \citet{Das2020} with APOGEE abundances, but is not well measured at low [Fe/H] by GALAH with less than 50\% detections below $\mathrm{[Fe/H]} < -1\dex$. For Cu and Ni the detection rate falls under 50\% below $\mathrm{[Fe/H]} < -1.4$, for Na below $\mathrm{[Fe/H]} < -1.8$, for Si below $\mathrm{[Fe/H]} < -2.0$ and for Mg as well as Mn below $\mathrm{[Fe/H]} < -2.4$.

Because of the limitations in detection and element precision within our sample, we limit ourselves to Mg (and neglect Si as well as [$\alpha$/Fe]), Na (and neglect Al), and Mn, Cu, and Ni subsequently.

\subsection{Dissecting the abundance space with Gaussian Mixture Models (GMM)} \label{sec:gaussian_mixture_models}

Assessing membership probabilities of an unknown number of underlying distributions from high-dimensional data with uncertainties is an increasingly important task in the era of large-scale surveys. Due to the complexity of the data, the selection of appropriate techniques from the plethora of methods available is non-trivial. In the case of accreted stars, both $k$-means \citep{Hayes2018, Mackereth2019} and GMMs \citep{Das2020} have been applied successfully to APOGEE data, but have not taken uncertainties of the data into account. $k$-means might suffer from inflexibilities in component shapes and lacks a probabilistic component assignment. GMMs, however, are more flexible and find a mixture of multi-dimensional Gaussian probability distributions \citep{VanderPlas2016}.

We emphasise that the aim of our study is to identify accreted stars, not to find subgroups among the accreted stars. To unravel the underlying true distribution from our noisy data to first order, we therefore can apply GMMs via Extreme Deconvolution \citep[XD,][]{Bovy2011}. In particular, we aim to use the extreme deconvolution Gaussian mixture modelling code \textsc{xdgmm} by \citet{Holoien2017}. We optimise the model likelihood using the iterative Expectation-Maximization algorithm \citep{Dempster1977} embedded in \textsc{xdgmm}'s implementation of \textsc{astroml} \citep{astroml}.

Our input to \textsc{xdgmm} is a matrix of features and their uncertainty matrix for $n$ stars. Features are different combinations of the 6 most promising elements Mg, Si, Na, Mn, Cu, and Ni, in different notations and combinations, such as [Mg/Na] or [Cu/Fe], and their uncertainties. Because \textsc{xdgmm}s are computationally expensive, we first use simple GMMs, that is, \textsc{scikit-learn}'s \textsc{GaussianMixture} \citep{scikit-learn}, to explore which combination of measurements is most promising. We discuss the possible combinations of measurements to features subsequently.

Both for simple GMMs and \textsc{xdgmm}, we estimate how many model components are preferable by using the Bayesian information criterion \citep[BIC,][]{Schwarz1978}, defined as 
\begin{equation}
\text{BIC} = \ln (n) k - 2 \log \mathcal{L},
\end{equation}
with $n$ being the number of stars/observations of $k$ components of the GMM yielding a maximised likelihood function $\mathcal{L}$. We test up to $k \leq 30$ components for the simple GMMs and $k \leq 10$ for the \textsc{xdgmm}. We select the model with the lowest BIC as the best one.

\subsubsection{Assessing abundance combinations with simple GMMs} \label{sec:sample_gmm}

There are several ways in which features from the 6 most promising elements Mg, Si, Na, Mn, Cu, and Ni can be used in order to assess which stars are most likely accreted: 1) feed all of them as individual features, 2) combine some element abundances either via their ratio or their sum, such as [Mg/Na] or [Mn+Cu/Fe] or 3) select only a subset to fit. Given our limited measurement precision, we try to find combinations with the clearest separations and Gaussian-like shape.

Due to the selection function of GALAH, the data of GALAH+ DR3 is dominated by observations of the low-$\alpha$ disk, which is not the focus of this study. Including these stars in a GMM would shift the focus of the algorithm away from the typically metal-poor accreted stars and we therefore implement an initial cut on the iron abundance of $\mathrm{[Fe/H] < -0.6}$. This does not affect the low-$\alpha$ halo stars and still leaves a significant part of the high-$\alpha$ halo as can be seen from the percentiles of [Fe/H] in Tab.~\ref{tab:xfe_percentiles}.

We plot the detection rates of the promising elements for low metallicities in Fig.~\ref{fig:Completeness_Combinations}a, showing a clear difference in the detectability of these elements towards the metal-poor regime. We are now concerned with combinations of them. In Fig.~\ref{fig:Completeness_Combinations}b we plot combinations of different pairs of groups. We see that the detection rate of Mg+Mn is similar to that of the less well measured Mg, and Mg+Cu is similar to that of Cu. In Fig.~\ref{fig:Completeness_Combinations}c we plot the most numerous combinations of 3 (Mg+Na+Mn), 4 (Mg+Na+Mn+Cu), and 5 (Mg+Si+Na+Mn+Ni+Cu) elements. The combination of Mg+Mn is the one with the highest detection rate, followed by that of Mg+Na+Mn (which is similar to Mg+Na and Na+Mn), followed by that of Mg+Na+Mn+Cu (similar to any combination of these elements with Cu), followed by the combination of all 6 elements.

\figuretextwidth{17cm}{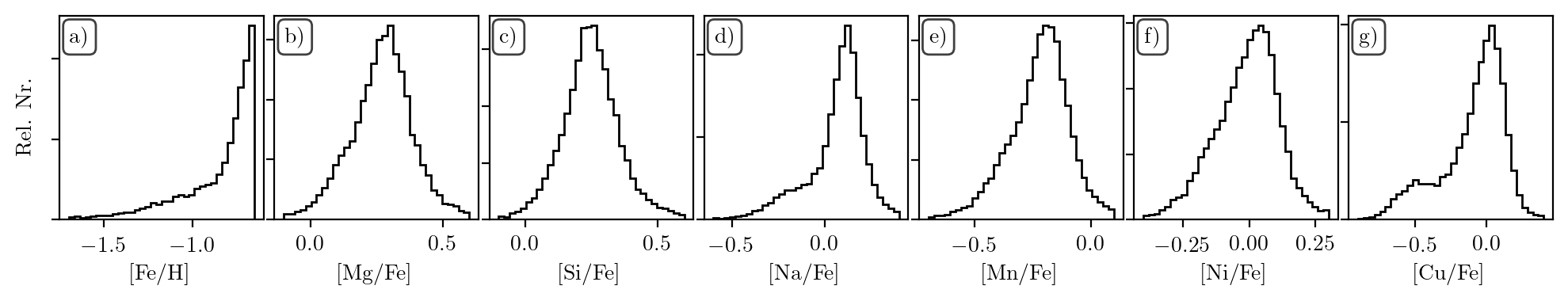}{gaussian_mixture_models}{
\textbf{Histograms of [Fe/H], [Mg/Fe], [Si/Fe], [Na/Fe], [Mn/Fe], [Ni/Fe], and [Cu/Fe] for stars with $\mathrm{[Fe/H]} < -0.6$ which passed the basic quality cuts (Eq.~\ref{eq:basic_cuts})}.
Only stars with unflagged measurements for all these elements are shown. Extensive \textsc{corner} plots are provided in the supplementary material.
}

To get a first impression of possible 2D combinations of the 6 most promising elements, we inspect the \textsc{corner} plot \citep{corner} both in abundance space as well as the difference with respect to the $50^\text{th}$ percentile (a robust representative of the high-$\alpha$ halo stars) in an uncertainty weighted version to identify again significant differences but this time in 2D space\footnote{Among the metal-poor stars, accreted stars stand out most significantly in Fe, Na, and Cu (3 $\sigma$ or more) and less in Mg, Si, Mn, and Ni (around 2 $\sigma$).}. Here, we only show the histograms in Fig.~\ref{fig:hist_high_separation_elements} and provide the corner plots in the supplementary material\footnote{In these 2D-density plots, accreted features are located in the bottom left.}. Looking at the histograms, we see clear double-peak structures for [Na/Fe] and even resolved for [Cu/Fe]. Asymmetries are visible for [Fe/H], [Mg/Fe], [Mn/Fe], and [Ni/Fe]. [Si/Fe] shows no clear asymmetry in the histogram.

For each of the combinations listed in Tab.~\ref{tab:sample_gmm}, we have fitted simple GMMs from \textsc{scikit-learn}'s \textsc{GaussianMixture} between 3 and 30 components. Because of the cut we employed in [Fe/H] as well as the complex, non-Gaussian, structure with respect to [Fe/H], it is not reasonable to include [Fe/H] itself as an input label. We use it, however, later-on as a label to assess the components. To limit high-confidence outliers, we have applied further cuts to the data via limits on the uncertainties ($\texttt{e\_X\_fe} < 0.25\dex$) as well as boundaries for the abundances ($ \mathrm{[Fe/H]} < -0.5 $, $-0.3 < \mathrm{[Mg/Fe]} < 0.7$,$ -0.3 < \mathrm{[Si/Fe]} < 0.7$, $ -0.7 < \mathrm{[Na/Fe]} < 0.7$, $ -0.3 < \mathrm{[Mn/Fe]} < 0.25$, $ -0.7 < \mathrm{[Ni/Fe]} < 0.25$, and $ -0.3 < \mathrm{[Cu/Fe]} < 0.7$) in addition to the basic cuts and abundance flags. We plot the distribution of BICs (normalised to the lowest BIC) in Fig.~\ref{fig:bic_stats}. All combinations are best recovered with simple GMMs with fewer than 10 components.

\figurecolumnwidth{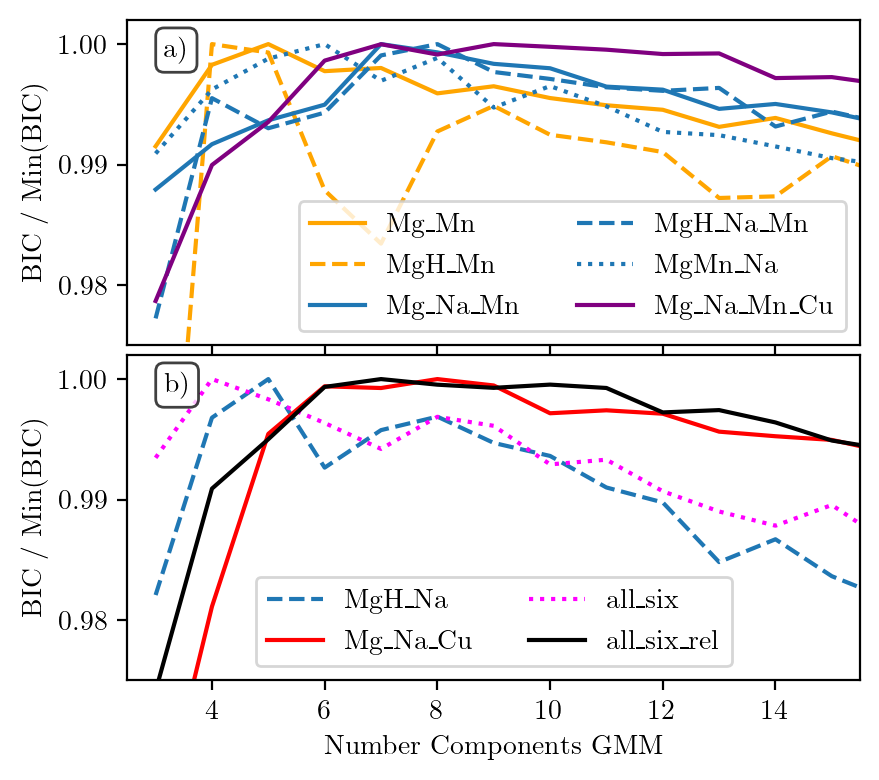}{gaussian_mixture_models}{
\textbf{Bayesian information criterions (BIC) normalised to the lowest value per realisation) for different simple Gaussian Mixture Models.} With our normalisation, better, that is, more negative BIC values result in higher normalised BIC. The GMMs are indicated in the legend and listed in Tab.~\ref{tab:sample_gmm}. Normalised BIC values for more than 15 components continuously fall.
}

\input{tables/gmm_sampling.tex} 

\input{tables/simple_gmm_selection.tex} 

\figuretextwidth{17cm}{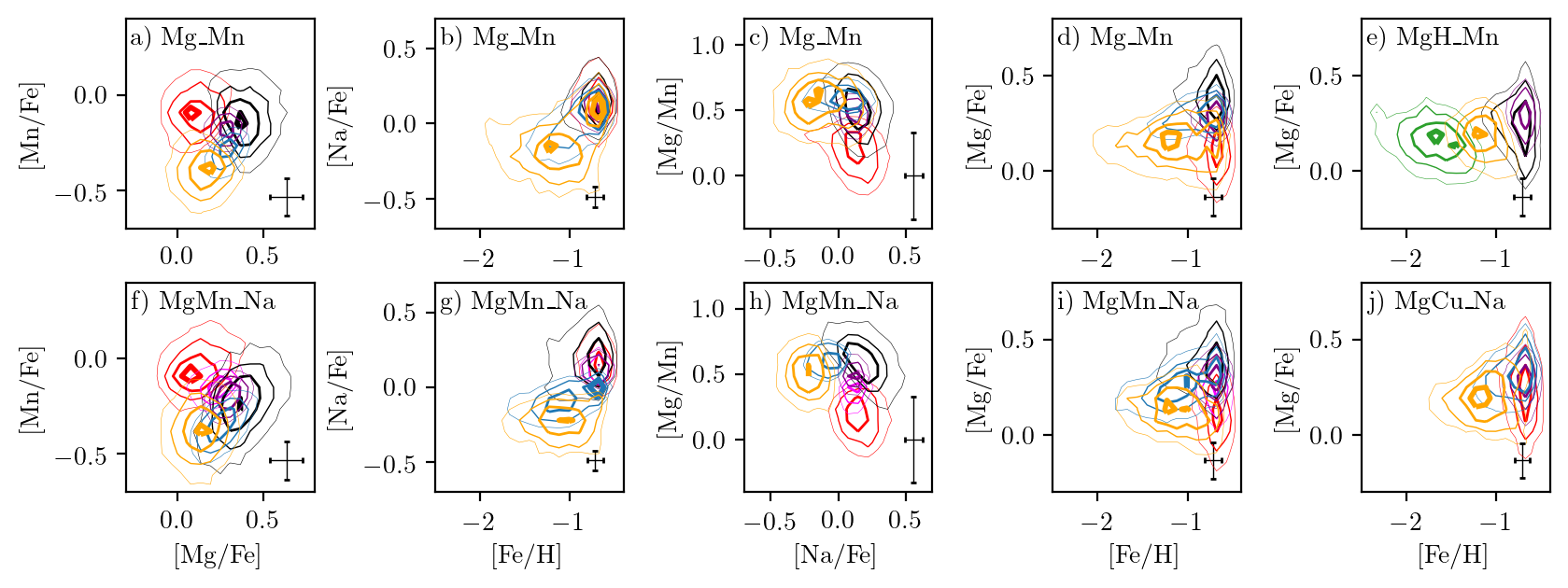}{gaussian_mixture_models}{
\textbf{Overview of input planes for the simple Gaussian Mixture Models.}
\textbf{Coloured densities} indicate probability-weighted distributions of the individual components. We colour similar components of different GMMs with similar colours (see text for details), but stress that the colours of the columns are independent of each other.
\textbf{Panel a)} shows [Mg/Fe] vs. [Mn/Fe] for the GMM \texttt{Mg\_Mn} (used as input plane).
\textbf{Panel b)} [Fe/H] vs. [Na/Fe] for the GMM \texttt{Mg\_Mn}, showing the orange component also extending towards super-Solar [Na/Fe].
\textbf{Panel c)} [Na/Fe] vs. [Mg/Mn] for the GMM \texttt{Mg\_Mn}.
\textbf{Panel d)} [Fe/H] vs. [Mg/Fe] for the GMM \texttt{Mg\_Mn}, showing the orange component overlapping with the red component.
\textbf{Panel e)} [Fe/H] vs. [Mg/Fe] for the GMM \texttt{MgH\_Mn}, showing the accreted stars fitted with two components.
\textbf{Panel f)} shows [Mg/Fe] vs. [Mn/Fe] for the GMM \texttt{MgMn\_Na}.
\textbf{Panel g)} [Fe/H] vs. [Na/Fe] for the GMM \texttt{MgMn\_Na}, showing a clear separation of the orange component from those with super-Solar [Na/Fe] via an intermediate blue component.
\textbf{Panel h)} [Na/Fe] vs. [Mg/Mn] for the GMM \texttt{MgMn\_Na} (used as input plane).
\textbf{Panel i)} [Fe/H] vs. [Mg/Fe] for the GMM \texttt{MgMn\_Na}, showing the orange component separated from the red component.
\textbf{Panel h)} [Fe/H] vs. [Mg/Fe] for the GMM \texttt{MgCu\_Na}.
We only plot data with posterior probabilities above 0.25 for the individual components.
}

In addition to testing different element combinations, we also explore the influence of the abundance notation. For example, we test both the use of [Mg/Fe] vs. [Mn/Fe] as well as [Mg/H] vs. [Mn/Fe]. The latter is motivated by the findings by \citet{Feuillet2021} who separated accreted stars in the [Mg/H] vs. [Al/Fe] plane \citep[see also][]{Ting2012}. They found [Mg/H] to be a cleaner tracer, as [Mg/Fe] is influenced by the onset of SNIa Fe contributions. Additionally, we fit a combination of ratios of nucleosynthesis pathway tracers. We use both [Mg/Mn] and [Mg/Cu], which are likely tracing SNe II contributions from massive stars and SNIa of low mass stars \citep{Kobayashi2020}. We further test the use of [Mg/Fe], [Na/Fe], [Mn/Fe], and [Cu/Fe] as input, as we expect differences for Na, Mn, and Cu because of the metallicity-dependence of hypernovae \citep{Kobayashi2006,Kobayashi2020}. Finally, we also test the combination of all six elements with 6 dimensions, as well as with a reduced dimensionality through [Mg/Cu], [Si/Cu], [Ni/Fe], and [Na/Fe].

Whilst we fit the GMMs to the data points without uncertainties, we take uncertainties into account when predicting membership probabilities via Monte Carlo sampling. For each datapoint, we sample the input abundances 1000 times with means and standard deviations from \texttt{X\_fe} and \texttt{e\_X\_fe} and calculate a mean membership posterior probability for the components. For our simple GMM plots, we require a probability of at least 0.25 and use the probability as weight for the density plots. We list the probabilities for the most important components (for this study) in Tab.~\ref{tab:simple_gmm_selection}. Reported percentiles of distributions are weighted by these probabilities.

We start our exploration with a simple input of [Mg/Fe] and [Mn/Fe] (\texttt{Mg\_Mn}) and recover the best result with 5 GMM components. These are shown in the top left panel of Fig.~\ref{fig:best_gmm_samplings_selection} via density contours. By inspecting the position of the components in this abundance plane, we can identify the 5 components and subsequently trace similar groups in the other projections\added{(including the action space provided in the supplementary material)} as the following:
\begin{enumerate}
\item Red \& Magenta -- low-$\alpha$ disk
\item Black \& Purple \& Rose -- high-$\alpha$ disk/halo
\item Blue -- metal-poor intermediate-$\alpha$; not clearly accreted nor high-$\alpha$ disk/halo (MP-i$\alpha$)
\item Orange -- mainly accreted stars
 \item Green -- mainly accreted stars ([Mg/H]-poor $< -1.3$)
\end{enumerate}

\figuretextwidth{17cm}{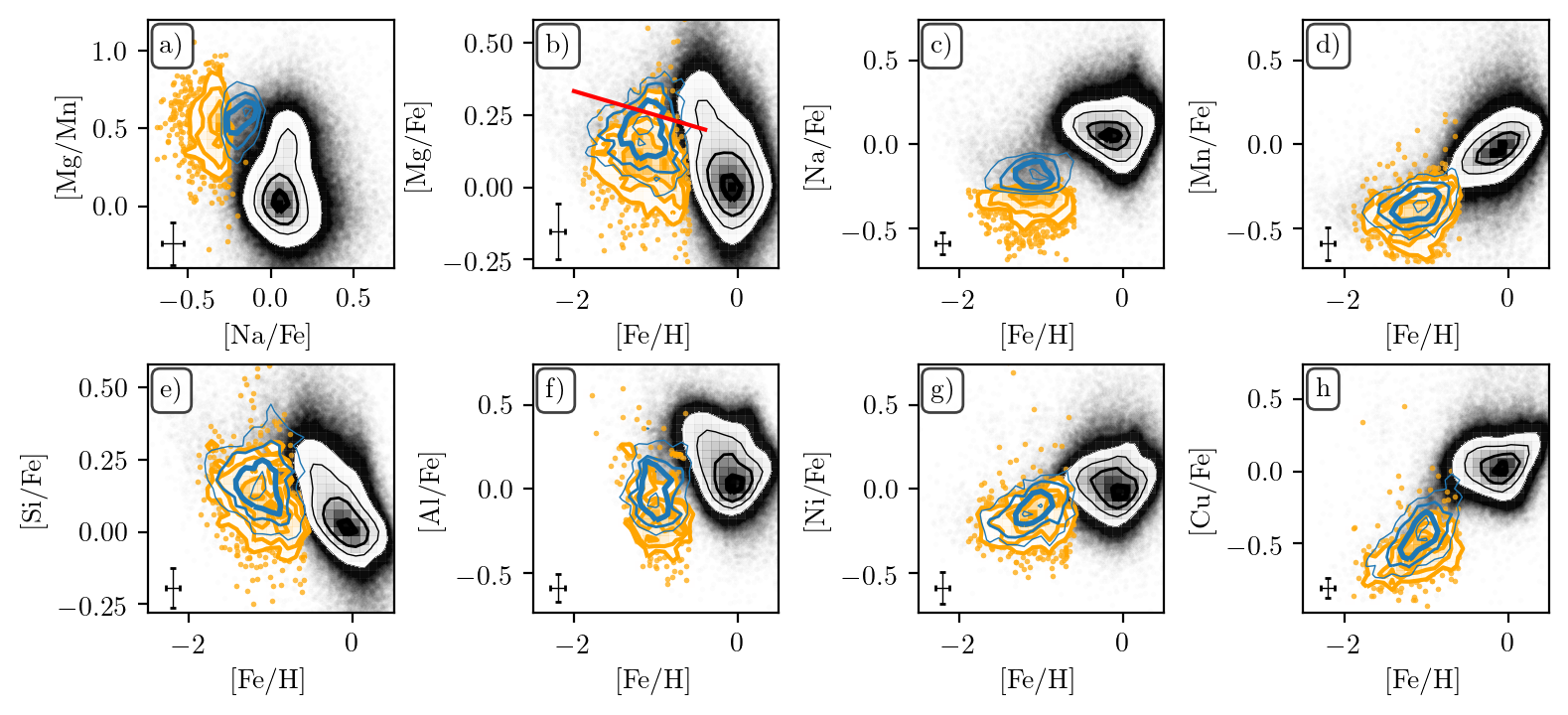}{chronochemodynamic_comparison}{
\textbf{Overview of two metal-poor components of the \textsc{xdgmm} in abundance planes that were identified as those with the highest separation significance in Sec~\ref{sec:choosing_chemical_selection}.}
Orange indicates the accreted component (with sub-Solar [Na/Fe]).
Blue indicates the in-situ component (with higher [Na/Fe]).
The red line in panel b) indicates the selection between low- and high-$\alpha$ halo suggested by \citet{Nissen2010}.
Only stars with probabilities above 0.45 for each component are shown, as suggested by the overlap analysis of Sec.~\ref{sec:overlap_planes}.
}

Stars of the red component (Fig.~\ref{fig:best_gmm_samplings_selection}a-d) have values closest to Solar [Mg/Fe] and [Mn/Fe] and are mostly [Fe/H]-rich stars in the sample. Stars of the black/purple/rose component also have the highest [Fe/H] values in the sample, but also the highest [Mg/Fe] ones, making them likely high-$\alpha$ disk/halo stars, with a possible contamination by low-$\alpha$ disk stars. Stars of the blue component differ from the black/purple/rose ones, because they have lower [Mn/Fe] values. These values are, however, not as low as those of the orange component, which is consistent with accreted stars, based on our intuition of the chemical composition of low-$\alpha$ halo stars. We are later concerned with the distribution of the individual abundances. Here we are interested to identify which abundances and abundance planes are needed to identify accreted stars. Especially for the orange component of \texttt{Mg\_Mn}, we notice a contamination from stars with Solar [Na/Fe] (Fig.~\ref{fig:best_gmm_samplings_selection}b), broadening the distribution to \input{depending_text/Mg_Mn_dist1_NaFe}.

Before adding [Na/Fe] as input to resolve this issue, we assess a slightly different input of [Mg/H] and [Mn/Fe] (Fig.~\ref{fig:best_gmm_samplings_selection}e). [Mg/H] is a purer tracer of SNII contributions \citep{Kobayashi2020, Feuillet2021}. We see that in the projections, the models are giving more weight to the [Mg/H] poor stars, and model them with two components - an [Mg/H]-poor (dark-orange around \input{depending_text/MgH_Mn_dist2_MgH}) and [Mg/H]-richer one (orange around \input{depending_text/MgH_Mn_dist1_MgH.tex}). Interestingly, both exhibit very similar [Mg/Fe] distributions with \input{depending_text/MgH_Mn_dist2_MgFe} and \input{depending_text/MgH_Mn_dist1_MgFe}, respectively. Further, the orange component is now slightly more confined to sub-Solar \input{depending_text/MgH_Mn_dist1_NaFe}. The GMM fails, however, to tell apart low- from high-$\alpha$ disk stars, which are modelled with two extended components with similar means.

When adding [Na/Fe] to the GMM, the models need typically between 6 and 8 components to fit the data well. We have tested different combinations of the 3 abundances as input (we attach a figure for the other GMMs similar to Fig.~\ref{fig:best_gmm_samplings_selection} in the supplementary material for a complete overview). They all include a component similar to the orange one from \texttt{Mg\_Mn}, but are not contaminated with Solar [Na/Fe] stars. This leads to a clearer separation between the accreted component and the other components, especially in [Na/Fe], with one intermediate (blue) component between them (compare Figs.~\ref{fig:best_gmm_samplings_selection}b and g). 

Inspired by the argument discussed in \citet{Hawkins2015}, we also test the abundance ratio [Mg/Mn]. Such ratios are excellent tracers of orthogonal nucleosynthesis pathways \citep[e.g.][]{Ting2012, Ting2015}. In this particular case, Mg is primarily produced by SNII \citep{Nomoto2013} and Mn by SNIa \citep{Gratton1989}. This idea was already applied by \citet{Das2020} for APOGEE data. They used [Mg/Mn] paired with [Al/Fe], the latter tracing SNII contributions while being sensitive to the progenitor C and N abundances. For GALAH+ DR3 data, however, we again turn to [Na/Fe] instead of [Al/Fe] due to the higher detection rate for the GMM \texttt{MgMn\_Na}. Similar to \texttt{Mg\_Na\_Mn}, we find an accreted component (orange) that is separated by the typical disk components through an intermediate component (blue). Both orange and blue components share similar [Mn/Fe] (\input{depending_text/MgMn_Na_dist1_Mn} and \input{depending_text/MgMn_Na_dist2_Mn} for orange and blue components respectively), but differ in their [Mg/Fe] and thus [Mg/Mn] values.

We further test adding the iron-peak element Cu to the GMM, both instead (\texttt{MgCu\_Na}) and in addition to Mn (\texttt{Mg\_Na\_Mn\_Cu}), but do not find more promising component separations than without Cu. In particular, the distribution of the accreted component (orange) is very similar to those of the other GMMs, but includes less stars due to the detection limit of Cu. We have further tested GMMs using all 6 elements Mg, Si, Na, Mn, Ni, and Cu with different input combinations - without any improvement (see online material).

Given the decreasing number of stars available for an increasing number of abundances used for the GMM, we decide to continue hereafter with \texttt{MgMn\_Na}. Although we already achieve remarkable separations only with \texttt{Mg\_Mn}, we are concerned by the contamination of stars with super-Solar [Na/Fe] for the latter GMM. The latter GMM would be promising, if for each star, a limit $\mathrm{[Na/Fe]} \ngtr 0$ could be estimated. As the simple GMMs do not take into account uncertainties, when fitting the components, we now use the input of [Na/Fe] and [Mg/Mn] with their uncertainties for the \textsc{xdgmm}.

\subsubsection{XDGMM with [Na/Fe] vs. [Mg/Mn]} \label{sec:xdgmm_MgMn_Na} \label{sec:xdgmm_selection}

For our final selection of accreted stars within the chemical planes, we apply the \textsc{xdgmm} introduced at the beginning of this section. We use the abundance plane of [Na/Fe] vs. [Mg/Mn], which we identify as the most promising one in terms of separation significance of elements (see Sec.~\ref{sec:choosing_chemical_selection}), detection rate towards low iron abundances (see Fig.~\ref{fig:Completeness_Combinations}) as well as our test of the possible abundance planes with simple GMMs in Sec.~\ref{sec:sample_gmm}.

We tested up to 10 Gaussian components and find the lowest BIC value for 5 components. Among these, we recover the component with low [Na/Fe] and high [Mg/Mn] values, that is, the accreted component. We plot the abundance overview of this component with orange contours in [Na/Fe] vs. [Mg/Mn] as well as the seven elements vs. [Fe/H] with the highest separation significance in Fig.~\ref{fig:nafe_mgmn_overview}. We further identify a component overlapping with the accreted component (plotted with blue contours in Fig.~\ref{fig:nafe_mgmn_overview}), which shows on average higher [Mg/Fe], [Na/Fe], [Al/Fe], and [Cu/Fe] values. As we calculate a probability of each source to belong to a component, we test which probability threshold to use subsequently (in Sec.~\ref{sec:overlap_planes}) and discuss the reliability of our selection of accreted stars in Sec.~\ref{sec:reliability_selection}.

\subsection{Dynamical selection of GSE stars for this study} \label{sec:dynamical_selection}

For the dynamical selection of accreted stars, and especially GSE stars, we resort to the literature, as reviewed in Sec.~\ref{sec:selection_techniques} and listed in Tab.~\ref{tab:selection_techniques}. Here we limit ourselves to the dynamical selection by \citet{Feuillet2021}, as this was shown to be least contaminated. Hereafter, we refer to the dynamical selection as the sample of stars that passes the basic cuts (Eq.~\ref{eq:basic_cuts}) and have angular momenta $-500 < L_Z < 500 \kpckms$ as well as radial actions $30 < \sqrt{J_R / \kpckms} < 55$, as suggested by \citet{Feuillet2021}. We plot the distribution of GALAH+ DR3 stars within the $L_Z$ vs. $\sqrt{J_R}$ plane in Fig.~\ref{fig:chemdyn_selection_plane}f in black and the clean selection box by \citet{Feuillet2021} with a red dashed rectangle. The stars of GALAH+ DR3 within this box are then shown in a red density contour plot. The majority of stars are located at the lower edge of the box, indicating that more stars would be selected with a lower threshold of $J_R$. Subsequently, we assess the overlap (and non-overlap) of the dynamical selection with our chemical one.

\section{Chrono-chemodynamic properties of the chemically and dynamically selected accreted stars} \label{sec:chronochemodynamics}

In this section, we compare a variety of properties of the chemically and dynamically selected accreted stars, including the metallicity distribution function, abundance distributions, dynamical properties, and stellar ages. Hereafter, our chemical selection refers to the orange selections in [Na/Fe] vs. [Mg/Mn] space, and the dynamical selection refers to the red selection in $L_Z$ vs. $\sqrt{J_R}$ space (see previous Section and Fig.~\ref{fig:chemdyn_selection_plane}). We begin by assessing the overlap of the different selections, both in numbers and in their respective selection planes and then extend the comparison to the other properties.

\subsection{Selection overlap} \label{sec:overlap_planes}

\figurecolumnwidth{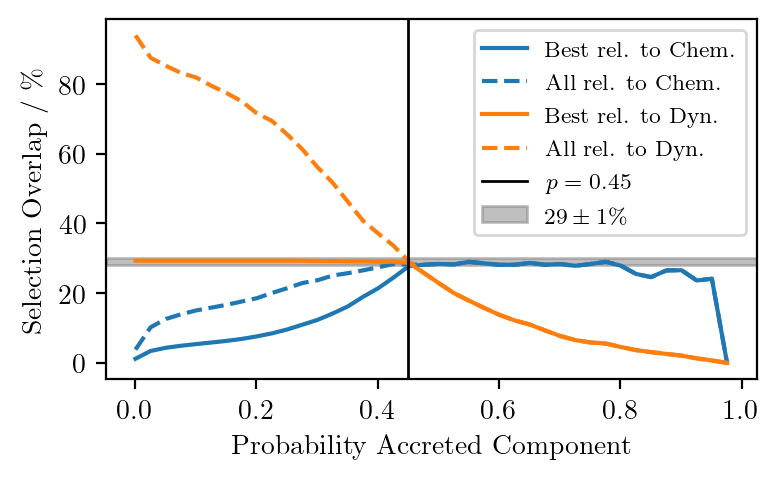}{chronochemodynamic_comparison}{
\textbf{Percentage of overlap between the chemically and dynamically selected stars as a function of the membership probability of stars belonging to the GMM component of accreted stars.} Lines indicate the percentage as a function of all chemically selected stars (blue) and all dynamically selected stars (orange). Solid lines require that the accreted component is the one with the highest ('best') probability. The grey area indicates an overlap of $(29\pm1)\%$, where the overlap plateaus with respect to the chemical selection. The black solid line indicates a normalised probability of 0.45, the location where both lines meet and where the overlap as a function of chemical selection does not increase for larger probabilities.
}

\figuretextwidth{17cm}{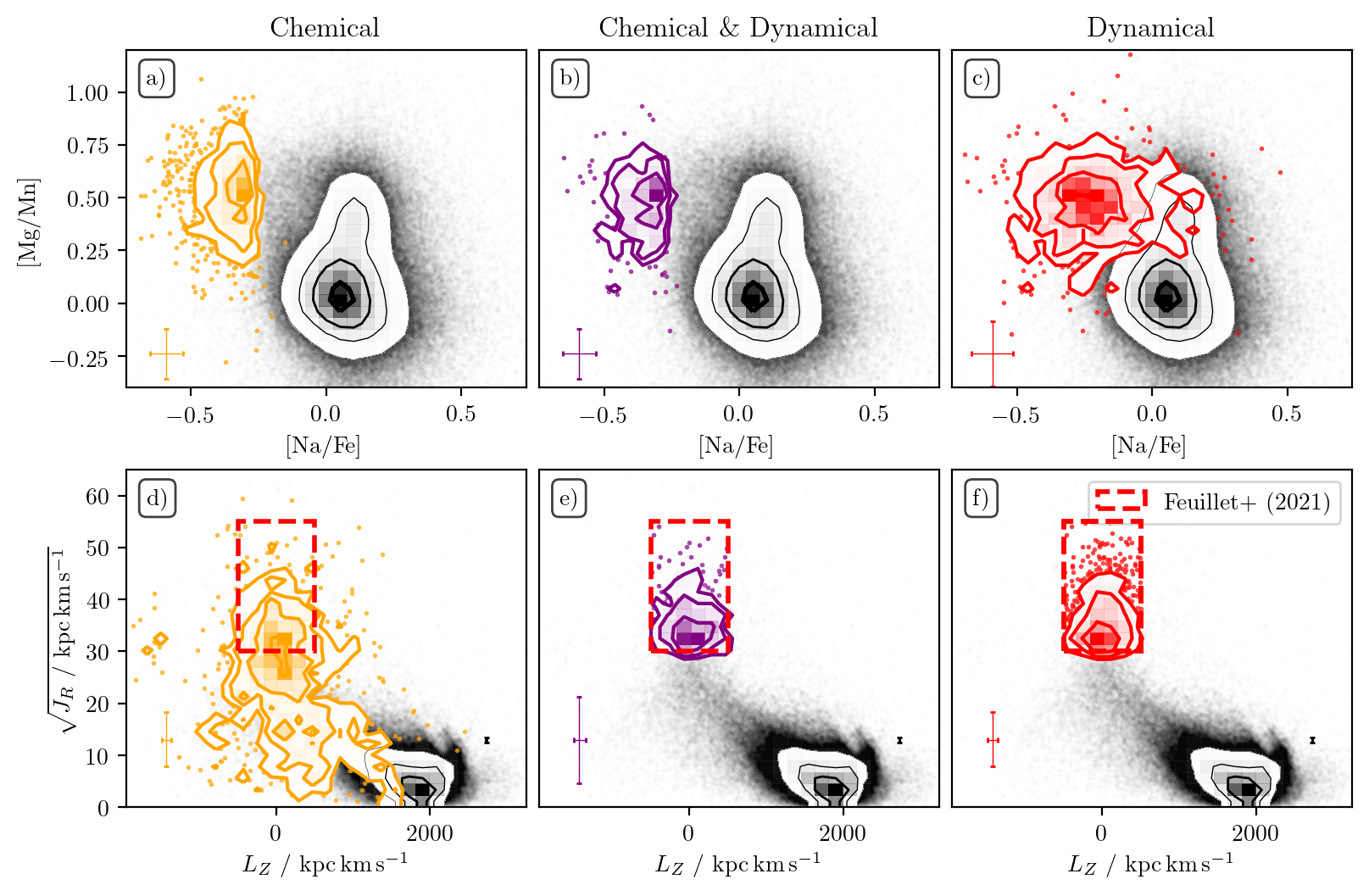}{chronochemodynamic_comparison}{
\textbf{Comparison of chemical and dynamical selections in their respective planes, [Na/Fe] vs. [Mg/Mn] (top panels) and $L_Z$ vs. $\sqrt{J_R}$, respectively.}
\textbf{Left panels (a and d):} Chemical selection (orange).
\textbf{Middle panels (b and e):} Overlap of chemical and dynamical selection (purple).
\textbf{Right panels (c and f):} Dynamical selection (red).
Black background contours show the GALAH+ DR3 sample.
The red dashed box indicates the clean selection of GSE stars by \citet{Feuillet2021}.
}

\input{tables/chemodyn_comparison} 

As we have computed the probabilities that the stars in GALAH+ DR3 belong to the accreted Gaussian component, the overlap with the dynamical selection box varies significantly as a function of this probability. We plot the selection overlap as a function of the probability of stars belonging to the accreted component in our chemical GMM in Fig.~\ref{fig:quantitative_overlap_chemdyn} in blue relative to the chemical selection and in orange relative to the dynamical selection (for which unflagged measurements of Mg, Na, and Mn were available). We further differentiate between the probability of a star belonging to the accreted component (solid lines) and this probability also being the highest among all possible components (dashed lines). Both are naturally the same above 0.5, but here we are also interested in studying the contamination by other components, for example when a star is equally likely or more likely to belong to another component. 

We find that the overlap reaches a plateau at $(29\pm1)\%$ above a probability of 0.45. That implies that, below this probability, our chemical selection might be either contaminated or that we are selecting from a accreted structure that is different to the dynamically selected (clean) one. Looking at the overlap with respect to the available stars within the dynamical selection, we see a decrease as we restrict our chemical selection towards a more and more reliable selection (higher probability). It falls below $(29\pm1)\%$ around a probability of 0.45. Under the assumption that we are selecting from the same population, we therefore find the best compromise between contamination and sample size for a probability of 0.45 or higher of stars belonging to the accreted (orange) model component and use this hereafter as a threshold for our chemical selection. In Tab.~\ref{tab:xdgmm_dynamical_selection}, we list the stars and their probability of belonging to the accreted component as well as the stars selected via the dynamical selection from \citet{Feuillet2021}. \added{We note that sampling the orbital parameters within their uncertainties and keeping only those stars that lie often enough (more than 50 or even 90\% of the time) within the dynamical selection box would reject roughly 90\% of likely true GSE stars and in particular those with low $\sqrt{J_R}$, but remove only roughly 10\% of false positives. We thus decided not to take this approach.}

\input{tables/selected_stars} 

With the maximum selection overlap of $(29\pm1)\%$ in mind, we now consider which of the selection planes show agreement and disagreement between the samples. In Fig.~\ref{fig:chemdyn_selection_plane}, we plot the chemical selection plane in the top panels and the dynamical selection plane in the bottom panels.

Comparing Fig.~\ref{fig:chemdyn_selection_plane}a and c, and with the help of the percentiles listed in Tab.~\ref{tab:chronochemodynamic_properties} for each selection, it becomes clear that the chemical selection exhibits lower [Na/Fe] values in a tighter distribution (\input{depending_text/chem_percentiles_Na_fe}) than the dynamical one (\input{depending_text/dyn_percentiles_Na_fe}). The distributions of [Mg/Mn] values, however, agree fairly well (\input{depending_text/chem_percentiles_MgMn_fe} vs. \input{depending_text/dyn_percentiles_MgMn_fe}). We discuss this disagreement in Sec.~\ref{sec:prospects_chem_tagg}, as it hints towards a limitation of our chemical selection to tell apart accreted stars from in-situ stars (see [Na/Fe] in Fig.~\ref{fig:nafe_mgmn_overview}, where the metal-poor in-situ component is located around $\mathrm{[Na/Fe]} > 0 \dex$).

In Fig.~\ref{fig:chemdyn_selection_plane}d we clearly see that the actions of stars from the chemical selection (orange contours) extend far outside of the clean dynamical selection (red dashed rectangle), that is $\sqrt{J_R / \kpckms}$ of \input{depending_text/chem_percentiles_sqrt_J_R} compared to the \input{depending_text/dyn_percentiles_sqrt_J_R}, which have to be within the clean selected box with $30 < \sqrt{J_R / \kpckms} < 55$. This selection was chosen by \citet{Feuillet2021} in order to avoid contamination by the high-$\alpha$ disk, whose dynamically hot tail extends towards these high radial actions \citep[e.g.][]{Feuillet2020, Das2020}. Both of our distributions in angular momentum $L_Z$ (\input{depending_text/chem_percentiles_L_Z}$\kpckms$ compared to the \input{depending_text/dyn_percentiles_L_Z}$\kpckms$) agree at high radial actions. The on average slightly prograde orbits of the chemical selection are mainly caused by the stars with low $J_R$ and positive $L_Z$. Based on the density contours in Fig.~\ref{fig:chemdyn_selection_plane}d, we see that these stars are however only a minority and their numbers drop significantly below $\sqrt{J_R / \kpckms} < 20$, that is \input{depending_text/chem_below_20_20}, \input{depending_text/chem_below_15_15}, and \input{depending_text/chem_below_10_10} below $\sqrt{J_R / \kpckms}$ of 20, 15, and 10. We also note that \input{depending_text/chem_above_500} of the chemically selected stars exceed the upper limit of $L_Z \sim 500\kpckms$ set by \citet{Feuillet2021}. \input{depending_text/chem_above_1000} even exceed $L_Z > 1000\kpckms$, that is, roughly half of the Sun's angular momentum. We find similar values for the sample by \citet{Das2020}, with average values of $\sqrt{J_R \kpckms} = 30_{-10}^{+11}$ and $L_Z = -10_{-590}^{+612} \kpckms$
in the same notation, and 19\% of the accreted stars in their sample exhibiting $L_Z > 500\kpckms$.

\figuretextwidth{17cm}{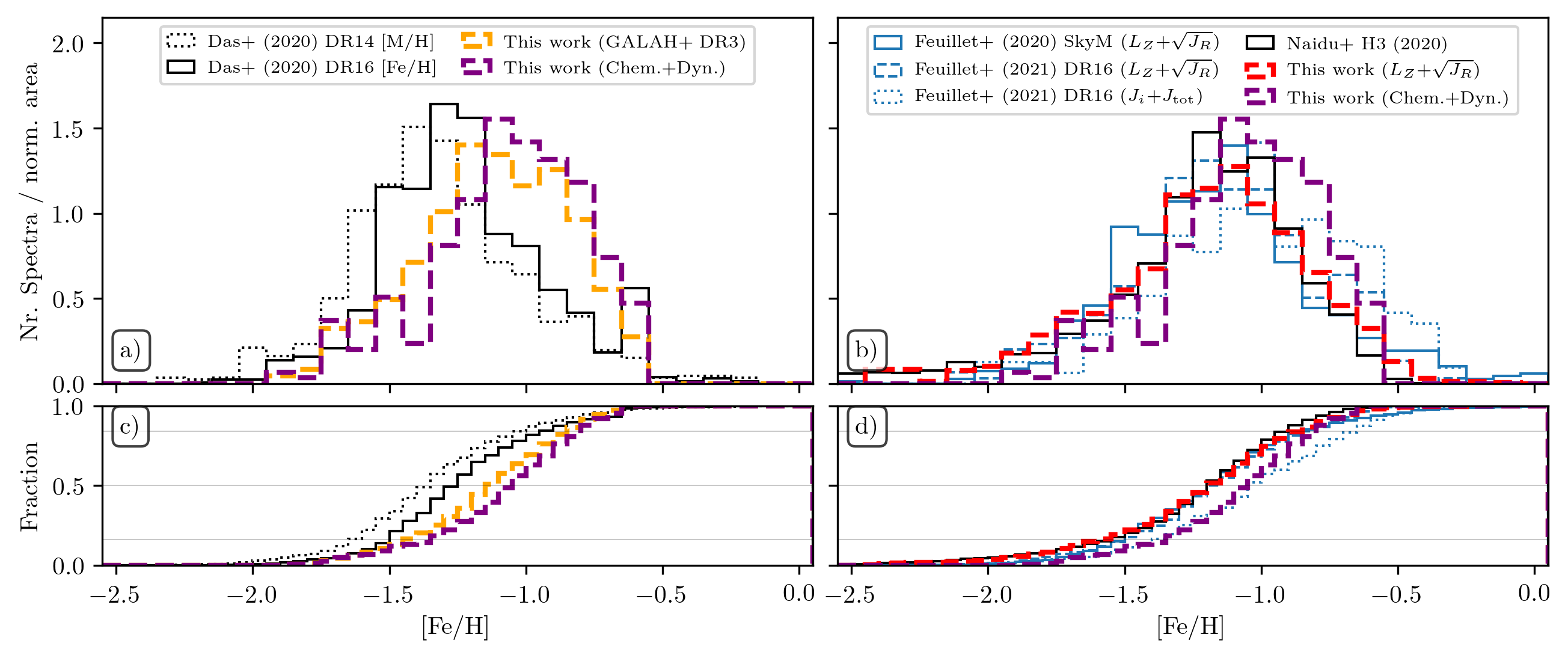}{chronochemodynamic_comparison}{
\textbf{Relative (top panels) and cumulative (bottom panels) distribution of iron abundances [Fe/H] for our samples of accreted stars.} \textbf{Left panels a and c} show chemically selected accreted stars and compare with the results by \citet{Das2020}.
\textbf{Right panels b and d} show the dynamical selections of the GSE by our work and \citet{Feuillet2020, Feuillet2021} as well as \citet{Naidu2020}.
}

We discuss these non-overlapping stars with low $J_R$ and/or high $L_Z$ in Sec.~\ref{sec:non_overlap}. Before that, we are interested in exploring the chrono-chemodynamic properties of the chemical, dynamical, and chemodynamic selection, this last selection being the overlap of the chemical and dynamical selection and thus a less complete, but even cleaner selection of the GSE than a purely dynamical one.

\subsection{Stellar chemistry} \label{sec:gse_stellar_chemistry}

The chemical properties of accreted stars, and especially the GSE, have only come into focus in the last decade \citep{Nissen2018,Helmi2020}. As spectroscopic surveys were only able to collect data in recent years, studies of the chemistry of accreted stars with a plethora of different instruments are often limited to the iron abundance, which we discuss first. We then briefly present the distributions of the other elements, discussed in major abundance groups, and compare to the literature where available. For this section, we will use both the distributions shown in Figs.~\ref{fig:fe_h_hist_cdf} and \ref{fig:chemdyn_chemistry} and quantified within Tab.~\ref{tab:chronochemodynamic_properties} with the aim to outline significant differences between the two selections.

\figuretextwidth{17cm}{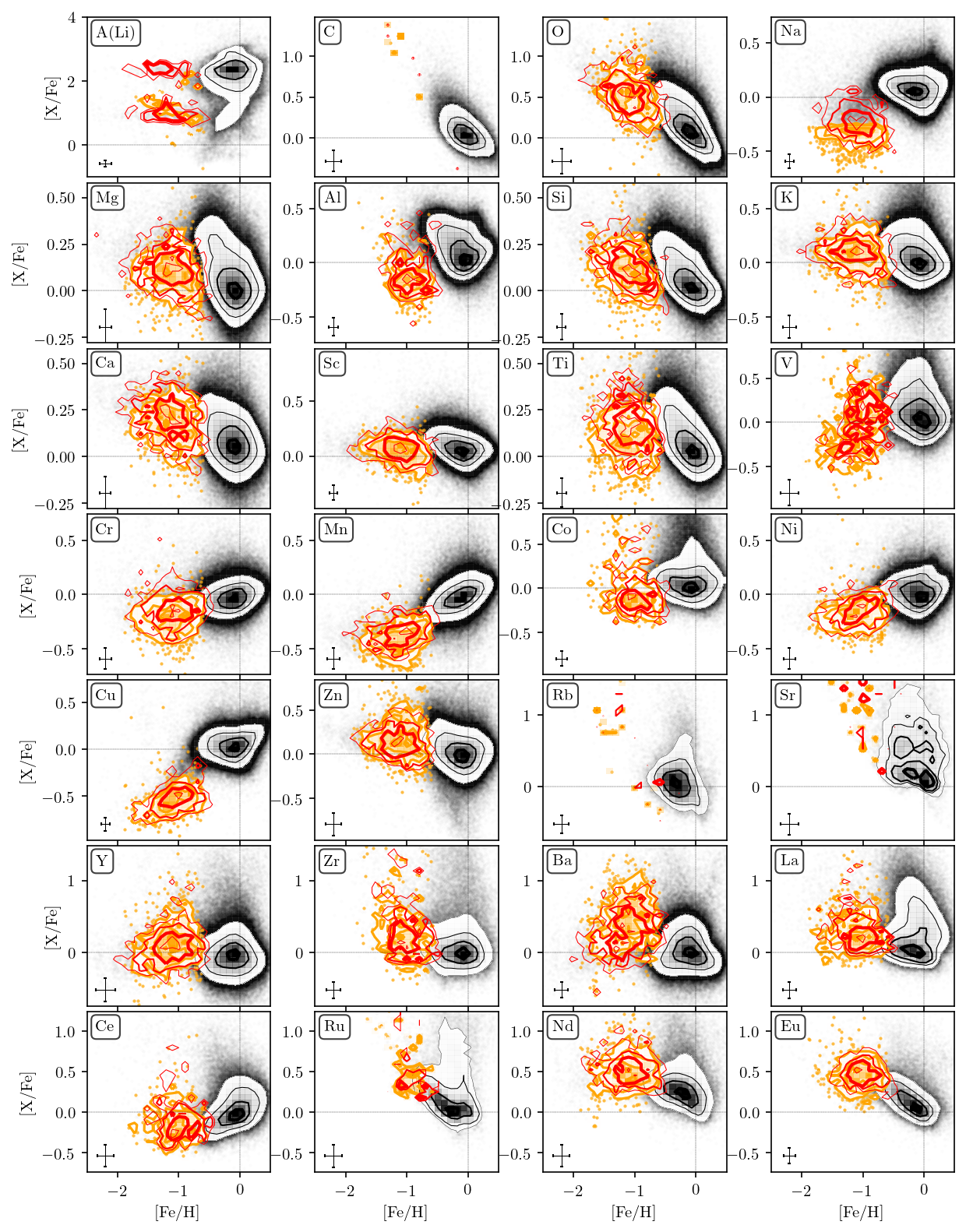}{chronochemodynamic_comparison}{
\textbf{Abundance distributions [X/Fe] (and absolute abundance for Li) as a function of iron abundance [Fe/H] for elements X (noted in each panel).} Shown are the distributions of all GALAH+ DR3 stars (black contours) as well as those of the chemically selected (orange contours) and dynamically selected (red contours) accreted stars. Quantities of each distributions are listed in Tab.~\ref{tab:chronochemodynamic_properties} together with the distribution of the stars within both the chemical and dynamical selection.
}

\subsubsection{Iron abundance [Fe/H] as metallicity tracer} \label{sec:gse_stellar_chemistry_fe_h}

Based on the study of the inner halo by \citet{Carollo2007} and \citet{Ivezic2008}, the iron abundance of (inner) halo stars ($<10\kpc$) could be expected to peak between $\mathrm{[Fe/H]} \sim -1.6$ and $\mathrm{[Fe/H]} \sim -1.45\pm0.32$ based on SEGUE data. \citet{FernandezAlvar2017} found a peak around $\mathrm{[Fe/H]} \sim -1.5$ in the inner halo when using APOGEE DR12. Using APOGEE DR13, \citet{Hayes2018} showed the distribution of stars with low [Fe/H] and [Mg/Fe] (consistent with the low-$\alpha$ halo) to peak around -1.2 and -1.3 (see their Fig.~2), similar to findings by \citet{Matsuno2019}. With data from APOGEE DR14, \citet{Das2020} found that their chemically selected accreted stars (with a similar chemical selection plane as ours) peak at $\mathrm{[M/H]} \sim -1.3$ (see Fig.~\ref{fig:fe_h_hist_cdf}a and c) and dominate in this [Fe/H] regime below $\mathrm{[Fe/H]} < -1.2$ by contributing up to 30\% of stars \citep[see also][]{Mackereth2019,Naidu2020}. Following up the identified GSE with SkyMapper and APOGEE DR16, \citet{Feuillet2020, Feuillet2021} found the distribution to be best described with a Gaussian around $\mathrm{[Fe/H]} \sim -1.17 \pm 0.34\dex$ and $\mathrm{[Fe/H]} \sim -1.15$ respectively (see Fig.~\ref{fig:fe_h_hist_cdf}b and d). A similar distribution around $\mathrm{[Fe/H]} \sim -1.15_{-0.33}^{+0.24}$ was recovered by \citet{Naidu2020}, who used data of the H3 Survey with a different dynamical selection. They found, however, a more extended tail towards metal-poor stars within their data (pink lines in Fig.~\ref{fig:fe_h_hist_cdf}b and d). Our data do not show such an extended tail, when comparing the cumulative distribution functions in Fig.~\ref{fig:fe_h_hist_cdf}d.

Using the same selection as \citet{Feuillet2020}, but data from the TOPoS Survey, \citet{Bonifacio2021} find a lower average metallicity of $\mathrm{[M/H]} \sim -1.45 \pm 0.3$ (estimated by us based on their Fig.~20). They report, however, [M/H] and not [Fe/H] and further assume [$\alpha$/Fe] = 0.4 for these low metallicities\footnote{Although their comparisons of [M/H] with [Fe/H] values show that their [M/H] traces [Fe/H] on average for the metal-rich part of their sample, we note that differences between both quantities could be as large as $0.3\dex$ because of changes in [$\alpha$/Fe].}. While their finding is more aligned with those by \citet{Das2020}, both when using the original [M/H] values from APOGEE DR14 used by them and the updated DR16 [Fe/H] values, they are much more metal-poor than all of the other distributions shown in Fig.~\ref{fig:fe_h_hist_cdf}.

The [Fe/H] distribution of the chemically selected sample (\input{depending_text/chem_percentiles_fe_h}) agrees well with the dynamically selected one (\input{depending_text/dyn_percentiles_fe_h}). However, although these distributions agree, their overlap (chemodynamical selection) is on average more metal-rich by \input{depending_text/diff_chem_chemdyn_fe_h} and \input{depending_text/diff_dyn_chemdyn_fe_h}, respectively. We find good agreement between the iron abundance distribution, shown in Fig.~\ref{fig:fe_h_hist_cdf}, of our dynamically selected GSE sample (\input{depending_text/dyn_percentiles_fe_h}) and the values by \citet{Naidu2020} and \citet{Feuillet2021}, both in terms of the mean/median value and dispersion (see Tab.~\ref{tab:chronochemodynamic_properties}).

We acknowledge note that different studies are potentially surveying different parts of the GSE \citep[see also the discussion by][]{Bonifacio2021}. The overlap between the different studies is currently minimal and more follow-up is needed. Our results and agreement with most studies suggests, however, that we are mainly selecting GSE stars. We assess this further with the individual abundances.

\subsubsection{Light elements: Li, C, O}

The Li abundances of both chemical and dynamical selections are distributed with very few stars along two sequences in Fig.~\ref{fig:chemdyn_chemistry}, agreeing between the selections. The higher A(Li) sequence (\input{depending_text/chem_high_A_Li.tex} for the chemical and \input{depending_text/dyn_high_A_Li.tex} for the dynamical selection) traces the Spite Plateau \citep{Spite1982}. It is sparsely populated by the few stars, mainly dwarfs. The lower sequences (\input{depending_text/chem_low_A_Li.tex} and \input{depending_text/dyn_low_A_Li.tex}) are close to the Solar abundance defined by GALAH+ DR3 \citep{Buder2021} and populated mainly by giant stars. However, measurements of Li are limited to a low number of stars, and we therefore do not compile quantitative distributions in Tab.~\ref{tab:chronochemodynamic_properties} for the chemical and chemodynamical selections. 

Due to the wavelength range of GALAH, we are not able to put constraints on either C or N. We therefore refer to the studies by \citet{Nissen2014} as well as \citet{Hawkins2015} and \citet{Hayes2018} for further insight.

We find [O/Fe] to agree between the different selections and our chemodynamical selection with $\mathrm{[O/Fe]} =$ \input{depending_text/chemdyn_percentiles_O_fe} to be slightly above the ratios found by \citet{Ramirez2012b} and \citet{Nissen2014} around 0.4 and much above the ratios found in APOGEE (around 0.3) by \citet{Hawkins2015} and \citet{Hayes2018}. The disagreement between the latter, however, is found for all metal-poor stars between GALAH and APOGEE \citep[e.g.][]{Buder2018} \added{and observed between most optical and near-IR O abundances \citep[e.g.,][]{Bensby2014}. It may arise due to 3D NLTE effects in the optical O triplet \citep[e.g.,][]{Amarsi2020} or systematics in modelling the molecular effects in the IR CO and OH lines \citep[e.g.,][]{Collet2007, Hayek2011}.}

\subsubsection{$\alpha$-process elements: Mg, Si, S, Ca, Ti} \label{sec:chronochemodynamics_alpha}

While both chemical and dynamical selections recover the low-$\alpha$ enhancement expected for accreted stars, we notice that the abundances of the dynamical selection typically extend towards higher values than the chemical selection. In particular, the median of the distributions increase in their deviation from Mg ($\input{depending_text/diff_chemdyn_Mg_fe.tex}\dex$) and Si ($\input{depending_text/diff_chemdyn_Si_fe.tex}\dex$) towards Ca ($\input{depending_text/diff_chemdyn_Ca_fe.tex}\dex$) and even more pronounced for Ti ($\input{depending_text/diff_chemdyn_Ti_fe.tex}\dex$). We find that a significant fraction of dynamically selected stars exhibit higher amounts of $\alpha$-enhancement than the chemically selected ones and this deviation increases from Mg to Ti. In particular, between \input{depending_text/above_84thchem_Mg_fe.tex} (for Mg) and \input{depending_text/above_84thchem_Ti_fe.tex} (for Ti) of the dynamically selected stars have [Mg/Fe] and [Ti/Fe] above the $84^\text{th}$ percentile of the chemically selected sample. It has to be noted though, that Mg was one of our the elements used for the chemical selection.

\subsubsection{Light odd-Z elements: Na, Al, K} \label{sec:chronochemodynamics_oddz}

We already pointed out the sub-Solar [Na/Fe] values of the accreted stars throughout this study. In Sec.~\ref{sec:overlap_planes}, we have, however, also identified higher [Na/Fe] abundances of the dynamical selection with respect to the chemical one. We find a similar difference for [Al/Fe], that is, \input{depending_text/chem_percentiles_Al_fe}$\dex$ for the chemical and \input{depending_text/dyn_percentiles_Al_fe}$\dex$ for the dynamical selection, the latter extending towards much higher and even super-Solar [Al/Fe] values. We find that K behaves different than Na and Al, in that [K/Fe] is found to be typically above $0\dex$.

\subsubsection{Iron-peak elements: Sc to Zn} \label{sec:chronochemodynamics_ironpeak}

The distribution of most iron-peak elements (with exception of Sc and Zn) within our chemical and dynamical selections are sub-Solar, with the lowest values for [Mn/Fe] around \input{depending_text/chemdyn_percentiles_Mn_fe}$\dex$ and [Cu/Fe] around \input{depending_text/chemdyn_percentiles_Cu_fe}$\dex$. Due to the difficulty associated to measure V in GALAH, we find a larger scatter for this element, similar to the results by \citep{Hawkins2015}. We note a slight upturn of [Ni/Fe] with increasing [Fe/H] for the dynamical selection, which differs both from our chemical selection and the decreasing trend found by \citet{Nissen2010} and \citet{Hawkins2015}. For [Zn/Fe], our values are higher than those around $0\dex$ found by \citet{Nissen2011}. Also for these elements, we find that the dynamically selected stars extend towards slightly higher abundances, most pronounced for Cr and Mn with differences around \input{depending_text/diff_chemdyn_Cr_fe.tex}$\dex$. While the dispersion of the distributions moves around the $0.1\dex$ level, we note outliers towards higher [X/Fe] for the elements V and Co. For both elements, possible systematic trends towards higher values have been cautioned for GALAH+ DR3 \citep{Buder2021}. However, also the results by \citet{Hawkins2015} showed a slightly larger scatter for [V/Fe].

\subsubsection{Neutron-capture elements: Rb to Eu}

Neutron-capture elements are the least well measured elements in GALAH+ DR3, especially at low metallicities. Because of the limited amount of measurements, we will not comment on the distributions for Rb, Sr \citep[see however][]{Aguado2021}, and Ru.
For the neutron-capture elements, we also see the largest dispersion in the distributions, that is, typically on the order of $0.2-0.3\dex$. Among these elements, we also find significant differences in the distributions. For Y and Ba, the abundances of the chemical selection are above those of the dynamical selection, whereas for Zr, La, Ce, Nd, and Eu this is not the case. We note that our distributions \citep[see also][]{Aguado2021} of Y and Ba are higher than those by \citet{Nissen2011}. Comparing our distribution with the measurements by \citet{Venn2004} for Ba and La as well as \citet{Fishlock2017} for Zn, Zr, La, and Nd strengthens the impression from Sec.~\ref{sec:enrichment_differences}, that the GALAH+ DR3 measurements for these elements are close to or below the detection limit. For Eu, we  refer the reader to the dedicated studies by \citet{Matsuno2021} and \citet{Aguado2021} with GALAH+ DR3 data.

\figuretextwidth{17cm}{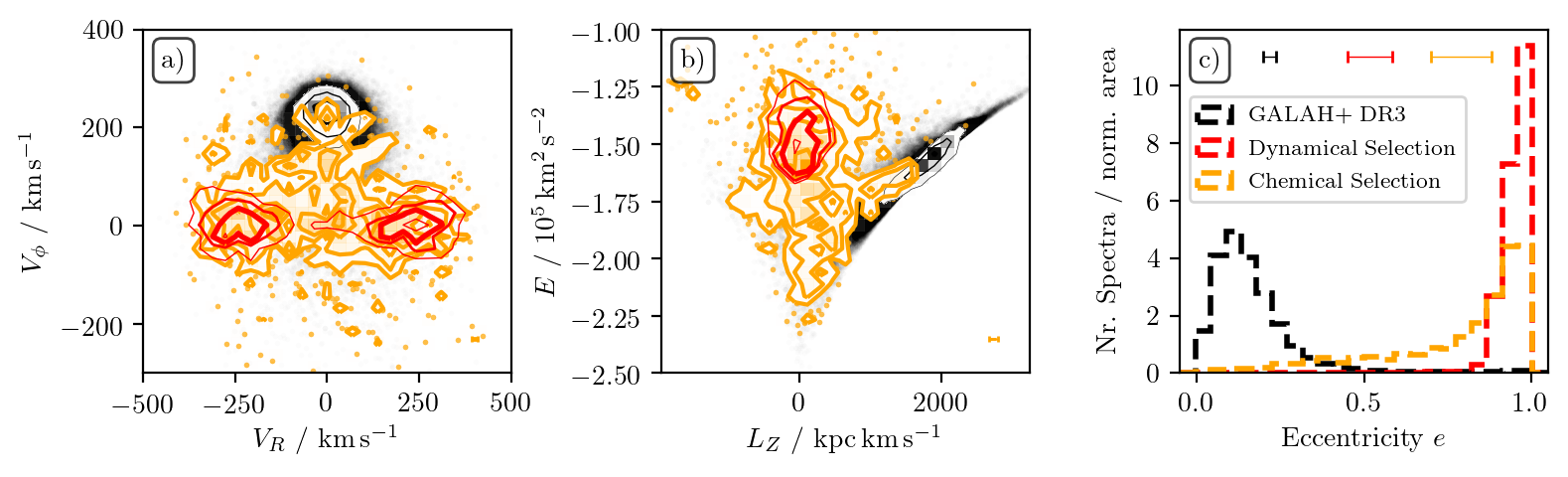}{chronochemodynamic_comparison}{
\textbf{Comparison of kinematic properties (Galactocentric velocities $V_R$ vs. $V_\phi$) as well as dynamic properties ($L_Z$, $E$, and $e$) for stars selected as accreted ones by means of chemistry (orange) and dynamics (red).} Black contours/lines denote the overall GALAH+ DR3 sample (mainly disk stars).
}

\subsection{Stellar kinematics/dynamics} \label{sec:gse_stellar_dynamics}

In Sec.~\ref{sec:overlap_planes}, we have already identified significant differences in the radial actions of chemically and dynamically selected accreted stars, while their angular momenta were on average similar around $L_Z \sim 0 \kpckms$. Here, we return to the plane of Galactocentric velocities, $V_R$ vs. $V_\phi$ (Fig.~\ref{fig:chemdyn_dynamics}a) as well as $L_Z$ vs. $E$ (Fig.~\ref{fig:chemdyn_dynamics}b), in which the Sausage \citep{Belokurov2018} and \Gaia-Enceladus \citep{Helmi2018} were initially discovered. As expected from the dynamical selection of GSE stars with the highest radial actions, these stars (red contours in Fig.~\ref{fig:chemdyn_dynamics}) also are restricted to the regions with highest $V_R$. The quantities of $V_R$ listed in Tab.~\ref{tab:chronochemodynamic_properties} are therefore not really descriptive, but still inform us about a slight asymmetry in $V_R$ for the GALAH+ DR3 sample. Dynamically selected GSE stars with the highest radial actions seem to show larger $V_R$ than negative ($V_R=$\input{depending_text/dyn_percentiles_vR_Rzphi}$\kms$). We do not notice this asymmetry in the chemically selected stars ($V_R=$\input{depending_text/chem_percentiles_vR_Rzphi}$\kms$). The location of chemically selected stars overlaps with the dynamically selected ones in this plane, but extends beyond it and covers the whole area of low $V_\phi \sim 0 \kms$ for $-400 < V_R < 400\kms$. We further notice a significant extension of \input{depending_text/chem_above_vphi_100kms} chemically selected stars with \added{$100 \kms < V_\phi < v_\circ$}. This velocity space is usually dominated by the \added{high-$\alpha$ or inner} stellar disk (black contours in Fig.~\ref{fig:chemdyn_dynamics}), thus suggesting that \input{depending_text/chem_above_vphi_100kms} of the chemically accreted stars show disk-like kinematics. Going back to the action space, we identify the stars with $V_\phi > 100 \kms$ as those that also exhibit lower radial actions, that is, $\sqrt{J_R / \kpckms} = $\input{depending_text/chem_above_vphi_100kms_sqrt_J_R.tex} in Fig.~\ref{fig:chemdyn_selection_plane}d, marking a significant overlap with low-$L_Z$ disk orbits.

In action-energy space (Fig.~\ref{fig:chemdyn_dynamics}b), we again identify GSE stars via their low $\vert L_Z \vert$. Accreted stars selected via their chemistry (orange) show a large distribution of energies (\input{depending_text/chem_percentiles_Energy_10_5.tex}$ \times 10^5 \kmkmss$). Comparing these values with those by \citet{Horta2021}, who used the same gravitational potential \citep{McMillan2017} within the same orbit calculation code \textsc{galpy} \citep{Bovy2015, Mackereth2018} suggests a non-negligible overlap with the Inner Galaxy Structure (IGS) / Heracles they identified \citep{Horta2021}. In a similar manner, we find most members of the GSE (\input{depending_text/chem_energy_above_m18}) tend to have orbit energies above $-1.8 \times 10^5 \kmkmss$. In addition, however, we also find \input{depending_text/chem_energy_above_m20} chemically selected stars with $E < -2.0 \times 10^5 \kmkmss$. Similar to the IGS/Heracles stars (within a $4\kpc$ sphere around the Galactic centre), these stars are located within the Inner Galaxy at $R = $\input{depending_text/chem_energy_above_m20_R}$ \kpc$ (but further away from the Galactic plane at $\vert z \vert =$\input{depending_text/chem_energy_above_m20_absZ}$\kpc$ because of GALAH's selection function with $\vert b \vert > 10\deg$). We discuss this further in Sec.~\ref{sec:reliability_selection}, as this raises the questions how reliable - or how contaminated - our selection is or if there is an actual connection between IGS/Heracles and the GSE.

Among the many possible dynamic parameters of accreted stars analysed in the literature \citep[e.g.][]{Schuster2012}, the eccentricity $e$ of orbits has been identified to be a rather distinctive property among the GSE \citep{Mackereth2019, Naidu2020}. Our sample supports this fact as well (Fig.~\ref{fig:chemdyn_dynamics}c), showing extraordinarily high eccentricities for both selections. Such high values ($e=$\input{depending_text/dyn_percentiles_ecc}) are introduced by the dynamical selection itself. However, the majority of chemically selected accreted stars also show such high values ($e=$\input{depending_text/chem_percentiles_ecc}), although with a much larger and significant tail towards lower values.

\figurecolumnwidth{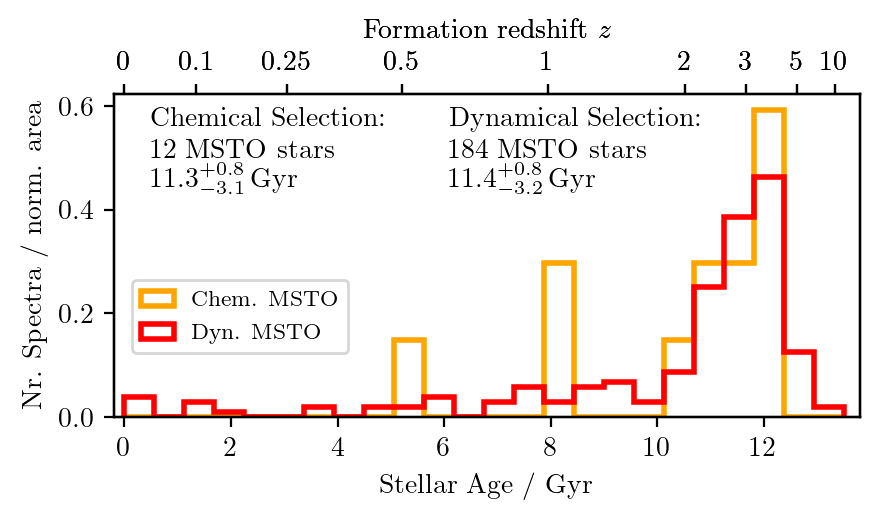}{stellar_ages}{
\textbf{Stellar Age distributions of our chemically selected accreted stars (orange) and dynamically selected GSE stars (red).} Ages are estimated via isochrone fitting, which is most precise for main-sequence turn-off (MSTO) stars. Their distribution (12 and \added{184} stars, respectively) is plotted with solid lines and we annotate their 16/50/$84^\text{th}$ percentiles. Uncertainties are on average 13\% for MSTO stars and 50\% for all stars (mainly giants).
}

\subsection{Stellar ages} \label{sec:gse_stellar_ages}

Stellar ages are likely the ultimate key to study the evolution of the Galaxy and it is therefore also essential to study the ages of accreted stars to place constraints on the beginning, duration, and end of accretion events. For GALAH+ DR3, stellar ages are provided as part of a value-added-catalogue estimated via isochrone fitting. For a detailed explanation of this analysis we refer to the DR3 release paper by \citet{Buder2021}.

As reliable stellar ages are still difficult to estimate, the best way for our sample to estimate ages is based on isochrone fitting of MSTO stars (Eq.~\ref{eq:msto}). This limits our sample to 12 and \added{184} MSTO stars for the chemically and dynamically selected samples (see Fig.~\ref{fig:age_histogram}). 

In general, stars of the GSE are very old, that is \input{depending_text/chem_age_msto}$\Gyr$ (chemically selected) and \input{depending_text/dyn_age_msto}$\Gyr$ (dynamically selected). We see a sharp drop of stellar ages below $10\Gyr$ and only few stars below this age. We note that of the three chemically selected MSTO stars with ages below $10\Gyr$, two have likely underestimated stellar ages, as their positions in the color-magnitude diagram are consistent with significantly older ages. Their dynamic properties, like $e > 0.9$, are consistent with the GSE.

These age distributions allow us to put constraints on the end of the accretion. We discuss this in Sec.~\ref{sec:age_timescale}, where we also put our estimates into the context of the literature. 

\section{Discussion} \label{sec:discussion}

The aim of our study is to find a way to best select accreted stars in the MW with chemical abundance data from GALAH+ DR3, and use those data to characterise accreted stars, especially of the GSE, chrono-chemodynamically. Here, we reflect upon this endeavour and several key aspects. Firstly, in Sec.~\ref{sec:prospects_chem_tagg} we discuss the prospect of chemically tagging accreted stars and telltale elements of accretion. We then discuss the differences found for chemical and dynamical selections (Sec.~\ref{sec:dissimilarity}). These differences inform our discussion on how to move forward towards a chemodynamical selection of accreted stars in Sec.~\ref{sec:towards_chemodyn}. Lastly, we briefly put our age estimates into the context of other studies and discuss the implications for timescales of star formation and accretion in Sec.~\ref{sec:age_timescale}.

\subsection{Prospects for chemically tagging the accreted halo} \label{sec:prospects_chem_tagg}

Why do we expect in-situ and accreted stars to show different chemical enrichment histories? If we accept the picture that the early Milky Way was assembled bottom-up via hierarchical aggregation of smaller elements and significant amount of accretion events \citep[e.g.][]{Searle1978}, the difference of the chemical evolution of such accreted stars depends significantly on the initial mass function, the mass of the accreted system, and star formation history of each accreted structure. Such differences will, for example, influence at which [Fe/H] we see the typical knee in the [$\alpha$/Fe] vs. [Fe/H] plane, at which SNIa kick in \citep[e.g.][]{Tinsley1979, GilmoreWyse1991, McWilliam1997, Matteucci2021}. The chrono-chemodynamic data of GALAH+ DR3 is so rich that we cannot address all questions here. We explicitly postpone a discussion of the resemblance of the accreted structures with dwarf Spheroidal galaxies to a follow-up, but refer to recent work by \citet{Hayes2018} and \citet{Monty2020} as well as the review by \citet{Nissen2018}. For this work, we concentrate on the following questions: Which telltale elements have we identified (Sec.~\ref{sec:tell_tale})? How reliable is our selection, that is, is our chemical selection actually selecting the GSE and how contaminated is this selection (Sec.~\ref{sec:reliability_selection}) in combination with the question, how chemically different are accreted stars from in-situ ones?

\subsubsection{Telltale elements of accretion based on GALAH+ DR3} \label{sec:tell_tale}

Among the 30 elements measured by GALAH+ DR3, not all increase the prospects of chemical tagging equally \citep{Ting2015} and are potentially useful to disentangle accreted stars from in-situ stars purely based on chemistry. We already elaborated on the separation significance between low- and high-$\alpha$ halo in Sec.~\ref{sec:enrichment_differences} (see also Tab.~\ref{tab:xfe_percentiles} as well as the detection rate towards lower metallicities in Fig.~\ref{fig:Completeness_Combinations}). We found a compromise between these criteria and the number of measurements when limiting ourselves to measurements of Mg, Na, and Mn. Here, we are now concerned with putting the separation significance in a nucleosynthetic context, to identify telltale elements of accretion based on GALH+ DR3.

Similar to \citet{Nissen2012}, the data of GALAH+ DR3 does not suggest a difference for Li between the accreted stars and the rest of the distribution. \citet{Simpson2021} already showed that the Li abundances of the GSE agrees with the in-situ stars in the metal-poor regime. Similar to \citet{Molaro2020}, they conclude that the Spite plateau is universal and cannot serve to identify accreted stars.

If we were able to measure C down to lower metallicities in GALAH+ DR3, this element (in combination with N) would certainly provide a powerful diagnostic. This was shown previously by \citet{Nissen2014} as well as \citet{Hawkins2015} and \citet{Hayes2018}, who find changes of [C/Fe] and [N/Fe] in low-$\alpha$ halo stars, but a conserved and lower [(C+N)/Fe] abundance relative to the canonical disk stars. Carbon is mainly produced by massive stars (especially through SNII) and Asymptotic Giant Branch (AGB) stars \citep{Kobayashi2020}. The low [C/Fe] values at low [Fe/H] thus suggest fewer contributions from both SNII and AGB stars in the birthplaces of accreted stars \citep[see][for further discussion]{Nissen2014}.

Similar to Mg, we see that the accreted stars of the GSE, independent of their selection, are lower in their $\alpha$-process element abundances than the high-$\alpha$ disk, as already found in previous studies \citep{Venn2004, Nissen2010, Hawkins2015, Hayes2018, Mackereth2019, Koppelman2019, Koppelman2021, Recio-Blanco2021, DiMatteo2020, Matsuno2021}. We have shown a decreasing separation significance $r$ in Tab.~\ref{tab:xfe_percentiles} from Mg to Ca, with exception of the more precisely measured element Ti. This confirms the decreasing difference between them as a function of $\alpha$-process element number, as shown by \citet{Hayes2018} and is expected based on the changing contribution of SNIa to these individual elements \citep{Tsujimoto1995,Kobayashi2020b}. In this respect, [Mg/Fe] - or better [Mg/H] \citep{Feuillet2021} - is the purest tracer of enrichment through SNII \citep[e.g.][]{Kobayashi2020}.

As discussed in the literature \citep{Nissen2010, Hawkins2015, Hayes2018}, the light odd-Z elements Na and Al are also enriched through SNII, but the yields show a very strong dependence on the metallicity of their progenitors, which influences a cascade of element production and recycling during He burning and the CNO cycle \citep[e.g.][]{Woosley2002, Kobayashi2006, Kobayashi2020}. As such, the abundances of Na and Al behave differently from the $\alpha$-process elements in the metal-poor regime and also for systems with different enrichment histories. This makes Na and Al (the latter less well measured in GALAH+ DR3) telltale elements for accretion \citep[see also][]{Kobayashi2011, Ting2012}.

For the iron-peak elements between Sc and Zn, we are facing a complex superposition of different nucleosynthetic processes causing significant differences especially between odd and even element abundance trends. Cr to Ni are expected to be formed mainly during thermonuclear explosions of SNe, as well as in incomplete or complete Si-burning during explosive burning of core-collapse SNe \citep{Kobayashi2006,Kobayashi2020}. We have found several of the iron-peak elements, such as Mn, Ni, and Cu to show significantly lower enrichment compared to the Galactic disk - in agreement with previous observations \citep{Nissen2010, Nissen2011, Hawkins2015, Hayes2018}. In particular, the behaviour of Mn and Ni, both mainly enriched by SNIa, can inform our understanding of the nucleosynthesis via SNIa, including those with sub-Chandrasekhar masses \citep{Kobayashi2020b,delosReyes2020,Sanders2021}. Within GALAH+ DR3, Mn is well measured down to lowest metallicities (see Fig.~\ref{fig:Completeness_Combinations}), whereas Ni is less well measured. The estimates for [Cu/Fe] suggest that Cu itself also has the potential of being a telltale element. Its enrichment is, similar to Na and Al, dependent on the metallicity of the progenitors, that is, massive stars that have exploded as SNII \citep{Kobayashi2006,Kobayashi2020}. A better detection rate for Cu and Ni would certainly place these elements among the rank of telltale elements.

The most difficult and, to a large extent, still enigmatic enrichment processes are found for neutron-capture elements. Chemical enrichment models nowadays model these elements with a combination of AGB stars, core-collapse SNe (including SNe II, HNe, electron-capture SNe and magneto-rotational SNe), $\nu$-driven winds, neutron star mergers, and black hole mergers \citep[see][and references therein]{Kobayashi2020}. But there remains significant uncertainty about the sites and yields of neutron-capture. Thanks to detailed individual studies as well as large spectroscopic surveys, more and more observational data for neutron-capture elements become available, and GALAH itself is delivering abundance estimates for up to 12 neutron-capture elements. Our measurements of neutron-capture elements show both higher scatter (see Fig.~\ref{fig:nafe_xfe_nissen_all_hah}) and are (with the exception of Ce) also on average higher than previous estimates by \citet{Nissen2011} and \citet{Fishlock2017}. This includes both the first peak s-process elements like Zr and Y as well as second peak s-process elements like Ba and La. \citet{Fishlock2017} especially find low [Y/Eu] abundances among the accreted stars. This difference with respect to the in-situ stars holds valuable information on the possible build-up of the Galactic halo by low-mass dwarf galaxies and more massive mergers \citep{Venn2004}. As metal-poor AGB stars are likely contributing to low [Y/Eu] abundances \citep{Venn2004}, these abundance ratios will help us to put more constraints on the origins of elements \citep[see also][]{Recio-Blanco2021}, including the amount of r-process enhancement \citep{Aguado2021} and the contribution of neutron-stars mergers and core-collapse SNe via [Eu/Mg] \citep{Matsuno2021}.

\subsubsection{Reliability and contamination of our chemical selection: Are we actually selecting accreted stars and especially the GSE?} \label{sec:reliability_selection}

After the identification of telltale elements in GALAH+ DR3, we apply GMMs in Sec.~\ref{sec:gaussian_mixture_models} to identify substructures in chemical space, which are (to first order) different from the disk. That is, with our applied GMM we are actually finding overdensities in the chemical space of [Na/Fe] vs. [Mg/Mn]. Here we are now concerned with the question how reliable such a selection is to identify accreted stars. In particular, we are interested in the question of wether we are actually selecting stars of the dynamically identified GSE or if our chemical selection is significantly different (or contaminated).

To what extent do previously identified accreted structures overlap with our chemical selection? \citet{Naidu2020} have already elaborated on this important key problem for their GSE selection, finding at least an overlap of Arjuna, Wukong/LMS-1, the Helmi streams, Aleph, and Sagittarius (Sgr).

\citet{Naidu2021} argue that Arjuna is the retrograde debris (with $L_Z < -700 \kpckms$) of the GSE with similar [Fe/H]. As such, we expect these stars to also appear in our selection. When inspecting the retrograde tail of our chemical selection, we find only a small portion of stars, that is, \input{depending_text/chem_L_Z_below_m500} below $L_Z < -500 \kpckms$ and \input{depending_text/chem_L_Z_below_m700} below $L_Z < -700 \kpckms$, in agreement with the value of $\sim 5\%$ from the study by \citet{Naidu2021} and confirming that Arjuna is likely contained in our selection, but not significantly contributing to it. A more detailed study of the \input{depending_text/chem_L_Z_below_m700_number} stars of this debris structure and its possible chemical differences with respect to the GSE as well as comparison with all the \input{depending_text/chem_L_Z_below_m700_all} stars with $L_Z < -700 \kpckms$ is therefore not necessary for this particular discussion.

We further do not anticipate a significant contamination by Wukong/LMS-1 \citep{Yuan2020a, Naidu2020}. This structure has been identified to be overlapping with the low [Fe/H] and low $e$ and prograde tail of the GSE \citep{Naidu2020}. We adopt a selection similar to the one by \citet{Naidu2020}, and find that \input{depending_text/chem_wukong1} of our chemical selection fulfills criteria for Wukong/LMS-1, with $\mathrm{[Fe/H]} < -1.45$ and $e < 0.7$. Further limiting the sample to prograde orbits with $200 < L_Z~/~\kpckms < 1000$ lowers this number to an insignificant \input{depending_text/chem_wukong2}, not even taking into account the orbit energy restriction by \citet{Naidu2020}.

Because GALAH+ DR3 is mainly observing stars in the Solar neighbourhood (81.2\% of the stars are within $2\kpc$), we do not expect a significant contamination by Sgr in our data set. While \citet{Hasselquist2017} found most of the Sgr core stars to be more metal-rich than our chemical selection, \citet{Hasselquist2019} found stars of the Sgr stream to overlap with accreted stars in chemical space. The latter stars exhibit eccentricities of 0.40-0.85 and apocentre radii $R_\text{apo} > 25\kpc$. After comparing the latter with our typically lower values of $R_\text{apo} = $\input{depending_text/chem_percentiles_R_ap}$\kpc$ for our chemical selection, we conclude that there is no significant contamination of our selection by Sgr and Sgr stream stars. We can exclude a significant contamination by the Helmi stream based on the low number of Helmi stream stars found in GALAH+ DR3 data by \citet{Limberg2021}. Due to our [Fe/H] cut, we further do not expect significant contamination in our chemical selection from Aleph. This overdensity was discovered by \citet{Naidu2020} as a prograde, highly circular, dynamic overdensity. It is yet to be classified but its chemistry ($\mathrm{[Fe/H]} \sim -0.51$ and $\mathrm{[\alpha/Fe]} \sim 0.19$) as well as location ($R = 11.1_{-1.6}^{+5.7}\kpc$) and high angular momentum resemble the hot tail of the outer low-$\alpha$ disk.

\figurecolumnwidth{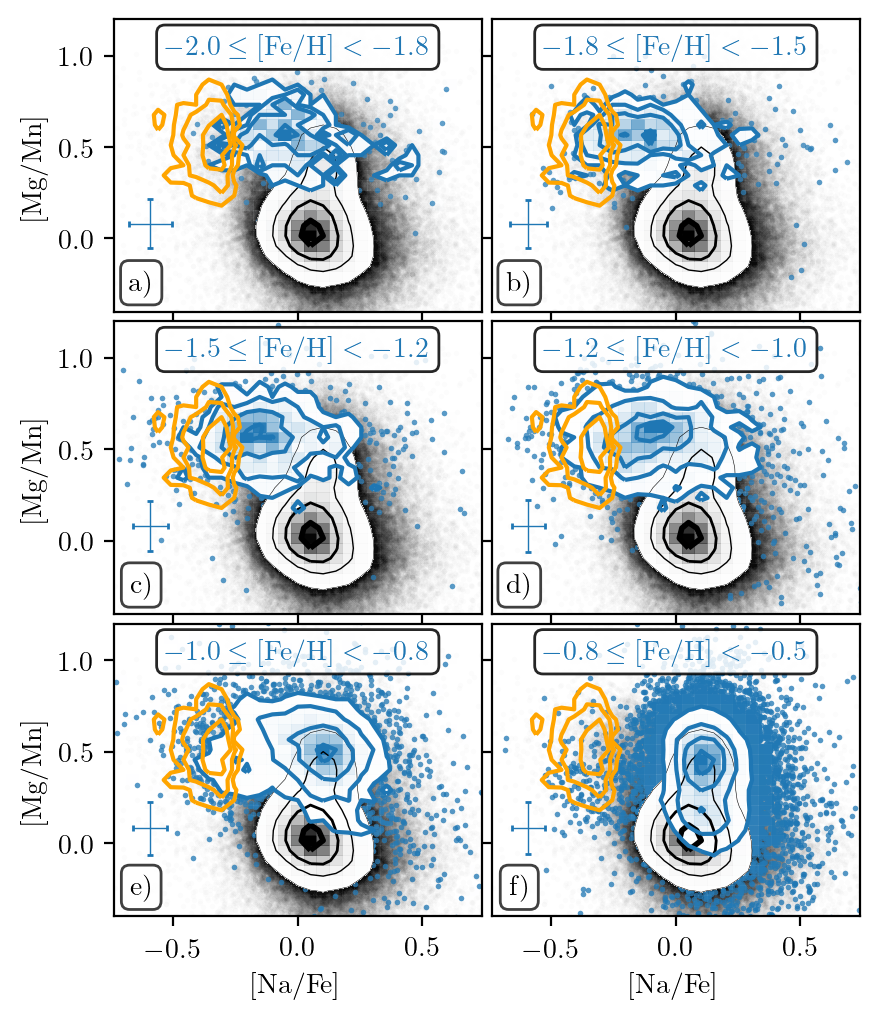}{chronochemodynamic_comparison}{
\textbf{Distribution of stars with different [Fe/H] values (blue contours) within the [Na/Fe] vs. [Mg/Mn] plane.} Stars of GALAH+ DR3 (mainly disk with $\mathrm{[Na/Fe]} \gtrsim 0$ are shown in black contours in the background. Accreted stars (see Fig.~\ref{fig:chemdyn_selection_plane}) are expected around (-0.3,0.5).}

We also come back to the possible contamination by the IGS/Heracles \citep{Horta2021} mentioned in Sec.~\ref{sec:gse_stellar_dynamics}. There we found \input{depending_text/chem_energy_above_m20} chemically selected stars with $E < -2.0 \times 10^5 \kmkmss$ located in the inner Galaxy at $R = $\input{depending_text/chem_energy_above_m20_R}$ \kpc$. Together with the portion of \input{depending_text/chem_above_500} stars with prograde orbits, typical of the high-velocity disk ($L_Z \sim 500\kpckms$), these are the two most significant (identified) sources of overlap/contamination. Similar to \citet{Horta2021}, we therefore have to discuss the question of whether we can tell apart accreted structures chemically both from other accreted structures (GSE and Heracles) as well as from in-situ stars (GSE and the in-situ disk). \citet{Horta2021} argue based on comparisons of chemical evolutions models from \citet{Andrews2017}, that both accreted and in-situ overlap in chemical space of [Al/Fe] vs. [Mg/Mn] and thus cannot, to our understanding, be completely separated.

We follow this question up, but with a slightly different angle, by looking at where stars with different [Fe/H] values are distributed within the [Na/Fe] vs. [Mg/Mn] diagram, see Fig.~\ref{fig:NaFe_MgMn_FeH_bins}. Although this projection is not separating accreted from in-situ stars, it is giving an idea of where these stars are distributed over different [Fe/H] ranges. In general, we see the trend that we cannot completely distinguish the abundances of the most metal-poor (panel a) stars from the high-$\alpha$ disk in this plane. However, going from $\mathrm{[Fe/H]} \sim -2$, we see that the distribution in [Na/Fe] widens towards $\mathrm{[Fe/H]} \sim -1$ (panels b-e), before it overlaps again with the high-$\alpha$ disk at $\mathrm{[Fe/H]} > - 0.9$. This suggests that there is a range in [Fe/H] where we can obtain a higher purity sample of accreted stars that are less contaminated by the in-situ disk. However, to fully understand the underlying structure and the completeness of the separation in this and other chemical planes, one needs to expand the comparison with chemical evolution models as done by \citet{Horta2021} towards models and cosmological hydrodynamical simulations that trace the chemical enrichment and include mergers \citep{Buck2020, Buck2021, Sestito2021}.

Here we aim to identify to what extent our chemical selection is truly identifying only GSE stars as the most dominant accreted structure in dynamical space. In Secs.~\ref{sec:chronochemodynamics_alpha} and \ref{sec:chronochemodynamics_ironpeak}, we find that our chemical selection of accreted stars tends to choose more Na- and Mn-poor stars than the dynamical selection of the GSE. Combining these effects, their [Mg/Mn] ratios again behave rather similarly, that is, \input{depending_text/chem_percentiles_MgMn_fe}$\dex$ for the chemical and \input{depending_text/dyn_percentiles_MgMn_fe}$\dex$ for the dynamical selection. This suggests that differences in Mg and Mn abundances are not driving the difference between the chemical and dynamical selections. We have, however, identified that our chemical selection is not selecting stars with $\mathrm{[Na/Fe]} \gtrsim 0\dex$, as it attributes these stars to an intermediate component (shown in blue in Fig.~\ref{fig:nafe_mgmn_overview}). Comparing our chemical and dynamical selections in terms of their [Na/Fe] coverage, thus constitutes a significant mismatch: \input{depending_text/above_84thchem_Na_fe.tex} of the dynamically selected GSE stars have higher [Na/Fe] values than the $84^\text{th}$ percentile of the chemically selected ones. To solve this issue in the future, multiple pathways are possible: a) increase the number of elemental abundance measurements and decrease their uncertainty in the hope that the differences between accreted and in-situ stars become detectable within the [Na/Fe] vs. [Mg/Mn] plane; b) find other abundance planes/combinations; c) combine chemical with dynamical information to select accreted stars. Option a) will only be available with new/better data, e.g. thanks to ongoing observations of GALAH Phase 2 as well as from upcoming surveys such as 4MOST \citep{deJong2019}, WEAVE \citep{WEAVE2018}, and SDSS-V \citep{Kollmeier2017}. We have already explored option b) throughout this study (see Sec.~\ref{sec:our_selection_techniques}).

A complicating factor regarding the validity of currently used orbit actions is the assumption of axisymmetry, that is, the neglect of the Galactic bar. Whilst it is beyond the scope of this study to perform quantitative comparisons, we note that the existence of the bar and the orbits that we calculated suggest a significant interaction of accreted GSE stars with the bar. For stars with small radial actions, an interaction with the bar (with a certain pattern speed $\Omega_b$) could for example scramble their initial action $J_R$ and $L_Z$, while shifting their position in the $E$ vs. $L_Z$ along a line of constant Jacobi integral $E_J = E - \Omega_b \cdot L_Z$ \citep{Binney2008, Sellwood2014}. This has to be tested in the future, but could explain both an underdensity of stars at low $L_Z \sim 0 \kpckms$ and $E \sim -2\times10^5 \kmkmss $ and an overdensity of accreted stars with larger (almost disk-like) $E$ and $L_Z$, as the interaction with the bar may have increased both quantities.

\figurecolumnwidth{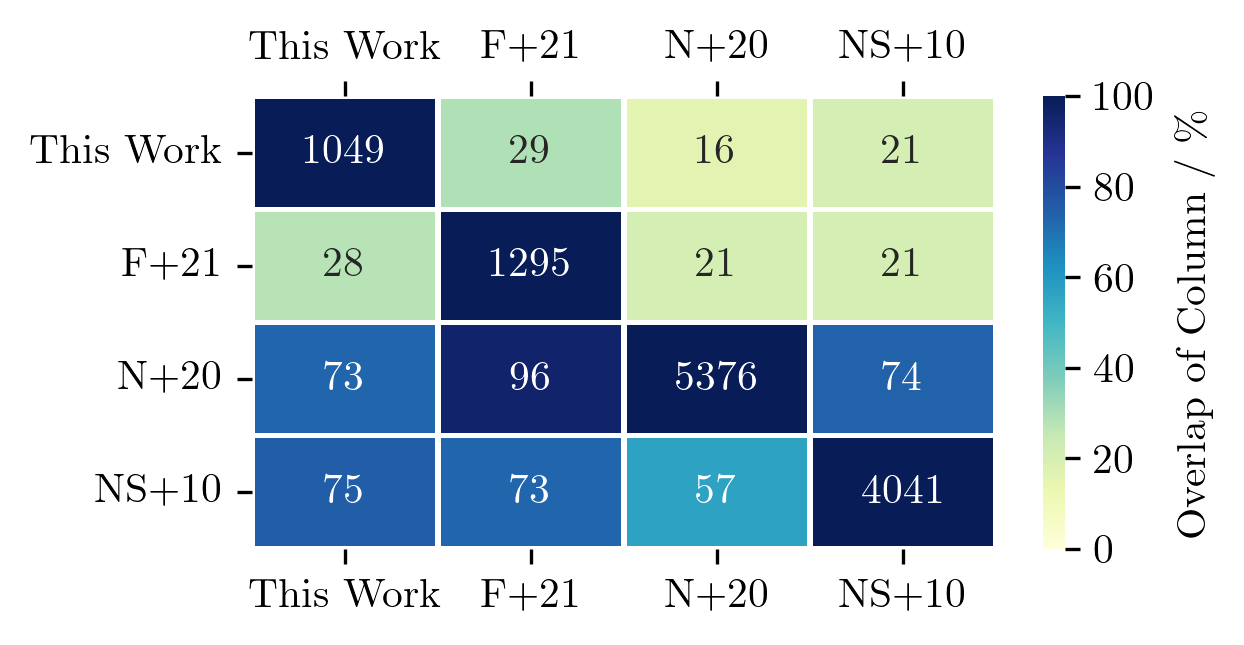}{chronochemodynamic_comparison}{
\textbf{Overlap of the different selections used throughout this study, that is via [Na/Fe] vs. [Mg/Mn] for ``This Work'', $L_Z$ vs. $\sqrt{J_R}$ for F+21 \citep{Feuillet2021}, $e$ and [Fe/H] vs. [$\alpha$/Fe] for N+20 \citep{Naidu2020}, and $v_\text{tot}$ and [Fe/H] vs. [Mg/Fe] for NS+10 \citep{Nissen2010}.} Diagonal entries show number of spectra per selection. Non-diagonal entries show overlap percentages relative to the stars per column. Percentages have been adjusted for possible flags to be independent of abundance detection limits (e.g. for [Na/Fe]). 
}

\subsection{The (dis-)similarity of samples based on different selection techniques and surveys} \label{sec:dissimilarity}

\subsubsection{Our selections vs. others}

As we set out to identify how similar chemical selection of accreted stars are to dynamical ones, in Fig.~\ref{fig:selection_overlap} we look at the actual overlap between these different selection techniques, when applied to GALAH+ DR3. This allows us to confirm independently that the low-$\alpha$ halo found by \citet{Nissen2010} is indeed significantly overlapping with the selection of the GSE by \citet{Naidu2020}, that is, 57\% and 74\% depending on what sample we use as denominator. We further see that the clean dynamical selection by \citet{Feuillet2021} indeed overlaps almost 100\% with the selection by \citet{Naidu2020}, but covers only 21\% of stars.

Comparing our selection with other techniques, we have already identified an overlap with the clean dynamical selection of \citet{Feuillet2021} of $(29\pm1_\%$ (see Sec.~\ref{sec:overlap_planes}) as this selection avoids low $\sqrt{J_R}$ regions possibly contaminated by the high-$\alpha$ halo and disk. Using additional chemical information to tell apart high-$\alpha$ from low-$\alpha$ stars has been pioneered by \citet{Nissen2010} among kinematic halo stars and optimised by \citet{Naidu2020} with eccentricities above $e > 0.7$. We find that our chemical selection overlaps significantly with both studies (75\% and 73\% respectively), although our selection only includes 16-21\% of the stars selected by both studies. This suggests that we are indeed only selecting a chemically defined subset of the low-$\alpha$ halo / GSE. We have therefore checked how these numbers change if we also include the metal-poor intermediate-$\alpha$ component (see blue contours in Fig.~\ref{fig:nafe_mgmn_overview}). This would lead to an increase in our numbers of accreted stars from 1049 to 4910 and an increase with all other selections from 16-29\% to 60-78\% (with respect to the latter selections). However, it would also increase the contamination, as the overlap with respect to our selection decreases from 28\% to 16\% compared to the selection by \citet{Feuillet2021} and even worse from 73-75\% to 54-47\% for the other two selections.

Comparing the metallicity distribution function of our chemical selection with the one by \citet{Das2020}, we have established in Sec.~\ref{sec:gse_stellar_chemistry_fe_h}, that the stars that they select as accreted are significantly more metal-poor (\input{depending_text/das2020_fe_h_dr16.tex}$\dex$ with updated APOGEE DR16 [Fe/H] values) than our selection (\input{depending_text/chem_percentiles_fe_h}$\dex$). We found however excellent agreement between our metallicity distribution function and the one from \citet{Naidu2020} and \citet{Feuillet2020,Feuillet2021} based on actions\added{; compare also to the values of $\mathrm{[Fe/H]} = -1.24 \pm 0.37$ from LAMOST \citep{Amarante2020b}.}

When looking at the actual overlap of our selection and the one by \citet{Das2020}, we find four and two (different) stars overlapping with our chemical and dynamical selection, respectively. According to APOGEE DR16, the iron abundances of the four stars ($\mathrm{[Fe/H]} = $\input{depending_text/das2020_feh_chem}) are similarly more metal-poor by \input{depending_text/das2020_feh_chem_diff} than our estimates, similar to the disagreement of our and their overall MDF (see Fig.~\ref{fig:fe_h_hist_cdf}). This is somewhat surprising, as we found a very similar MDF with the dynamical selection of APOGEE DR16 by \citet{Feuillet2021}. Contrary to this, the [Fe/H] estimates of the two dynamically selected stars (with $\mathrm{[Fe/H]} = -1.59$ and $-1.12\dex$) are more similar, with only $0.1\dex$ lower values for APOGEE DR16.

The discrepancy of different GSE selections was already discussed by \citet{Bonifacio2021}, as they also found lower metallicities than \citet{Naidu2020} and \citet{Feuillet2020}. While different metallicity scales of the different surveys could be the source of this disagreement, in this work, we demonstrate that the new selection within APOGEE DR16 by \citet{Feuillet2021} and the selection by \citet{Das2020}, but with updated values from APOGEE DR16, show a disagreement. This is an important finding and suggests that the chemical and dynamical selections of the same survey are selecting slightly different samples, that is, the chemical one by \citet{Das2020} is more metal-poor within APOGEE DR16 than the one by \citet{Feuillet2021}. It should also be mentioned, that the selection suggested by \citet{Myeong2019} in the $J_\phi/J_\text{tot}$ vs. $\left(J_R - J_Z\right)/J_\text{tot}$ plane with APOGEE DR14 stars resulted in a more metal-rich sample ($\mathrm{[Fe/H]} \sim -1.0$). Again, this suggests that different chemical and dynamical selections - already within APOGEE DR16 - result in slightly different selections.

Moving forward, it will be important to compare the different selections of the GSE also spatially and while taking the selection functions into account \citep[e.g.][]{Lane2021}, as different surveys probe different parts of the Galaxy and differences within the surveys might also reflect spatial differences of the GSE. It should also be assessed in more detail whether the chemical selection by \citet{Das2020} did for example also select a larger amount of IGS/Heracles stars, for which \citet{Horta2021} find mean [Fe/H] around $-1.3\dex$ in the inner Galaxy, thus possibly contaminating the selection by \citet{Das2020}. \added{Analysing the target selection of the sample from \citet{Das2020}, we identified that roughly 18\% of the stars were targeted by APOGEE during observations  observations of more metal-poor globular cluster and streams.} While we cannot draw significant conclusions from these differences, we note that stellar surveys suffer from the ability to sufficiently benchmark iron abundances in the metal-poor regime due to still low numbers of benchmark stars. More efforts similar to those of \citet{Hawkins2016} and \citet{Karovicova2020} will be needed to fully validate the iron-abundances in the metal-poor regime. One very metal-poor turn-off star found as part of the GSE by \citet{Naidu2020} was observed by GALAH, but without parameters reported within GALAH+ DR3 due to its high $T_\text{eff}$, large distance, and thus low signal-to-noise GALAH+ DR3 spectrum.

\subsubsection{Non-overlapping stars as key} \label{sec:non_overlap}

\figurecolumnwidth{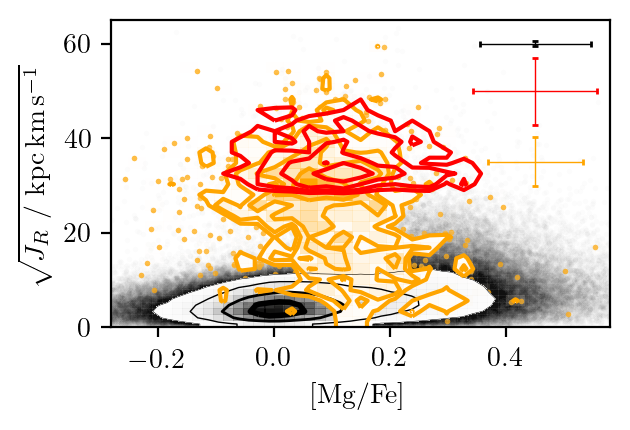}{chronochemodynamic_comparison}{
\textbf{Distribution of [Mg/Fe] vs. $\sqrt{J_R}$ for GALAH+ DR3 (black contours). Overlaid are the dynamically (red) and chemically (orange) selected accreted stars.} The latter extend towards lower $\sqrt{J_R}$ at increasing [Mg/Fe] down to the region populated by the high-$\alpha$ stellar disk.
}

In Secs.~\ref{sec:overlap_planes} and \ref{sec:gse_stellar_dynamics}, we found stars within the chemically selected accreted stars with $\sqrt{J_R~/~\mathrm{kpc\,km\,s^{-1}}} < 30$ and higher $V_\phi > 100\,\mathrm{km\,s^{-1}}$. In particular the stars with low $\sqrt{J_R~/~\mathrm{kpc\,km\,s^{-1}}} < 20$ are also those with higher $L_Z = 380_{-550}^{+710}\kpckms$. We plot the change of [Mg/Fe] with radial action $\sqrt{J_R}$ in Fig.~\ref{fig:overlap_mgfe_sqrtjr}. Further, we find that the dynamically selected GSE extends towards higher $\alpha$-enhancement into the region, where we would expect the in-situ stars to be situated. This can be further appreciated not only when looking at our dynamical selection of the $\alpha$-process element abundances in Fig.~\ref{fig:chemdyn_chemistry}, but also Fig.~11 by \citet{Naidu2020}.

Future studies should model the distribution of accreted stars in the [Mg/Fe] vs. $\sqrt{J_R}$ plane, especially in terms of time scales. We sadly have no stars within the lower right quadrant of Fig.~\ref{fig:overlap_mgfe_sqrtjr} among the chemically selected accreted MSTO stars. These stars would otherwise allow us to study if there is an age gradient in this plane. Our prediction for future studies with reliable stellar ages is that stars (born during the merger) with lower $\sqrt{J_R}$ should not only be more enriched in [Mg/Fe], but also younger, as their birth material was likely mixed with the $\alpha$-enhanced material of the Milky Way (contrary to their older accreted siblings) or exhibited a burst of star formation \citep{GilmoreWyse1991}. Finding such a gradient would allow us to estimate how much mixing happened over the time-scale of the merger and also put limits on the time-scale of the merger. Comparisons of their ages with those of the high-$\alpha$ disk and halo would also aid the necessary estimation of false-positive chemical selection (contamination) of disk stars as accreted stars.

\subsection{Towards a chemodynamical selection of the GSE} \label{sec:towards_chemodyn}

We have identified that the chemical selection extends significantly outside the clean dynamical one in dynamical space and vice-versa in chemical space. Only $(29\pm1)\%$ of the stars of our chemical selection were found within the clean dynamical selection box (Sec.~\ref{sec:overlap_planes} and Fig.~\ref{fig:chemdyn_selection_plane}d). There are three avenues to solve this disagreement:
(i) loosen constraints on the chemical abundances for the chemical selection (to also include the high [Na/Fe] stars of the GSE), 
(ii) loosen constraints on the dynamical selection (to also include the low $\sqrt{J_R}$ stars found for the GSE), or 
(iii) combine less strict constraints of the chemical and dynamical selections.

Concerning option (i), this would lead to adding stars from the intermediate Gaussian component (see blue component in Fig.~\ref{fig:nafe_mgmn_overview}) to our selection. As we are not able to separate the contamination by in-situ stars within this blue component, it is however beyond the scope of this paper to estimate the contamination and we resort to the other options here.

\figurecolumnwidth{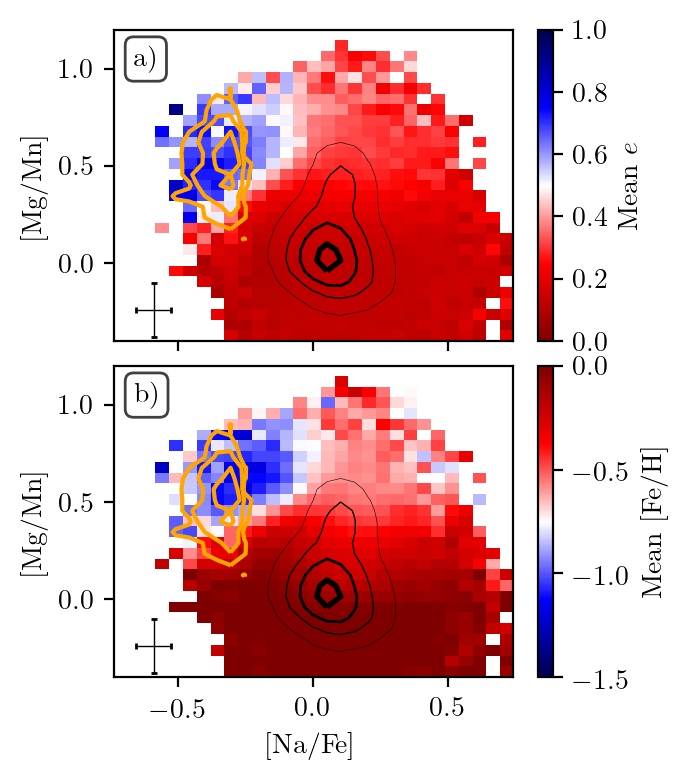}{chronochemodynamic_comparison}{
\textbf{Mean eccentricity (panel a) and [Fe/H] (panel b) in different regions of the [Na/Fe] vs. [Mg/Mn] plane for stars of GALAH+ DR3.}
Only bins with more than 5 stars have been populated.
Density contours correspond to those from Fig.~\ref{fig:chemdyn_selection_plane}a) with the chemically selected accreted stars (orange) and all stars of GALAH+ DR3 (black).
}

\figuretextwidth{16.2cm}{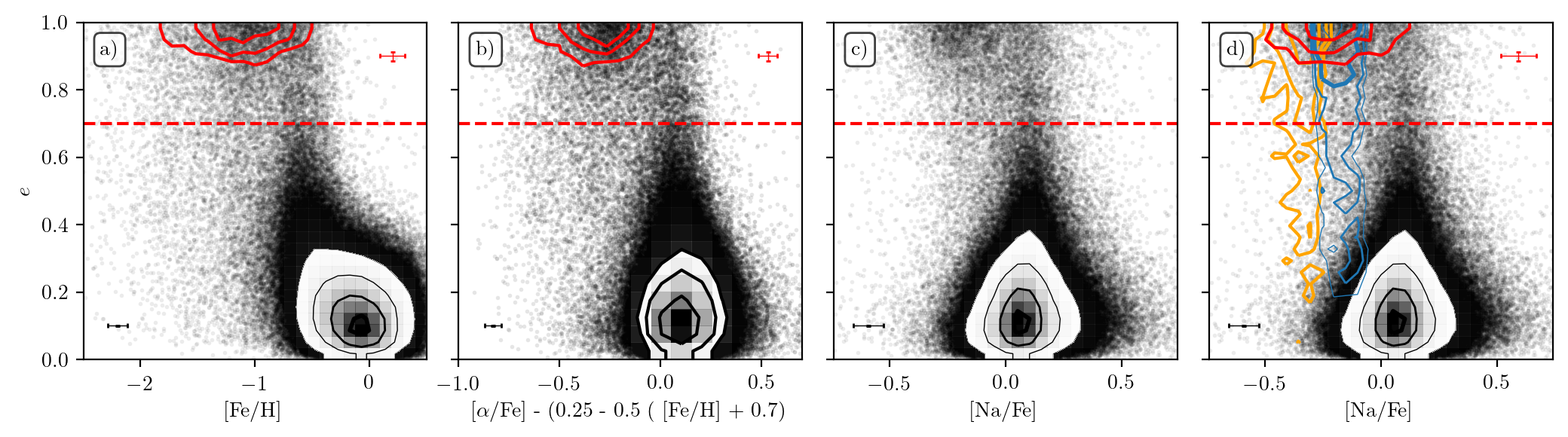}{chronochemodynamic_comparison}{
\textbf{Distribution of eccentricity as a function of different abundances of GALAH+ DR3 (black contours) and the dynamically selected stars (red contours).}
\textbf{Panel a)} as a function of [Fe/H].
\textbf{Panel b)} as a function of an adjusted difference between [$\alpha$/Fe] and [Fe/H] as suggested by \citet{Naidu2020}.
\textbf{Panel c)} as a function of [Na/Fe].
\textbf{Panel d)} as a function of [Na/Fe] with additional contours indicating our chemically selected accreted (orange) and intermediate (blue) components.
Red dashed lines indicates the $e$ limited as suggested by \citet{Naidu2020}.
}

Option (ii) was already tested by \citet{Feuillet2020} and suggested that below $\sqrt{J_R / \kpckms} < 30$ a significant contamination by the disk is starting to dominate a dynamical selection.

Finally, we are interested in combining less-strict chemical with informative dynamical properties towards a chemodynamical selection. The literature is already rich in suggested selections (see Sec.~\ref{sec:selection_techniques}). Inspired by the promising analysis of eccentricity $e$ by \citet{Mackereth2019} and \citet{Naidu2020}, we assess the possible combination of this orbit parameter with chemical abundances. In Fig.~\ref{fig:NaFe_MgMn_Fe_H_ecc}, we plot our chemical selection plane [Na/Fe] vs. [Mg/Mn], but coloured by mean eccentricities (panel a) and coloured by mean [Fe/H] (panel b). We see a very sharp transition between typically low eccentricity stars (red colours in Fig.~\ref{fig:NaFe_MgMn_Fe_H_ecc}a) for the disk stars (black contours) and high eccentricities ($e > 0.6$) in the upper left quartile. From this figure, it also becomes evident that we are only selecting the low [Na/Fe] stars (orange contours showing the chemically selected stars) of the high eccentricity stars. We remind ourselves that \citet{Mackereth2019} found $\sim 2/3$ of nearby halo stars have $e > 0.8$, and \citet{Naidu2020} selected stars based on their high eccentricities with $e > 0.7$. Which values of eccentricity should now be favoured? Whilst we stress that finding the best chemodynamic selection is beyond the scope of this paper, a first look at the distribution of chemical parameters as a function of eccentricity $e$, see Fig.~\ref{fig:nafe_e}, can inform future studies. Here, we see clear overdensities around $e \ll 0.5$ and $e \gg 0.8$. The latter also coincides with the position of our dynamically identified GSE stars (red contours). To allow the comparison with different chemical abundances, we plot [Fe/H], the selection of low- and high-$\alpha$ stars by \citet{Naidu2020} - similar to the cut by \citet{Nissen2010} - and [Na/Fe] in the different panels. Depending on what a certain survey is able to measure, it thus could be explored to combine eccentricity with any of these combinations. Fig.~\ref{fig:nafe_e}c also suggests, that already upper limits on [Na/Fe], like $\mathrm{[Na/Fe]} \not> 0$, could suffice to select accreted stars and overcome detection limits. We thus suggest to further assess the combination of abundance limits - such as $\mathrm{[Na/Fe]} \not> 0$ or $\mathrm{[Al/Fe]} \not> 0$ - with orbit limits - such as $e > 0.7$ as suggested by \citet{Mackereth2019} and \citet{Naidu2020}.

\subsection{Timescales of star formation and accretion} \label{sec:age_timescale}

Stellar ages help us to limit accretion timescales (also of the GSE) and trace back, which events shaped the formation of the Milky Way, including the merging history of our Galaxy \citep{Wyse2001}. Multiple studies \citep[e.g.][]{Jofre2010, Schuster2012, Hawkins2014, Gallart2019, Das2020, Montalban2021} have delivered estimates of stellar ages of accreted/GSE stars. As we are still unable to both estimate very reliable ages and further disentangle the MW halo from the disk reliably, the age estimates of different samples are in disagreement on several fronts.

Looking at the kinematic halo stars, \citet{Gallart2019} found that the accreted stars, selected as the blue sequence (in photometric colours) of the kinematic halo were coeval with their redder counterpart, and both significantly older than the average MW high-$\alpha$ (thick) disk star. While \citet{Schuster2012} and \citet{Hawkins2014} also found the metal-poor accreted stars to be coeval with the old high-$\alpha$ halo and disk stars, they identified the metal-rich end of the accreted population to be younger than the majority of the high-$\alpha$ halo/disk.

With the help of asteroseismically aided observations, \citet{Montalban2021} were able to estimate some of the most precise ages of GSE stars to date. They confirmed that the average GSE star is likely slightly younger (or coeval within uncertainties) than the average old and nearly coeval high-$\alpha$ stars \citep{Miglio2021} with robust asteroseismically aided age estimates. They thus concluded that a significant part of the MW high-$\alpha$ disk was already in place before the infall of the GSE at around $10\Gyr$, echoing the conclusions of several other earlier papers, based on more limited data \citep[e.g.][and references therein]{Wyse2001}. In particular, the age of one in-situ high-$\alpha$ star that was already in place and likely dynamically heated by the merger allowed \citep{Chaplin2020} to infer at 68\% confidence that the earliest the merger could have begun was $11.6\Gyr$ ago at 68.

Our average ages of \input{depending_text/chem_age_msto}$\Gyr$ and \input{depending_text/dyn_age_msto}$\Gyr$, respectively, coincide with this estimate, but appear to be older (although consistent within uncertainties) than the average of the distribution found by \citet{Montalban2021} at $9.7\pm0.6\Gyr$. As the accuracy of stellar age estimation is subject to several complex factors, such as atomic diffusion \citep[see e.g.][]{Jofre2011}, a discussion of this disagreement, including the contentious ages of young $\alpha$-enhanced stars with large masses \citep[e.g.][]{Chiappini2015,Zhang2021}, is beyond the scope of this paper.

Looking at the lower end of stellar ages, \citet{Bonaca2020} found both the star formation rate of the high-$\alpha$ disk and in-situ halo stars to truncate $8.3 \pm 0.1 \Gyr$ ago ($z \simeq 1$), whereas they find the star formation of accreted stellar halo to truncate $10.2_{-0.1}^{+0.2} \Gyr$ ago ($z \simeq 2$). While small in size, 75\% (9/12) and \added{75\% (138/184)} of the MSTO stars of our chemically and dynamically identified accreted stars also show ages above $10\Gyr$ (Fig.~\ref{fig:age_histogram}).

While these observations suggest that the GSE has not significantly influenced the formation of the thick disk, it has fuelled the hypothesis that the last major merger of the GSE is chronologically not only consistent with the decrease of star formation in the high-$\alpha$ disk \citep{Bonaca2020} and beginning of star formation of the low-$\alpha$ disk, but that there is actually a causal connection \citep[see e.g.][]{Buck2020}. If true, this allows us to put constraints on the merger timescale and subsequent onset of the low-$\alpha$ disk formation \citep[e.g.][]{Wyse2001,DiMatteo2019,Belokurov2020}. 

The jury is still out as to whether the GSE is responsible for the decrease in thick disk star formation. But it seems that at least the timing is plausible in this eventful Galactic epoch. Future studies of the chrono-chemodynamic properties of the accreted and in-situ stars promise to shed light on the circumstances of drastic changes in the Galactic structure.

\section{Conclusions} \label{sec:conclusions}

In this study we set out to identify which elements are best used to identify accreted stars in our Milky Way based on elemental abundances from the third data release of the GALAH Survey \citep{Buder2021} and to compare this chemical selection with dynamical ones from the literature. The key findings of our study are:
\begin{itemize}
\item To identify the best set of elements for this task, we follow the approach by \citet{Nissen2010} to identify accreted stars (in their paper called low-$\upalpha$ halo stars) via their high total velocities and low [Mg/Fe] compared to the (thick) disk stars. We find several elements showing a significant separation in GALAH+ DR3 between abundances of the accreted stars and the disk (or in-situ stars), including Mg, Si, Na, Al, Mn, Ni, and Cu. Their detection rates as a function of [Fe/H] vary strongly, and we find the best compromise of significance of separation and detection rates for Mg, Mn, and Na - ranking them as the best telltale elements of accretion based on GALAH+ DR3 (see discussion in Sec.~\ref{sec:tell_tale}).
\item We test the identification of accreted stars based on these elements in different abundance planes via Gaussian Mixture Models and find the best results from [Na/Fe] vs. [Mg/Mn], similar to the selection via [Al/Fe] vs. [Mg/Mn] used by \citet{Hawkins2015} and \citet{Das2020} for data from the APOGEE Survey. 
\item We compare the chrono-chemodynamic properties of stars identified via this chemical selection with those of the accreted stars of the GSE, selected via a box in $L_Z$ vs. $\sqrt{J_R}$ space as suggested by \citet{Feuillet2021}. We discuss the implications for chemical tagging of accreted stars as well as how we can interpret the difference between chemical and dynamical selections. Our main points are: Firstly, values of [Na/Fe] of stars identified through dynamical  selection are typically $0.1\dex$ higher than those of the chemical selection and secondly, the radial actions $J_R$ of the chemical selection extend well below the clean selection in dynamical space suggested by \citet{Feuillet2021}. In particular, only $(29\pm1)\%$ of the chemically selected stars fall within the clean dynamical selection box. \item We discuss the reliability and contamination of our selection in Sec.~\ref{sec:reliability_selection}, finding that our chemical selection is possibly - but insignificantly - contaminated by the IGS/Heracles (\input{depending_text/chem_energy_above_m20}) and other accreted structures.
\item We find \input{depending_text/chem_above_500} of the chemically selected accreted stars on prograde orbits $L_Z > 500 \kpckms$, that is, overlapping with the hot disk. More follow-up is necessary to identify if this is caused by contamination or an actual overlap in dynamical space, thus suggesting changes of orbits of accreted stars (see our discussion in Sec.~\ref{sec:non_overlap}).
\item If one is interested in the chemical properties of the GSE it is favourable to use the quantities estimated based on the dynamical selection. To analyse the dynamical properties of the GSE, however, those estimated from the chemical selection should be preferred. Again, we caution that we expect a small contamination by the IGS/Heracles (\input{depending_text/chem_energy_above_m20}) as well as possibly the dynamically hot stellar disk. \input{depending_text/chem_above_500} of our chemically selected stars exhibit $L_Z > 500\kpckms$. This is an upper limit of our contamination, as it could be either caused by contamination of our selection or changed orbits of truly accreted stars. We do not suggest to use the overlap of both selections, as we have identified significant differences due to the overlap of accreted and in-situ stars in [Na/Fe], which prohibit us to distinguish the accreted stars with high [Na/Fe] from the in-situ ones within our uncertainties. In particular, \input{depending_text/above_84thchem_Na_fe} of the stars of the dynamically selected GSE have [Na/Fe] above the $84^\text{th}$ percentile of the chemically selected ones.
\item We therefore also discuss how we can find a better selection of accreted stars in chemodynamical space. In Sec.~\ref{sec:towards_chemodyn}, we thus show how the previously suggested orbit eccentricity \citep[see e.g.][]{Mackereth2019, Naidu2020} in combination with chemistry can help future studies to find appropriate selections.
\item \added{To allow reproducibility of results and better comparison between different studies and selections, as well as their measured properties, we encourage researchers to report all assumptions going into the calculation of orbit parameters and to report uncertainties.}
\item Finally, we use age estimates of MSTO stars to find typical ages of \input{depending_text/chem_age_msto}$\Gyr$ (chemically selected) and \input{depending_text/dyn_age_msto}$\Gyr$ (dynamically selected) for the accreted stars in GALAH+ DR3. \added{We see a significant drop below $10\Gyr$ in our sample and a tentative agreement with the finding by \citet{Bonaca2020} of a truncation of star formation of accreted stars around $10\Gyr$. However, our distributions only include 12 and 184 MSTO stars for the chemical and dynamical selections, respectively, and are thus prone to outliers. Due to the small numbers of stars younger than $10\Gyr$, we cannot draw strong conclusions concerning the cessation of  star formation.}
\end{itemize}

\section{Outlook} \label{sec:Outlook}

With the ongoing development of new instruments and the beginning of the era of large-scale stellar surveys \citep[see][for reviews]{Nissen2018, Jofre2019}, the bulge and halo have now also come into reach and we start to see streams and substructures in the halo \citep[see e.g.][for a review]{Helmi2020}, which are evidence of ongoing and past accretion events. How significant these events were is still under investigation: How many mergers happened? Where are their remnants now? How (dis-)similar are their properties to the in-situ stars that were already in the Galaxy? Which of these were major mergers? How much (primordial) gas did they bring into the Galaxy? What is the connection between mergers with the pause in star formation and different chemistry that we observe between the high- and low-$\alpha$ disk? \citet{Helmi2020} concludes that, to be able to interpret various structures, we need detailed chemical abundances of stars with full phase-space information, which in-turn motivates the continuing efforts of ongoing and upcoming surveys.

With the availability of astrometric information from the \Gaia satellite mission and its Data Releases 1 \citep{Brown2016}, DR2 \citep{Brown2018}, and eDR3 \citep{Brown2021} as well as the industrial revolution of stellar spectroscopic surveys, delivering millions of chemical abundance measurements \cite[for a review see][]{Jofre2019}, our selection techniques of accreted stars start to shift from kinematic towards chemodynamic or even purely chemical properties.

We are, however, just at the beginning of truly understanding the interplay of kinematics/dynamics, chemistry, and ages of the different substructures. We know that, when it comes to these different properties, kinematic properties change on short timescales, whereas dynamic properties (in a slowly evolving potential) are conserved for a longer period. But we hope that chemical abundances, as locked in the stellar atmosphere at birth, do not change significantly over cosmic time for individual stars, and are furthermore significantly different for different Galactic and extra-galactic birth places. Stellar ages (which are difficult to extract from our observables) are our best hope to narrow down the formation scenarios of our Galaxy.

Two major question that need to be answered in more holistic studies are: When can we actually identify a star as accreted? And how can we tell it apart from other accreted stars? In this study, we have been able to answer this question in terms of the most extreme cases, that is the significantly different enrichment of some stars for example in Na (and Al). However, we clearly are struggling at the overlap of accreted structures themselves (e.g. GSE and IGS/Heracles or GSE vs. Arjuna) as well as accreted stars and in-situ ones. When do we actually call stars in-situ, given that the Milky Way itself likely started from several smaller structures that then kept accreting and star formation also takes place during mergers from material of both in-situ and accreted material? More research is needed to push our understanding of the underlying (accreted) structure of our Milky Way and its building blocks. Possible clues might also be found by studying the spatial distribution of these stars compared to the older GSE stars. Due to the low number of identified stars, this should, however, be done by combining the stars identified by the various different and complementary surveys.

As a follow-up study we also strongly propose to attempt to associate the substructures detected in dynamical space that are not overlapping with the clean GSE box by \citet{Feuillet2021} in detail. This would be an application of the methodology similar to the one by \citet{Naidu2020}, as already done for the Helmi streams for GALAH+ DR3 by \citet{Limberg2021}, but is beyond the scope of this study. Along a similar line of thought, we also suggest to continue the efforts of searching for associations between accreted structures with other substructure. Such studies include the search for associations between globular clusters and accreted structures \citep{Massari2019}, stellar streams with globular clusters as potential progenitors \citep{Bonaca2021} and moving groups with accreted structures \citep{Schuler2021}. These studies made use of the data from different surveys, including data provided by \cite{Helmi2018b} in combination with the work by \citet{Vasiliev2019} and the H3 Survey \citep{Conroy2019}. The data from the GALAH survey is very complementary to these surveys, as it probes different regions of the Galaxy and/or adds the high-dimensional chemical perspective and thus allows to confirm found accreted associations even stronger.

In the future it will be vital to continue the effort of comparing present-day observations with both higher redshift observations as well as potential formation scenarios. High redshift observations may allow us to observe major mergers as they happen and inform us on their importance. Was the MW high-$\alpha$ (thick) disk for example heated up by major mergers like those of the GSE, or was it already born hot as the correlation of higher gas velocity dispersions at higher redshifts would suggest \citep{Wisnioski2015, Leaman2017}? Can we find a consistent story over all redshifts?

Favouring or excluding formation scenarios will need to go hand-in-hand with the comparison to (cosmological hydrodynamical) simulations \citep[e.g.][]{Mackereth2018, Bonaca2017, Wu2021}, which allow us to time accretion events and trace accreted stars spatially, dynamically, and now also chemically. Much progress has been made in recent years through studies of the in-situ and ex-situ fractions \citep[e.g.][]{Pillepich2015}, the influence of mergers on the $\alpha$-enhancement \citep[e.g.][]{Zolotov2010, Grand2020, Buck2020, Renaud2021}, the estimation of infall scenarios and parameters of the GSE \citep[e.g.][]{Villalobos2008, Koppelman2021, Naidu2021}, including the amount and importance of gas-rich and gas-poor mergers \citep[e.g.][]{Fensch2017, Renaud2021b} and telling apart different components of simulated galaxies \citep[e.g.][]{Obreja2019}. We expect great progress here and an iterative convergence on deciphering the origin of the elements, as elemental abundance measurements - especially of environments different than the already well-studied disk - inform the constraints on chemical enrichment processes and yields \citep[e.g.][]{FernandezAlvar2018b, Vincenzo2019, Eitner2020, Sanders2021, Ishigaki2021}.

\section*{Acknowledgements}

We acknowledge the traditional owners of the land on which the AAT and ANU stand, the Gamilaraay, the Ngunnawal and Ngambri people. We pay our respects to elders past, present, and emerging and are proud to continue their tradition of surveying the night sky in the Southern hemisphere.
This work was supported by the Australian Research Council Centre of Excellence for All Sky Astrophysics in 3 Dimensions (ASTRO 3D), through project number CE170100013.
KL acknowledges funds from the European Research Council (ERC) under the European Union's Horizon 2020 research and innovation programme (Grant agreement No. 852977).
TB acknowledges support from the European Research Council under ERC-CoG grant CRAGSMAN-646955.
JK and TZ acknowledge financial support of the Slovenian Research Agency (research core funding No. P1-0188) and the European Space Agency (PRODEX Experiment Arrangement No. C4000127986).
KCF acknowledges support from the the Australian Research Council under award number DP160103747.
CK acknowledges funding from the UK Science and Technology Facility Council (STFC) through grant ST/R000905/1 and ST/V000632/1, and the Stromlo Distinguished Visitorship at the ANU.
We thank Rohan Naidu for sharing the [Fe/H] values from \citet{Naidu2020}. Hyperlink figures to code access are inspired by Rodrigo Luger.

\section*{Facilities}

\textbf{AAT with 2df-HERMES at Siding Spring Observatory:}
The GALAH Survey is based data acquired through the Australian Astronomical Observatory, under programs: A/2013B/13 (The GALAH pilot survey); A/2014A/25, A/2015A/19, A2017A/18 (The GALAH survey phase 1), A2018 A/18 (Open clusters with HERMES), A2019A/1 (Hierarchical star formation in Ori OB1), A2019A/15 (The GALAH survey phase 2), A/2015B/19, A/2016A/22, A/2016B/10, A/2017B/16, A/2018B/15 (The HERMES-TESS program), and A/2015A/3, A/2015B/1, A/2015B/19, A/2016A/22, A/2016B/12, A/2017A/14, (The HERMES K2-follow-up program). This paper includes data that has been provided by AAO Data Central (datacentral.aao.gov.au).
\textbf{\Gaia: } This work has made use of data from the European Space Agency (ESA) mission \Gaia (\url{http://www.cosmos.esa.int/gaia}), processed by the \Gaia Data Processing and Analysis Consortium (DPAC, \url{http://www.cosmos.esa.int/web/gaia/dpac/consortium}). Funding for the DPAC has been provided by national institutions, in particular the institutions participating in the \Gaia Multilateral Agreement. 
\textbf{Other facilities:} This publication makes use of data products from the Two Micron All Sky Survey \citep{Skrutskie2006} and the CDS VizieR catalogue access tool \citep{Vizier2000}.

\section*{Software}

The research for this publication was coded in \textsc{python} (version 3.7.4) and included its packages
\textsc{astropy} \citep[v. 3.2.2;][]{Robitaille2013,PriceWhelan2018},
\textsc{corner} \citep[v. 2.0.1;][]{corner},
\textsc{galpy} \citep[version 1.6.0;][]{Bovy2015},
\textsc{IPython} \citep[v. 7.8.0;][]{ipython},
\textsc{matplotlib} \citep[v. 3.1.3;][]{matplotlib},
\textsc{NumPy} \citep[v. 1.17.2;][]{numpy},
\textsc{scipy} \citep[version 1.3.1;][]{scipy},
\textsc{sklearn} \citep[v. 0.21.3;][]{scikit-learn},
\textsc{statsmodels} (v. 0.10.1),
\textsc{xdgmm} (v. 1.1).
We further made use of \textsc{topcat} \citep[version 4.7;][]{Taylor2005};

\section*{Data Availability}

The data used for this study is published by \citet{Buder2021} and can be accessed publicly via \url{https://docs.datacentral.org.au/galah/dr3/overview/}.
We provide full tables for Tables~\ref{tab:simple_gmm_selection} and \ref{tab:xdgmm_dynamical_selection} in the supplementary material. All code to reproduce the analysis and figures can be accessed via \url{https://github.com/svenbuder/Accreted-stars-in-GALAH-DR3} and is also marked behind each figure with a link icon to this repository.

\bibliographystyle{mnras}
\bibliography{bib} 

{\noindent \rule{8.5cm}{1pt}
\noindent
$^{1}$Research School of Astronomy \& Astrophysics, Australian National University, Canberra, ACT 2611, Australia\\
$^{2}$ARC Centre of Excellence for All Sky Astrophysics in 3 Dimensions (ASTRO 3D), Australia\\
$^{3}$Department of Astronomy, Stockholm University, AlbaNova University Centre, SE-106 91 Stockholm, Sweden\\
$^{4}$Department of Astronomy, Columbia University, Pupin Physics Laboratories, New York, NY 10027, USA\\
$^{5}$Center for Computational Astrophysics, Flatiron Institute, 162 Fifth Avenue, New York, NY 10010, USA\\
$^{6}$Lund Observatory, Department of Astronomy \& Theoretical Physics, Box 43, SE-221 00 Lund, Sweden\\
$^{7}$Astrophysics Research Institute, Liverpool John Moores University, 146 Brownlow Hill, Liverpool L3 5RF, UK \\
$^{8}$Leibniz-Institut f{\"u}r Astrophysik Potsdam (AIP), An der Sternwarte 16, D-14482 Potsdam, Germany\\
$^{9}$Sydney Institute for Astronomy, School of Physics, A28, The University of Sydney, NSW 2006, Australia\\
$^{10}$Monash Centre for Astrophysics, Monash University, Australia\\
$^{11}$School of Physics \& Astronomy, Monash University, Australia\\
$^{12}$Department of Physics \& Astronomy, Macquarie University, Sydney, NSW 2109, Australia\\
$^{13}$Istituto Nazionale di Astrofisica, Osservatorio Astronomico di Padova, vicolo dell'Osservatorio 5, 35122, Padova, Italy\\
$^{14}$Faculty of Mathematics \& Physics, University of Ljubljana, Jadranska 19, 1000 Ljubljana, Slovenia\\
$^{15}$School of Physics, UNSW, Sydney, NSW 2052, Australia\\
$^{16}$Stellar Astrophysics Centre, Aarhus University, Ny Munkegade 120, DK-8000 Aarhus C, Denmark\\
$^{17}$Macquarie University Research Centre for Astronomy, Astrophysics \& Astrophotonics, Sydney, NSW 2109, Australia\\
$^{18}$Mullard Space Science Laboratory, University College London, Holmbury St. Mary, Dorking, Surrey, RH5 6NT, UK\\
$^{19}$Centre for Astrophysics, University of Southern Queensland, Toowoomba, QLD 4350, Australia\\
$^{20}$Centre for Astrophysics Research, University of Hertfordshire, Hatfield, AL10 9AB, UK\\
$^{21}$Research School of Computer Science, Australian National University, Acton ACT 2601, Australia\\
$^{22}$Center for Astrophysical Sciences \& Department of Physics \& Astronomy, The Johns Hopkins University, Baltimore, MD 21218
}

\appendix

\input{tables/selection_techniques} 

\section{Selection and visualisation techniques of halo/accreted stars} \label{sec:selection_techniques}

Many selections of major substructures in the halo via spatial, photometric, kinematic, dynamic, and chemical properties (or a combination of those) exist - we list several of them, categorised by the used selection properties in Table~\ref{tab:selection_techniques}.
\newpage
We acknowledge that this list is incomplete and not all studies that used the listed techniques are included. We hope this list is of use for giving the reader an overview of the existing techniques to uncover major merger events in the early history of the Milky Way \citep[see][for a review]{Helmi2020}.

\bsp	
\label{lastpage}
\end{document}

%% file: depending_text/chem_percentiles_sqrt_J_R.tex
$26_{-14}^{+9}$%

%% file: depending_text/percentage_vtot.tex
3.1\% (13296 spectra)%

%% file: depending_text/percentage_vtan.tex
2.3\% (9894 spectra)%

%% file: depending_text/median_uncertainties_fehmgfealphafe.tex
0.09, 0.10, and 0.04%

%% file: depending_text/high_alpha_halo_fehmgfealphafe.tex
$\mathrm{[Fe/H]} = -0.65_{-0.43}^{+0.24}$, $\mathrm{[Mg/Fe]} = 0.29_{-0.11}^{+0.11}$, and$\mathrm{[\alpha/Fe]} = 0.27_{-0.07}^{+0.08}$

%% file: depending_text/nr_preliminary_low_alpha_halo.tex
4164

%% file: depending_text/nr_preliminary_low_alpha_halo_na.tex
3838

%% file: tables/xfe_percentiles.tex
\begingroup
\renewcommand{\arraystretch}{1.19}
\begin{table*}
\centering
\caption{\textbf{Numbers of measurements and statistic properties of element abundances [X/Y] of the preliminary selected low-$\alpha$ ($l$) and high-$\alpha$ ($h$) halo stars.} For each abundance ratio, we report 16/50/$84^\text{th}$ percentiles. We further calculate mean $\mu_i$, standard deviation $\sigma_i$, and skewness $\tilde{\mu}_{i,3}$ after performing 2-$\sigma$-clipping (removing the top/bottom $2.275\%$ of the sample). In addition to the difference of the means we report their significance $r$. Major element groups are separated by horizontal lines: firstly [Fe/H] followed by light, $\alpha$-process elements, light odd Z, iron-peak, and neutron-capture elements.  \added{We caution that the values of $r$ for both $\alpha$ and Mg depend on where the line is drawn between low- and high-$\alpha$ samples.}}
\label{tab:xfe_percentiles}
\begin{tabular}{c|cccc|cccc|cc}
\hline \hline
\multirow{2}{*}{$\mathrm{[X/Y]}$} & \multicolumn{4}{c}{Prel. low-$\alpha$ halo (Eq.~\protect\ref{eq:prelim_low_alpha_halo})} & \multicolumn{4}{c}{Prel. high-$\alpha$ halo (Eq.~\protect\ref{eq:prelim_high_alpha_halo})} & \multirow{2}{*}{$\mu_l - \mu_h$} & \multirow{2}{*}{$r = \frac{\vert \mu_l - \mu_h \vert}{\sqrt{\sigma_l^2 + \sigma_h^2}}$}\\
 & Nr. & Perc. 16/50/84 & $\mu_l \pm \sigma_l$ & $\tilde{\mu}_{l,3}$ & Nr. & Perc. 16/50/84 & $\mu_h \pm \sigma_h$ & $\tilde{\mu}_{h,3}$ & &  \\
\hline
$\mathrm{[Fe/H]}$ & 3838 & $-1.15_{-0.37}^{+0.39}$ & $-1.15 \pm 0.33$ & $-0.04$ & 5230 & $-0.66_{-0.29}^{+0.16}$ & $-0.70 \pm 0.20$ & $-0.93$ & $-0.45$ & $1.16$  \\
\hline
$\mathrm{[Li/Fe]}$ & 525 & $1.28_{-0.58}^{+0.82}$ & $1.33 \pm 0.63$ & $0.18$ & 548 & $0.92_{-0.85}^{+1.08}$ & $1.00 \pm 0.82$ & $0.16$ & $0.34$ & $0.33$  \\
$\mathrm{[C/Fe]}$ & 25 & $0.77_{-0.26}^{+0.61}$ & $0.87 \pm 0.36$ & $0.38$ & 62 & $0.60_{-0.29}^{+0.39}$ & $0.66 \pm 0.28$ & $0.18$ & $0.21$ & $0.47$  \\
$\mathrm{[O/Fe]}$ & 3090 & $0.53_{-0.23}^{+0.26}$ & $0.54 \pm 0.22$ & $0.28$ & 4929 & $0.57_{-0.18}^{+0.20}$ & $0.58 \pm 0.18$ & $0.34$ & $-0.04$ & $0.15$  \\
\hline
$\mathrm{[\alpha/Fe]}$ & 3838 & $0.15_{-0.08}^{+0.07}$ & $0.15 \pm 0.07$ & $-0.24$ & 5230 & $0.28_{-0.05}^{+0.07}$ & $0.29 \pm 0.06$ & $0.59$ & $-0.14$ & $1.58$  \\
$\mathrm{[Mg/Fe]}$ & 3838 & $0.12_{-0.11}^{+0.08}$ & $0.12 \pm 0.09$ & $-0.53$ & 5230 & $0.33_{-0.06}^{+0.10}$ & $0.34 \pm 0.08$ & $0.83$ & $-0.23$ & $1.98$  \\
$\mathrm{[Si/Fe]}$ & 3750 & $0.14_{-0.10}^{+0.10}$ & $0.14 \pm 0.09$ & $0.12$ & 5174 & $0.27_{-0.08}^{+0.11}$ & $0.28 \pm 0.09$ & $0.63$ & $-0.14$ & $1.09$  \\
$\mathrm{[Ca/Fe]}$ & 3716 & $0.21_{-0.11}^{+0.10}$ & $0.20 \pm 0.10$ & $-0.26$ & 5045 & $0.26_{-0.11}^{+0.11}$ & $0.26 \pm 0.10$ & $0.13$ & $-0.06$ & $0.42$  \\
$\mathrm{[Ti/Fe]}$ & 3543 & $0.17_{-0.12}^{+0.14}$ & $0.18 \pm 0.13$ & $0.67$ & 5015 & $0.27_{-0.09}^{+0.11}$ & $0.28 \pm 0.10$ & $0.73$ & $-0.10$ & $0.62$  \\
\hline
$\mathrm{[Na/Fe]}$ & 3838 & $-0.18_{-0.14}^{+0.18}$ & $-0.17 \pm 0.15$ & $0.31$ & 5230 & $0.10_{-0.11}^{+0.10}$ & $0.10 \pm 0.10$ & $-0.03$ & $-0.27$ & $1.52$  \\
$\mathrm{[Al/Fe]}$ & 1580 & $-0.01_{-0.18}^{+0.25}$ & $0.01 \pm 0.20$ & $0.53$ & 4777 & $0.31_{-0.14}^{+0.12}$ & $0.30 \pm 0.12$ & $-0.29$ & $-0.29$ & $1.26$  \\
$\mathrm{[K/Fe]}$ & 3769 & $0.11_{-0.14}^{+0.12}$ & $0.10 \pm 0.12$ & $0.00$ & 5142 & $0.17_{-0.15}^{+0.16}$ & $0.17 \pm 0.14$ & $0.22$ & $-0.07$ & $0.37$  \\
\hline
$\mathrm{[Sc/Fe]}$ & 3805 & $0.06_{-0.09}^{+0.09}$ & $0.07 \pm 0.08$ & $0.02$ & 5198 & $0.14_{-0.08}^{+0.09}$ & $0.15 \pm 0.08$ & $0.27$ & $-0.08$ & $0.72$  \\
$\mathrm{[V/Fe]}$ & 1310 & $0.02_{-0.29}^{+0.32}$ & $0.04 \pm 0.30$ & $0.69$ & 2841 & $0.22_{-0.22}^{+0.33}$ & $0.27 \pm 0.26$ & $0.73$ & $-0.23$ & $0.57$  \\
$\mathrm{[Cr/Fe]}$ & 3586 & $-0.15_{-0.13}^{+0.13}$ & $-0.15 \pm 0.12$ & $0.08$ & 5101 & $-0.06_{-0.10}^{+0.10}$ & $-0.06 \pm 0.10$ & $0.29$ & $-0.09$ & $0.62$  \\
$\mathrm{[Mn/Fe]}$ & 3811 & $-0.36_{-0.12}^{+0.14}$ & $-0.36 \pm 0.12$ & $0.16$ & 5172 & $-0.19_{-0.12}^{+0.12}$ & $-0.19 \pm 0.11$ & $0.08$ & $-0.17$ & $1.05$  \\
$\mathrm{[Co/Fe]}$ & 1587 & $-0.07_{-0.13}^{+0.35}$ & $0.03 \pm 0.30$ & $1.85$ & 3844 & $0.09_{-0.12}^{+0.12}$ & $0.11 \pm 0.16$ & $2.08$ & $-0.08$ & $0.23$  \\
$\mathrm{[Ni/Fe]}$ & 3066 & $-0.15_{-0.12}^{+0.12}$ & $-0.14 \pm 0.11$ & $0.12$ & 4813 & $0.04_{-0.11}^{+0.09}$ & $0.04 \pm 0.09$ & $-0.08$ & $-0.18$ & $1.27$  \\
$\mathrm{[Cu/Fe]}$ & 2613 & $-0.49_{-0.14}^{+0.28}$ & $-0.45 \pm 0.19$ & $0.86$ & 4875 & $0.01_{-0.19}^{+0.13}$ & $-0.01 \pm 0.14$ & $-0.53$ & $-0.44$ & $1.82$  \\
$\mathrm{[Zn/Fe]}$ & 3629 & $0.16_{-0.15}^{+0.18}$ & $0.17 \pm 0.16$ & $0.37$ & 4824 & $0.21_{-0.16}^{+0.23}$ & $0.23 \pm 0.19$ & $0.38$ & $-0.07$ & $0.28$  \\
\hline
$\mathrm{[Rb/Fe]}$ & 124 & $0.12_{-0.22}^{+0.86}$ & $0.34 \pm 0.48$ & $0.75$ & 905 & $0.13_{-0.16}^{+0.19}$ & $0.15 \pm 0.18$ & $0.84$ & $0.19$ & $0.37$  \\
$\mathrm{[Sr/Fe]}$ & 126 & $1.02_{-0.58}^{+0.48}$ & $0.97 \pm 0.44$ & $-0.11$ & 386 & $0.74_{-0.39}^{+0.59}$ & $0.81 \pm 0.42$ & $0.40$ & $0.16$ & $0.26$  \\
$\mathrm{[Y/Fe]}$ & 3582 & $0.08_{-0.22}^{+0.25}$ & $0.09 \pm 0.22$ & $0.48$ & 4813 & $0.11_{-0.25}^{+0.31}$ & $0.13 \pm 0.27$ & $0.68$ & $-0.04$ & $0.12$  \\
$\mathrm{[Zr/Fe]}$ & 1311 & $0.26_{-0.25}^{+0.41}$ & $0.34 \pm 0.36$ & $1.33$ & 2653 & $0.20_{-0.22}^{+0.36}$ & $0.26 \pm 0.30$ & $1.15$ & $0.08$ & $0.17$  \\
$\mathrm{[Ba/Fe]}$ & 3822 & $0.31_{-0.30}^{+0.33}$ & $0.32 \pm 0.29$ & $0.41$ & 5216 & $0.14_{-0.28}^{+0.37}$ & $0.18 \pm 0.30$ & $0.67$ & $0.14$ & $0.34$  \\
$\mathrm{[La/Fe]}$ & 2441 & $0.25_{-0.18}^{+0.31}$ & $0.31 \pm 0.25$ & $1.16$ & 3497 & $0.17_{-0.16}^{+0.30}$ & $0.23 \pm 0.25$ & $1.40$ & $0.08$ & $0.22$  \\
$\mathrm{[Ce/Fe]}$ & 1083 & $-0.16_{-0.14}^{+0.25}$ & $-0.11 \pm 0.21$ & $1.39$ & 2140 & $-0.17_{-0.12}^{+0.19}$ & $-0.13 \pm 0.20$ & $2.09$ & $0.02$ & $0.05$  \\
\hline
$\mathrm{[Ru/Fe]}$ & 242 & $0.42_{-0.21}^{+0.46}$ & $0.51 \pm 0.33$ & $1.26$ & 850 & $0.36_{-0.18}^{+0.31}$ & $0.42 \pm 0.26$ & $1.36$ & $0.10$ & $0.22$  \\
$\mathrm{[Nd/Fe]}$ & 2954 & $0.47_{-0.16}^{+0.20}$ & $0.49 \pm 0.18$ & $0.65$ & 3765 & $0.34_{-0.14}^{+0.19}$ & $0.37 \pm 0.17$ & $1.22$ & $0.12$ & $0.48$  \\
$\mathrm{[Eu/Fe]}$ & 1841 & $0.44_{-0.16}^{+0.18}$ & $0.44 \pm 0.16$ & $0.25$ & 3045 & $0.30_{-0.11}^{+0.12}$ & $0.31 \pm 0.11$ & $0.43$ & $0.13$ & $0.70$  \\
\hline
\end{tabular}
\end{table*}
\endgroup

%% file: depending_text/lah_p16_p84_Li.tex
0.71$  and $2.10%

%% file: depending_text/lah_p16_p84_O.tex
0.30$  and $0.79%

%% file: depending_text/skewness_O_lah.tex
0.28%

%% file: depending_text/alpha_scatter.tex
0.07, 0.09, 0.09, 0.10, and 0.13 for [$\alpha$/Fe], [Mg/Fe], [Si/Fe], [Ca/Fe], and [Ti/Fe], compared to the much higher value of 0.22 for [O/Fe]%

%% file: depending_text/skewness_Mg_lah.tex
-0.53%

%% file: depending_text/skewness_Ti_lah.tex
0.67%

%% file: depending_text/lah_Ti_detection_percentage.tex
92\%%

%% file: depending_text/low_alpha_halo_nafe_0.tex
602 spectra, that is, 16\%)%

%% file: depending_text/lah_Al_detection_percentage.tex
41\%%

%% file: depending_text/skewness_Al_lah.tex
0.53%

%% file: depending_text/lah_K_detection_percentage.tex
98\%%

%% file: depending_text/lah_Ni_detection_percentage.tex
80\%%

%% file: depending_text/lah_Cu_detection_percentage.tex
68\%%

%% file: depending_text/lah_Co_detection_percentage.tex
41\%%

%% file: depending_text/lah_V_detection_percentage.tex
34\%%

%% file: depending_text/skewness_Cu_lah.tex
0.86%

%% file: depending_text/skewness_Co_lah.tex
1.85%

%% file: depending_text/skewness_V_lah.tex
0.69%

%% file: depending_text/lah_p16_p84_V.tex
-0.27$  and $0.34%

%% file: depending_text/lah_mul_muh_diff_Cu.tex
0.44%

%% file: depending_text/lah_mul_sigmal_Cr.tex
-0.15 \pm 0.12%

%% file: depending_text/lah_mul_sigmal_Mn.tex
-0.36 \pm 0.12%

%% file: depending_text/lah_mul_sigmal_Ni.tex
-0.14 \pm 0.11%

%% file: depending_text/lah_mul_sigmal_Zn.tex
0.17 \pm 0.16%

%% file: depending_text/lah_mul_sigmal_Y.tex
0.09 \pm 0.22%

%% file: depending_text/lah_mul_muh_diff_Y.tex
0.04%

%% file: depending_text/lah_mul_sigmal_Ba.tex
0.32 \pm 0.29%

%% file: depending_text/lah_Zr_detection_percentage.tex
34\%%

%% file: depending_text/lah_La_detection_percentage.tex
64\%%

%% file: depending_text/lah_Ce_detection_percentage.tex
28\%%

%% file: depending_text/skewness_Y_lah.tex
0.48%

%% file: depending_text/skewness_Zr_lah.tex
1.33%

%% file: depending_text/skewness_Ba_lah.tex
0.41%

%% file: depending_text/skewness_La_lah.tex
1.16%

%% file: depending_text/skewness_Ce_lah.tex
1.39%

%% file: depending_text/lah_mul_sigmal_Ce.tex
-0.11 \pm 0.21%

%% file: depending_text/lah_mul_sigmal_Nd.tex
0.49 \pm 0.18%

%% file: depending_text/lah_mul_sigmal_Eu.tex
0.44 \pm 0.16%

%% file: depending_text/lah_mul_muh_diff_Nd.tex
0.12%

%% file: depending_text/lah_mul_muh_diff_Eu.tex
0.13%

%% file: depending_text/pearsonr_alpha_Mg.tex
$\alpha$-Mg (0.63)%

%% file: depending_text/pearsonr_alpha_Si.tex
$\alpha$-Si (0.63)%

%% file: depending_text/pearsonr_Na_Al.tex
Na-Al (0.71)%

%% file: depending_text/pearsonr_Mn_Cu.tex
Mn-Cu (0.56)%

%% file: depending_text/pearsonr_Ni_Cu.tex
Ni-Cu (0.64)%

%% file: depending_text/pearsonr_Mg_Si.tex
Mg-Si (0.36)%

%% file: depending_text/pearsonr_Mn_Ni.tex
Mn-Ni (0.40)%

%% file: tables/gmm_sampling.tex
\begin{table}
\centering
\caption{Overview of the combinations used for the Simple Gaussian Mixture Models to estimate the number of components to sample out. The GMM input, consisting of the number of data points with each combination as input array has yielded the lowest BIC score for the number of components lists.}
\label{tab:sample_gmm}
\begin{tabular}{cccc}
\hline \hline
Set & \multicolumn{2}{c}{Input for simple GMMs (see Sec.~\ref{sec:sample_gmm})} & Comp. \\
 & Combination & Data Points & Nr. \\
\hline
\texttt{Mg\_Mn}   & [Mg/Fe], [Mn/Fe]  & 26810 & 5 \\
\texttt{MgH\_Mn}  & [Mg/H], [Mn/Fe]   & 26810 & 4 \\
\texttt{Mg\_Na\_Mn}  & [Mg/Fe], [Na/Fe], [Mn/Fe] & 26057 & 7 \\
\texttt{MgH\_Na\_Mn} & [Mg/H], [Na/Fe], [Mn/Fe] & 26057 & 8 \\
\texttt{MgMn\_Na}  & [Mg/Mn], [Na/Fe]  & 26057 & 6 \\
\texttt{MgCu\_Na}  & [Mg/Cu], [Na/Fe]  & 20974 & 4 \\
\hline
\texttt{MgH\_Na}  & [Mg/H], [Na/Fe]   & 26670 & 5 \\
\texttt{Mg\_Na\_Cu}  & [Mg/Fe], [Na/Fe], [Cu/Fe] & 20974 & 8 \\
\texttt{Mg\_Na\_Mn\_Cu} & [Mg/Fe], [Na/Fe],   & 20693 & 9 \\
     & [Mn/Fe], [Cu/Fe]  &   &  \\
\texttt{all\_6}   & [Mg/Fe], [Si/Fe], [Na/Fe], & 18544 & 7 \\
     & [Mn/Fe], [Ni/Fe], [Cu/Fe] &   &  \\
\texttt{all\_6\_rel}  & [Mg/Mn], [Si/Cu],  & 18544 & 5 \\
     & [Na/Fe], [Ni/Fe]   &   &  \\
   \hline
\end{tabular}
\end{table}

%% file: tables/simple_gmm_selection.tex
\begin{table*}
\centering
\caption{Sources selected via the different chemical selections. We highlight the probability in bold face, if it is the largest among the fitted components. The full table (including all GMM components) is available online together with a crossmatch with the GALAH+DR3 main and value-added-catalogs in a FITS file.}
\label{tab:simple_gmm_selection}
\setlength{\tabcolsep}{0.6em}
\begin{tabular}{cccccccccccccc}
\hline
GALAH+ DR3 & \multicolumn{2}{c}{\texttt{Mg\_Mn}} & \multicolumn{2}{c}{\texttt{MgH\_Mn}} & \multicolumn{2}{c}{\texttt{Mg\_Na\_Mn}} & \multicolumn{3}{c}{\texttt{MgH\_Na\_Mn}} & \multicolumn{2}{c}{\texttt{MgMn\_Na}} & \multicolumn{2}{c}{\texttt{MgCu\_Na}} \\
sobject\_id & Ac. MR & MP-i$\alpha$ & Ac. MR & Ac. MP & Ac. MR & MP-i$\alpha$ & Ac. MR & Ac. MP & MP-i$\alpha$ & Ac. MR & MP-i$\alpha$ & Ac. MR & MP-i$\alpha$ \\
\hline
131116000501004 & \textbf{0.65} & 0.09 & \textbf{0.54} & 0.27 & 0.3 & 0.13 & 0.27 & 0.17 & 0.25 & \textbf{0.33} & 0.25 & nan & nan \\
131116000501008 & 0.11 & 0.07 & 0.21 & 0.21 & 0.0 & 0.0 & 0.0 & 0.08 & 0.01 & 0.0 & 0.0 & nan & nan \\
131116000501014 & \textbf{0.41} & 0.07 & \textbf{0.45} & 0.34 & 0.4 & 0.1 & \textbf{0.31} & 0.16 & 0.24 & \textbf{0.52} & 0.13 & nan & nan \\
131116000501018 & \textbf{0.23} & 0.16 & \textbf{0.45} & 0.18 & 0.1 & 0.16 & 0.15 & 0.06 & 0.31 & 0.15 & \textbf{0.21} & \textbf{0.65} & 0.23 \\
131116000501022 & 0.01 & 0.14 & 0.01 & 0.0 & 0.0 & 0.02 & 0.0 & 0.0 & 0.0 & 0.0 & 0.07 & 0.03 & 0.35 \\
\dots  & \dots  & \dots  & \dots  & \dots  & \dots  & \dots  & \dots  & \dots  & \dots  & \dots  & \dots  & \dots  & \dots \\
\hline
\end{tabular}
\end{table*}

%% file: depending_text/Mg_Mn_dist1_NaFe.tex
$\mathrm{[Na/Fe]} = -0.12_{-0.19}^{+0.22}$%

%% file: depending_text/MgH_Mn_dist2_MgH.tex
$\mathrm{[Mg/H]} = -1.52_{-0.34}^{+0.24}$%

%% file: depending_text/MgH_Mn_dist1_MgH.tex
$\mathrm{[Mg/H]} = -0.97_{-0.23}^{+0.18}$%

%% file: depending_text/MgH_Mn_dist2_MgFe.tex
$\mathrm{[Mg/Fe]} = 0.15_{-0.12}^{+0.11}$%

%% file: depending_text/MgH_Mn_dist1_MgFe.tex
$\mathrm{[Mg/Fe]} = 0.18_{-0.13}^{+0.10}$%

%% file: depending_text/MgH_Mn_dist1_NaFe.tex
$\mathrm{[Mg/Fe]} = -0.14_{-0.16}^{+0.17}$%

%% file: depending_text/MgMn_Na_dist1_Mn.tex
$-0.37_{-0.13}^{+0.12}$%

%% file: depending_text/MgMn_Na_dist2_Mn.tex
$-0.35_{-0.10}^{+0.09}$%

%% file: tables/chemodyn_comparison.tex
\begingroup
\renewcommand{\arraystretch}{1.14}
\begin{table}
\centering
\caption{Chronochemodynamic properties (shown as $16^\text{th}$/$50^\text{th}$/$84^\text{th}$ percentiles) of the chemical and dynamical selection of accreted stars. We further list the properties of the stars that overlap between both selections. The selection criteria are explained in detail in Secs.~\ref{sec:xdgmm_selection} and \ref{sec:dynamical_selection}, respectively. Only distributions with more than 100 measurements are shown. Values in parentheses may be biased because they were used for the selection.}
\label{tab:chronochemodynamic_properties}
\begin{tabular}{cccc}
\hline
Property & Chemical & Chemodynamical & Dynamical \\
& Selection & Selection & Selection \\
\hline \hline
$\mathrm{[Fe/H]}$ & $-1.11_{-0.30}^{+0.28}$ & $-1.03_{-0.29}^{+0.26}$ & $-1.16_{-0.41}^{+0.32}$ \\
$\mathrm{[\alpha/Fe]}$ & $0.11_{-0.08}^{+0.07}$ & $0.10_{-0.06}^{+0.07}$ & $0.16_{-0.09}^{+0.12}$ \\
$\mathrm{[Li/Fe]}$ & - & - & $1.94_{-1.06}^{+0.97}$ \\
$\mathrm{[C/Fe]}$ & - & - & - \\
$\mathrm{[O/Fe]}$ & $0.54_{-0.23}^{+0.24}$ & $0.51_{-0.21}^{+0.17}$ & $0.55_{-0.23}^{+0.27}$ \\
$\mathrm{[Na/Fe]}$ & ($-0.35_{-0.09}^{+0.05}$) & ($-0.35_{-0.09}^{+0.06}$) & $-0.22_{-0.13}^{+0.17}$ \\
$\mathrm{[Mg/Fe]}$ & ($0.09_{-0.09}^{+0.09}$) & ($0.07_{-0.09}^{+0.09}$) & $0.12_{-0.10}^{+0.12}$ \\
$\mathrm{[Mg/Mn]}$ & ($0.52_{-0.17}^{+0.15}$) & ($0.48_{-0.17}^{+0.12}$) & $0.47_{-0.16}^{+0.14}$ \\
$\mathrm{[Al/Fe]}$ & $-0.18_{-0.12}^{+0.18}$ & $-0.18_{-0.12}^{+0.11}$ & $-0.12_{-0.13}^{+0.26}$ \\
$\mathrm{[Si/Fe]}$ & $0.10_{-0.10}^{+0.11}$ & $0.09_{-0.09}^{+0.11}$ & $0.13_{-0.10}^{+0.14}$ \\
$\mathrm{[K/Fe]}$ & $0.08_{-0.15}^{+0.13}$ & $0.08_{-0.15}^{+0.12}$ & $0.10_{-0.14}^{+0.14}$ \\
$\mathrm{[Ca/Fe]}$ & $0.17_{-0.11}^{+0.09}$ & $0.17_{-0.10}^{+0.11}$ & $0.23_{-0.13}^{+0.13}$ \\
$\mathrm{[Sc/Fe]}$ & $0.05_{-0.11}^{+0.09}$ & $0.05_{-0.11}^{+0.08}$ & $0.07_{-0.10}^{+0.11}$ \\
$\mathrm{[Ti/Fe]}$ & $0.10_{-0.12}^{+0.13}$ & $0.10_{-0.13}^{+0.14}$ & $0.19_{-0.13}^{+0.22}$ \\
$\mathrm{[V/Fe]}$ & $-0.06_{-0.29}^{+0.33}$ & $-0.07_{-0.28}^{+0.30}$ & $0.04_{-0.32}^{+0.32}$ \\
$\mathrm{[Cr/Fe]}$ & $-0.22_{-0.13}^{+0.12}$ & $-0.20_{-0.13}^{+0.11}$ & $-0.13_{-0.12}^{+0.14}$ \\
$\mathrm{[Mn/Fe]}$ & $-0.43_{-0.12}^{+0.12}$ & $-0.40_{-0.10}^{+0.13}$ & $-0.34_{-0.12}^{+0.14}$ \\
$\mathrm{[Co/Fe]}$ & $-0.11_{-0.11}^{+0.31}$ & $-0.14_{-0.11}^{+0.39}$ & $-0.08_{-0.13}^{+0.50}$ \\
$\mathrm{[Ni/Fe]}$ & $-0.18_{-0.10}^{+0.11}$ & $-0.19_{-0.10}^{+0.11}$ & $-0.15_{-0.12}^{+0.12}$ \\
$\mathrm{[Cu/Fe]}$ & $-0.57_{-0.12}^{+0.12}$ & $-0.58_{-0.10}^{+0.12}$ & $-0.52_{-0.13}^{+0.16}$ \\
$\mathrm{[Zn/Fe]}$ & $0.17_{-0.18}^{+0.23}$ & $0.14_{-0.17}^{+0.22}$ & $0.15_{-0.16}^{+0.19}$ \\
$\mathrm{[Rb/Fe]}$ & - & - & - \\
$\mathrm{[Sr/Fe]}$ & - & - & - \\
$\mathrm{[Y/Fe]}$ & $0.12_{-0.24}^{+0.25}$ & $0.10_{-0.22}^{+0.21}$ & $0.09_{-0.22}^{+0.25}$ \\
$\mathrm{[Zr/Fe]}$ & $0.19_{-0.22}^{+0.39}$ & $0.17_{-0.23}^{+0.61}$ & $0.27_{-0.24}^{+0.64}$ \\
$\mathrm{[Ba/Fe]}$ & $0.49_{-0.33}^{+0.28}$ & $0.43_{-0.31}^{+0.29}$ & $0.25_{-0.29}^{+0.36}$ \\
$\mathrm{[La/Fe]}$ & $0.23_{-0.17}^{+0.27}$ & $0.24_{-0.15}^{+0.27}$ & $0.28_{-0.17}^{+0.45}$ \\
$\mathrm{[Ce/Fe]}$ & $-0.17_{-0.15}^{+0.20}$ & $-0.16_{-0.17}^{+0.20}$ & $-0.12_{-0.17}^{+0.41}$ \\
$\mathrm{[Ru/Fe]}$ & - & - & - \\
$\mathrm{[Nd/Fe]}$ & $0.48_{-0.15}^{+0.20}$ & $0.49_{-0.12}^{+0.17}$ & $0.52_{-0.15}^{+0.21}$ \\
$\mathrm{[Eu/Fe]}$ & $0.46_{-0.15}^{+0.16}$ & $0.46_{-0.12}^{+0.16}$ & $0.48_{-0.13}^{+0.17}$ \\
\hline
$\sqrt{J_R~/~\mathrm{kpc\,km\,s^{-1}}}$ & $26_{-14}^{+9}$ & ($35_{-3}^{+6}$) & ($35_{-3}^{+6}$) \\
$L_Z~/~\mathrm{kpc\,km\,s^{-1}}$ & $100_{-430}^{+510}$ & ($10_{-250}^{+280}$) & ($10_{-230}^{+250}$) \\
$J_Z~/~\mathrm{kpc\,km\,s^{-1}}$ & $200_{-130}^{+240}$ & $160_{-110}^{+240}$ & $140_{-100}^{+230}$ \\
$V_R~/~\mathrm{km\,s^{-1}}$ & $-0_{-210}^{+190}$ & $-130_{-160}^{+390}$ & $20_{-300}^{+250}$ \\
$V_\phi~/~\mathrm{km\,s^{-1}}$ & $20_{-70}^{+100}$ & $2_{-32}^{+36}$ & $2_{-30}^{+34}$ \\
$e$ & $0.88_{-0.36}^{+0.09}$ & $0.96_{-0.04}^{+0.03}$ & $0.96_{-0.04}^{+0.03}$ \\
$E~/~10^5\,\mathrm{km^{2}\,s^{-2}}$ & $-1.61_{-0.26}^{+0.22}$ & $-1.45_{-0.10}^{+0.15}$ & $-1.46_{-0.11}^{+0.15}$ \\
$R_\text{ap}~/~\mathrm{kpc}$ & $11.7_{-5.2}^{+6.7}$ & $16.5_{-3.1}^{+5.7}$ & $16.2_{-3.1}^{+5.7}$ \\
$R_\text{peri}~/~\mathrm{kpc}$ & $0.78_{-0.59}^{+1.98}$ & $0.37_{-0.26}^{+0.45}$ & $0.34_{-0.25}^{+0.41}$ \\
$z_\text{max}~/~\mathrm{kpc}$ & $4.7_{-2.4}^{+5.4}$ & $6.7_{-3.6}^{+5.0}$ & $6.0_{-3.5}^{+5.4}$ \\
\hline
\end{tabular}
\end{table}
\endgroup

%% file: tables/selected_stars.tex
\begin{table}
\centering
\caption{Overview of sources selected as accreted. We list the normalised probability $p$ of sources to be selected chemically via the \textsc{xdgmm} of $\mathrm{[Na/Fe]}$ vs. $\mathrm{[Mg/Mn]}$ as well as those dynamically selected based on the suggestion by \citet{Feuillet2021} in the $L_Z$ vs. $\sqrt{J_R}$ plane. The selection criteria are explained in detail in Secs.~\ref{sec:xdgmm_selection} and \ref{sec:dynamical_selection}, respectively. Chemically selected stars, as selected for the analysis throughout this study with $p > 0.45$ are marked in bold.}
\label{tab:xdgmm_dynamical_selection}
\setlength{\tabcolsep}{0.6em}
\begin{tabular}{ccc}
\hline
GALAH+ DR3 & Chemical selection & Dynamical Selection \\
sobject\_id & $p (\mathrm{[Na/Fe]},\mathrm{[Mg/Mn]})$ & $p (L_Z\text{ vs. }\sqrt{J_R})$ \\
\hline
131116000501004 & 0.12 & 0 \\
131116000501201 & 0.03 & 1 \\
140209001701097 & \textbf{0.58} & 1 \\
140209003701238 & \textbf{0.89} & 0 \\
\dots  & \dots  & \dots  \\
\hline
\end{tabular}
\end{table}

%% file: depending_text/chem_percentiles_Na_fe.tex
$-0.35_{-0.09}^{+0.05}$%

%% file: depending_text/dyn_percentiles_Na_fe.tex
$-0.22_{-0.13}^{+0.17}$%

%% file: depending_text/chem_percentiles_MgMn_fe.tex
$0.52_{-0.17}^{+0.15}$%

%% file: depending_text/dyn_percentiles_MgMn_fe.tex
$0.47_{-0.16}^{+0.14}$%

%% file: depending_text/dyn_percentiles_sqrt_J_R.tex
$35_{-3}^{+6}$%

%% file: depending_text/chem_percentiles_L_Z.tex
$100_{-430}^{+510}$%

%% file: depending_text/dyn_percentiles_L_Z.tex
$10_{-230}^{+250}$%

%% file: depending_text/chem_below_20_20.tex
33\%%

%% file: depending_text/chem_below_15_15.tex
22\%%

%% file: depending_text/chem_below_10_10.tex
12\%%

%% file: depending_text/chem_above_500.tex
20\%%

%% file: depending_text/chem_above_1000.tex
8\%%

%% file: depending_text/chem_percentiles_fe_h.tex
$-1.11_{-0.30}^{+0.28}$%

%% file: depending_text/dyn_percentiles_fe_h.tex
$-1.16_{-0.41}^{+0.32}$%

%% file: depending_text/diff_chem_chemdyn_fe_h.tex
0.08%

%% file: depending_text/diff_dyn_chemdyn_fe_h.tex
0.13%

%% file: depending_text/chem_high_A_Li.tex
$2.20_{-0.07}^{+0.30}$%

%% file: depending_text/dyn_high_A_Li.tex
$2.37_{-0.16}^{+0.14}$%

%% file: depending_text/chem_low_A_Li.tex
$0.98_{-0.16}^{+0.15}$%

%% file: depending_text/dyn_low_A_Li.tex
$0.97_{-0.18}^{+0.24}$%

%% file: depending_text/chemdyn_percentiles_O_fe.tex
$0.51_{-0.21}^{+0.17}$%

%% file: depending_text/diff_chemdyn_Mg_fe.tex
0.03%

%% file: depending_text/diff_chemdyn_Si_fe.tex
0.03%

%% file: depending_text/diff_chemdyn_Ca_fe.tex
0.06%

%% file: depending_text/diff_chemdyn_Ti_fe.tex
0.09%

%% file: depending_text/above_84thchem_Mg_fe.tex
26\%%

%% file: depending_text/above_84thchem_Ti_fe.tex
39\%%

%% file: depending_text/chem_percentiles_Al_fe.tex
$-0.18_{-0.12}^{+0.18}$%

%% file: depending_text/dyn_percentiles_Al_fe.tex
$-0.12_{-0.13}^{+0.26}$%

%% file: depending_text/chemdyn_percentiles_Mn_fe.tex
$-0.40_{-0.10}^{+0.13}$%

%% file: depending_text/chemdyn_percentiles_Cu_fe.tex
$-0.58_{-0.10}^{+0.12}$%

%% file: depending_text/diff_chemdyn_Cr_fe.tex
0.08%

%% file: depending_text/dyn_percentiles_vR_Rzphi.tex
$20_{-300}^{+250}$%

%% file: depending_text/chem_percentiles_vR_Rzphi.tex
$-0_{-210}^{+190}$%

%% file: depending_text/chem_above_vphi_100kms.tex
21\%%

%% file: depending_text/chem_above_vphi_100kms_sqrt_J_R.tex
$12.4_{-7.1}^{+16.6}$%

%% file: depending_text/chem_percentiles_Energy_10_5.tex
$-1.61_{-0.26}^{+0.22}$%

%% file: depending_text/chem_energy_above_m18.tex
79\%%

%% file: depending_text/chem_energy_above_m20.tex
7\%%

%% file: depending_text/chem_energy_above_m20_R.tex
$2.6_{-1.1}^{+1.6}$%

%% file: depending_text/chem_energy_above_m20_absZ.tex
$1.6_{-0.4}^{+0.8}$%

%% file: depending_text/dyn_percentiles_ecc.tex
$0.96_{-0.04}^{+0.03}$%

%% file: depending_text/chem_percentiles_ecc.tex
$0.88_{-0.36}^{+0.09}$%

%% file: depending_text/chem_age_msto.tex
$11.3_{-3.1}^{+0.8}$%

%% file: depending_text/dyn_age_msto.tex
$11.4_{-3.2}^{+0.8}$%

%% file: depending_text/chem_L_Z_below_m500.tex
9\%%

%% file: depending_text/chem_L_Z_below_m700.tex
6\%%

%% file: depending_text/chem_L_Z_below_m700_number.tex
58%

%% file: depending_text/chem_L_Z_below_m700_all.tex
507%

%% file: depending_text/chem_wukong1.tex
4.3\%%

%% file: depending_text/chem_wukong2.tex
1.4\%%

%% file: depending_text/chem_percentiles_R_ap.tex
$11.7_{-5.2}^{+6.7}$%

%% file: depending_text/above_84thchem_Na_fe.tex
72\%%

%% file: depending_text/das2020_fe_h_dr16.tex
$-1.25_{-0.24}^{+0.33}$ %

%% file: depending_text/das2020_feh_chem.tex
$-1.42_{-0.16}^{+0.07}$%

%% file: depending_text/das2020_feh_chem_diff.tex
$0.22_{-0.01}^{+0.03}$%

%% file: tables/selection_techniques.tex
\begin{table*}    \centering
    \caption{
    \textbf{A compilation of different techniques to identify major accretion structures.}
    The list includes photometric information used in colour-magnitude diagrams (CMD), stellar kinematic properties such as Galactic longitude $l$ and latitude $b$, radial velocity $v_\text{rad}$, tangential velocity ($V_T$), total velocity ($V_\text{tot}$), Galactocentric Cartesian velocities ($V_X$, $V_Y$, and $V_Z$), Galactocentric cylindrical velocities ($V_R$, $V_\phi$, and $V_Z$), stellar dynamic properties such as maximum Galactocentric radius ($R_\text{max}$), actions ($J_R$, $J_\phi = L_Z$, $J_Z$, and total $J_\text{tot}$), eccentricity $e$, orbit energy $E$, as well as stellar chemical information such as the iron abundances relative to hydrogen [Fe/H], and element abundances of element X relative to iron [X/Fe]. $k$-means and Gaussian Mixture Models (GMM) are \texttt{scikit-learn} clustering algorithms \citep{scikit-learn}, whereas \textsc{StarGo} is a neutral-network-based clustering method \citet{Yuan2018}.
    We note that the references are not necessarily the first ones finding these properties, but examples of their application. In the case of [Na/Fe] vs. [Ni/Fe] for stars with high $V_\text{tot}$, the correlation has e.g. found by \citet{Nissen1997,Nissen2010} and discussed by \citet{Venn2004} before being applied explicitly by \citet{Bensby2014}.
    }
    \begin{tabular}{c|c|c}
        \hline \hline
        Category & Properties & Example Reference(s) \\
        \hline
        Stellar photometry & $m_i$ and/or $m_i - m_j$ & \citet{Belokurov2006} \\
        \hline
        Stellar kinematics & $V_X$, $V_Y$, $V_Z$, and $\sqrt{V_X^2+V_Z^2}$ & \citet{Koppelman2018} \\
		& $V_R$, $V_\phi$, $V_Z$ ellipsoid membership probability & \citet{Carollo2010} \\
		& \dots & \citet{Ishigaki2012, Ishigaki2013} \\
		& two-point velocity correlation function & \citet{ReFiorentin2015} \\
		& Neural-network based classification with \Gaia DR2 6D & \citet{Ostdiek2020} \\
		& (same as the previous) & \citet{Necib2020} \\
        \hline
        Stellar dynamics & $V_\phi$ and $R_\text{max}$ & \citet{Gratton2003} \\
        & $J_Z$ and $J_\perp = \sqrt{J_X^2 + J_Y^2}$ & \citet{Helmi1999} \\
        & $L_Z$ and $E$ & \citet{Helmi2017, Helmi2018} \\
        & $L_Z$, $E$, and $L_Z/\vert L_{Z,\text{circ}} \vert$ & \citet{Koppelman2019} \\
        & $J_\phi / J_\text{tot}$ and $(J_Z - J_R) / J_\text{tot}$ & \citet{Myeong2019} \\
        & $L_Z$ and $J_R$ & \citet{Feuillet2020} \\
        & $E$, $L$, $\theta = \arccos{L_Z/L}$, $\phi = \arctan{L_X/L_Y}$ via \textsc{StarGo} & \citet{Yuan2020} \\
        \hline
        Stellar kinematics and photometry & $v_\text{rad}$ and $m_i$ & \citet{Ibata1994} \\
        & $l$, $b$, $\mu_l$, $\mu_b$, $m_i$ via \textsc{streamfinder} & \citet{Malhan2018} \\
        & $V_T$ and sequences in the CMD & \citet{Babusiaux2018} \\
        & (same as the previous) & \citet{Haywood2018b} \\
        & (same as the previous) & \citet{Gallart2019} \\
        \hline
        Stellar chemokinematics & $V_\text{tot}$, [Fe/H], and [Mg/Fe] & \citet{Nissen2010} \\
        & \dots & \citet{Navarro2011} \\
        & $V_\text{tot}$, [Na/Fe], [Ni/Fe] & \citet{Bensby2014} \\
        & $l$, $b$, $v_\text{rad}$, [Fe/H], and [$\alpha$/Fe] & \citet{Hawkins2015} \\
        & $V_\phi$ and [Fe/H] & \citet{Belokurov2020} \\
        & (same as the previous) & \citet{An2021b} \\
        & Galactocentric spherical $V_\rho$, $V_\phi$, and [Fe/H] & \citet{Belokurov2018} \\
        & Galactocentric spherical $V_\rho$, $V_\phi$, $V_\theta$ and [Fe/H] via GMM & \citet{Myeong2018c} \\
        & $V_X$, $V_Y$, $V_Z$, and [Fe/H] via GMM & \citet{Nikakhtar2021} \\
        \hline
        Stellar chemodynamics & $e$, [Fe/H], [Mg/Fe], [Al/Fe], [Ni/Fe] via $k$-means & \citet{Mackereth2019} \\
        & [Fe/H], $J_R$, $L_Z$, and $J_Z$ & \citet{Myeong2018b} \\
        & $e$, [Fe/H], and [$\alpha$/Fe] & \citet{Naidu2020} \\
        & $E$ and $e$ informed by [Al/Fe] and [Mg/Mn] & \citet{Horta2021} \\
        \hline
        Stellar chemistry & [Fe/H] and [Mg/Fe] & \citet{DiMatteo2019, DiMatteo2020} \\
        & [Fe/H] and [$\alpha$/Fe] & \citet{Carollo2021} \\
        & [Al/Fe], [Mg/Mn] via GMM & \citet{Das2020} \\
        & [Al/Fe], [Mg/H] & \citet{Feuillet2021} \\
        & [Fe/H], [X/Fe] for (C+N), O, Mg, Al, Si, K, Ca, Cr, Mn, and Ni via $k$-means & \citet{Hayes2018} \\
        \hline
    \end{tabular}
    \label{tab:selection_techniques}
\end{table*}

%% file: buder_galah_accretion_chem.bbl
\begin{thebibliography}{}
\makeatletter
\relax
\def\mn@urlcharsother{\let\do\@makeother \do\$\do\&\do\#\do\^\do\_\do\%\do\~}
\def\mn@doi{\begingroup\mn@urlcharsother \@ifnextchar [ {\mn@doi@}
  {\mn@doi@[]}}
\def\mn@doi@[#1]#2{\def\@tempa{#1}\ifx\@tempa\@empty \href
  {http://dx.doi.org/#2} {doi:#2}\else \href {http://dx.doi.org/#2} {#1}\fi
  \endgroup}
\def\mn@eprint#1#2{\mn@eprint@#1:#2::\@nil}
\def\mn@eprint@arXiv#1{\href {http://arxiv.org/abs/#1} {{\tt arXiv:#1}}}
\def\mn@eprint@dblp#1{\href {http://dblp.uni-trier.de/rec/bibtex/#1.xml}
  {dblp:#1}}
\def\mn@eprint@#1:#2:#3:#4\@nil{\def\@tempa {#1}\def\@tempb {#2}\def\@tempc
  {#3}\ifx \@tempc \@empty \let \@tempc \@tempb \let \@tempb \@tempa \fi \ifx
  \@tempb \@empty \def\@tempb {arXiv}\fi \@ifundefined
  {mn@eprint@\@tempb}{\@tempb:\@tempc}{\expandafter \expandafter \csname
  mn@eprint@\@tempb\endcsname \expandafter{\@tempc}}}

\bibitem[\protect\citeauthoryear{{Aguado} et~al.,}{{Aguado}
  et~al.}{2021}]{Aguado2021}
{Aguado} D.~S.,  et~al., 2021, \mn@doi [\apjl] {10.3847/2041-8213/abdbb8},
  \href {https://ui.adsabs.harvard.edu/abs/2021ApJ...908L...8A} {908, L8}

\bibitem[\protect\citeauthoryear{{Amarante}, {Smith}  \& {Boeche}}{{Amarante}
  et~al.}{2020}]{Amarante2020b}
{Amarante} J. A.~S.,  {Smith} M.~C.,   {Boeche} C.,  2020, \mn@doi [\mnras]
  {10.1093/mnras/staa077}, \href
  {https://ui.adsabs.harvard.edu/abs/2020MNRAS.492.3816A} {492, 3816}

\bibitem[\protect\citeauthoryear{{Amarsi}, {Asplund}, {Collet}  \&
  {Leenaarts}}{{Amarsi} et~al.}{2015}]{Amarsi2015}
{Amarsi} A.~M.,  {Asplund} M.,  {Collet} R.,   {Leenaarts} J.,  2015, \mn@doi
  [\mnras] {10.1093/mnrasl/slv122}, \href
  {http://adsabs.harvard.edu/abs/2015MNRAS.454L..11A} {454, L11}

\bibitem[\protect\citeauthoryear{{Amarsi}, {Asplund}, {Collet}  \&
  {Leenaarts}}{{Amarsi} et~al.}{2016}]{Amarsi2016b}
{Amarsi} A.~M.,  {Asplund} M.,  {Collet} R.,   {Leenaarts} J.,  2016, \mn@doi
  [\mnras] {10.1093/mnras/stv2608}, \href
  {http://adsabs.harvard.edu/abs/2016MNRAS.455.3735A} {455, 3735}

\bibitem[\protect\citeauthoryear{{Amarsi}, {Nissen}, {Asplund}, {Lind}  \&
  {Barklem}}{{Amarsi} et~al.}{2019}]{Amarsi2019b}
{Amarsi} A.~M.,  {Nissen} P.~E.,  {Asplund} M.,  {Lind} K.,   {Barklem} P.~S.,
  2019, \mn@doi [\aap] {10.1051/0004-6361/201834480}, \href
  {http://adsabs.harvard.edu/abs/2019A%26A...622L...4A} {622, L4}

\bibitem[\protect\citeauthoryear{{Amarsi} et~al.,}{{Amarsi}
  et~al.}{2020}]{Amarsi2020}
{Amarsi} A.~M.,  et~al., 2020, \mn@doi [\aap] {10.1051/0004-6361/202038650},
  \href {https://ui.adsabs.harvard.edu/abs/2020A&A...642A..62A} {642, A62}

\bibitem[\protect\citeauthoryear{{Amorisco}}{{Amorisco}}{2017}]{Amorisco2017}
{Amorisco} N.~C.,  2017, \mn@doi [\mnras] {10.1093/mnras/stw2229}, \href
  {https://ui.adsabs.harvard.edu/abs/2017MNRAS.464.2882A} {464, 2882}

\bibitem[\protect\citeauthoryear{{An} \& {Beers}}{{An} \&
  {Beers}}{2021}]{An2021b}
{An} D.,  {Beers} T.~C.,  2021, \mn@doi [\apj] {10.3847/1538-4357/ac07a4},
  \href {https://ui.adsabs.harvard.edu/abs/2021ApJ...918...74A} {918, 74}

\bibitem[\protect\citeauthoryear{{Andrews}, {Weinberg}, {Sch{\"o}nrich}  \&
  {Johnson}}{{Andrews} et~al.}{2017}]{Andrews2017}
{Andrews} B.~H.,  {Weinberg} D.~H.,  {Sch{\"o}nrich} R.,   {Johnson} J.~A.,
  2017, \mn@doi [\apj] {10.3847/1538-4357/835/2/224}, \href
  {https://ui.adsabs.harvard.edu/abs/2017ApJ...835..224A} {835, 224}

\bibitem[\protect\citeauthoryear{{Astropy Collaboration} et~al.,}{{Astropy
  Collaboration} et~al.}{2013}]{Robitaille2013}
{Astropy Collaboration} et~al., 2013, \mn@doi [\aap]
  {10.1051/0004-6361/201322068}, \href
  {http://adsabs.harvard.edu/abs/2013A%26A...558A..33A} {558, A33}

\bibitem[\protect\citeauthoryear{{Astropy Collaboration} et~al.,}{{Astropy
  Collaboration} et~al.}{2018}]{PriceWhelan2018}
{Astropy Collaboration} et~al., 2018, \mn@doi [\aj] {10.3847/1538-3881/aabc4f},
  \href {https://ui.adsabs.harvard.edu/abs/2018AJ....156..123A} {156, 123}

\bibitem[\protect\citeauthoryear{{Bailer-Jones}, {Rybizki}, {Fouesneau},
  {Demleitner}  \& {Andrae}}{{Bailer-Jones} et~al.}{2021}]{BailerJones2021}
{Bailer-Jones} C.~A.~L.,  {Rybizki} J.,  {Fouesneau} M.,  {Demleitner} M.,
  {Andrae} R.,  2021, \mn@doi [\aj] {10.3847/1538-3881/abd806}, \href
  {https://ui.adsabs.harvard.edu/abs/2021AJ....161..147B} {161, 147}

\bibitem[\protect\citeauthoryear{{Barb{\'a}}, {Minniti}, {Geisler},
  {Alonso-Garc{\'\i}a}, {Hempel}, {Monachesi}, {Arias}  \&
  {G{\'o}mez}}{{Barb{\'a}} et~al.}{2019}]{Barba2019}
{Barb{\'a}} R.~H.,  {Minniti} D.,  {Geisler} D.,  {Alonso-Garc{\'\i}a} J.,
  {Hempel} M.,  {Monachesi} A.,  {Arias} J.~I.,   {G{\'o}mez} F.~A.,  2019,
  \mn@doi [\apjl] {10.3847/2041-8213/aaf811}, \href
  {https://ui.adsabs.harvard.edu/abs/2019ApJ...870L..24B} {870, L24}

\bibitem[\protect\citeauthoryear{{Belokurov} et~al.,}{{Belokurov}
  et~al.}{2006}]{Belokurov2006}
{Belokurov} V.,  et~al., 2006, \mn@doi [\apjl] {10.1086/504797}, \href
  {http://adsabs.harvard.edu/abs/2006ApJ...642L.137B} {642, L137}

\bibitem[\protect\citeauthoryear{{Belokurov}, {Erkal}, {Evans}, {Koposov}  \&
  {Deason}}{{Belokurov} et~al.}{2018}]{Belokurov2018}
{Belokurov} V.,  {Erkal} D.,  {Evans} N.~W.,  {Koposov} S.~E.,   {Deason}
  A.~J.,  2018, \mn@doi [\mnras] {10.1093/mnras/sty982}, \href
  {https://ui.adsabs.harvard.edu/abs/2018MNRAS.478..611B} {478, 611}

\bibitem[\protect\citeauthoryear{{Belokurov}, {Sanders}, {Fattahi}, {Smith},
  {Deason}, {Evans}  \& {Grand}}{{Belokurov} et~al.}{2020}]{Belokurov2020}
{Belokurov} V.,  {Sanders} J.~L.,  {Fattahi} A.,  {Smith} M.~C.,  {Deason}
  A.~J.,  {Evans} N.~W.,   {Grand} R. J.~J.,  2020, \mn@doi [\mnras]
  {10.1093/mnras/staa876}, \href
  {https://ui.adsabs.harvard.edu/abs/2020MNRAS.494.3880B} {494, 3880}

\bibitem[\protect\citeauthoryear{{Bensby}, {Feltzing}  \& {Oey}}{{Bensby}
  et~al.}{2014}]{Bensby2014}
{Bensby} T.,  {Feltzing} S.,   {Oey} M.~S.,  2014, \mn@doi [\aap]
  {10.1051/0004-6361/201322631}, \href
  {http://adsabs.harvard.edu/abs/2014A%26A...562A..71B} {562, A71}

\bibitem[\protect\citeauthoryear{{Binney}}{{Binney}}{2012}]{Binney2012}
{Binney} J.,  2012, \mn@doi [\mnras] {10.1111/j.1365-2966.2012.21757.x}, \href
  {http://adsabs.harvard.edu/abs/2012MNRAS.426.1324B} {426, 1324}

\bibitem[\protect\citeauthoryear{{Binney} \& {Tremaine}}{{Binney} \&
  {Tremaine}}{2008}]{Binney2008}
{Binney} J.,  {Tremaine} S.,  2008, {Galactic Dynamics: Second Edition}.
Princeton University Press

\bibitem[\protect\citeauthoryear{{Bird}, {Kazantzidis}, {Weinberg}, {Guedes},
  {Callegari}, {Mayer}  \& {Madau}}{{Bird} et~al.}{2013}]{Bird2013}
{Bird} J.~C.,  {Kazantzidis} S.,  {Weinberg} D.~H.,  {Guedes} J.,  {Callegari}
  S.,  {Mayer} L.,   {Madau} P.,  2013, \mn@doi [\apj]
  {10.1088/0004-637X/773/1/43}, \href
  {http://adsabs.harvard.edu/abs/2013ApJ...773...43B} {773, 43}

\bibitem[\protect\citeauthoryear{{Bland-Hawthorn} \&
  {Gerhard}}{{Bland-Hawthorn} \& {Gerhard}}{2016}]{BlandHawthorn_Gerhard2016}
{Bland-Hawthorn} J.,  {Gerhard} O.,  2016, \mn@doi [\araa]
  {10.1146/annurev-astro-081915-023441}, \href
  {http://adsabs.harvard.edu/abs/2016ARA%26A..54..529B} {54, 529}

\bibitem[\protect\citeauthoryear{{Bland-Hawthorn} et~al.,}{{Bland-Hawthorn}
  et~al.}{2019}]{BlandHawthorn2019}
{Bland-Hawthorn} J.,  et~al., 2019, \mn@doi [\mnras] {10.1093/mnras/stz217},
  \href {http://adsabs.harvard.edu/abs/2019MNRAS.486.1167B} {486, 1167}

\bibitem[\protect\citeauthoryear{{Bonaca}, {Conroy}, {Wetzel}, {Hopkins}  \&
  {Kere{\v s}}}{{Bonaca} et~al.}{2017}]{Bonaca2017}
{Bonaca} A.,  {Conroy} C.,  {Wetzel} A.,  {Hopkins} P.~F.,   {Kere{\v s}} D.,
  2017, \mn@doi [\apj] {10.3847/1538-4357/aa7d0c}, \href
  {http://adsabs.harvard.edu/abs/2017ApJ...845..101B} {845, 101}

\bibitem[\protect\citeauthoryear{{Bonaca} et~al.,}{{Bonaca}
  et~al.}{2020}]{Bonaca2020}
{Bonaca} A.,  et~al., 2020, \mn@doi [\apjl] {10.3847/2041-8213/ab9caa}, \href
  {https://ui.adsabs.harvard.edu/abs/2020ApJ...897L..18B} {897, L18}

\bibitem[\protect\citeauthoryear{{Bonaca} et~al.,}{{Bonaca}
  et~al.}{2021}]{Bonaca2021}
{Bonaca} A.,  et~al., 2021, \mn@doi [\apjl] {10.3847/2041-8213/abeaa9}, \href
  {https://ui.adsabs.harvard.edu/abs/2021ApJ...909L..26B} {909, L26}

\bibitem[\protect\citeauthoryear{{Bonifacio} et~al.,}{{Bonifacio}
  et~al.}{2021}]{Bonifacio2021}
{Bonifacio} P.,  et~al., 2021, \mn@doi [\aap] {10.1051/0004-6361/202140816},
  \href {https://ui.adsabs.harvard.edu/abs/2021A&A...651A..79B} {651, A79}

\bibitem[\protect\citeauthoryear{{Bovy}}{{Bovy}}{2015}]{Bovy2015}
{Bovy} J.,  2015, \mn@doi [\apjs] {10.1088/0067-0049/216/2/29}, \href
  {http://adsabs.harvard.edu/abs/2015ApJS..216...29B} {216, 29}

\bibitem[\protect\citeauthoryear{{Bovy}}{{Bovy}}{2016}]{Bovy2016b}
{Bovy} J.,  2016, \mn@doi [\apj] {10.3847/0004-637X/817/1/49}, \href
  {http://adsabs.harvard.edu/abs/2016ApJ...817...49B} {817, 49}

\bibitem[\protect\citeauthoryear{{Bovy}, {Hogg}  \& {Roweis}}{{Bovy}
  et~al.}{2011}]{Bovy2011}
{Bovy} J.,  {Hogg} D.~W.,   {Roweis} S.~T.,  2011, \mn@doi [Ann. Appl. Stat.]
  {10.1214/10-AOAS439}, \href
  {https://ui.adsabs.harvard.edu/abs/2011AnApS...5.1657B} {5, 1657}

\bibitem[\protect\citeauthoryear{{Bovy}, {Rix}  \& {Hogg}}{{Bovy}
  et~al.}{2012a}]{Bovy2012b}
{Bovy} J.,  {Rix} H.-W.,   {Hogg} D.~W.,  2012a, \mn@doi [\apj]
  {10.1088/0004-637X/751/2/131}, \href
  {http://adsabs.harvard.edu/abs/2012ApJ...751..131B} {751, 131}

\bibitem[\protect\citeauthoryear{{Bovy}, {Rix}, {Liu}, {Hogg}, {Beers}  \&
  {Lee}}{{Bovy} et~al.}{2012b}]{Bovy2012}
{Bovy} J.,  {Rix} H.-W.,  {Liu} C.,  {Hogg} D.~W.,  {Beers} T.~C.,   {Lee}
  Y.~S.,  2012b, \mn@doi [\apj] {10.1088/0004-637X/753/2/148}, \href
  {http://adsabs.harvard.edu/abs/2012ApJ...753..148B} {753, 148}

\bibitem[\protect\citeauthoryear{{Bovy}, {Rix}, {Schlafly}, {Nidever},
  {Holtzman}, {Shetrone}  \& {Beers}}{{Bovy} et~al.}{2016}]{Bovy2016}
{Bovy} J.,  {Rix} H.-W.,  {Schlafly} E.~F.,  {Nidever} D.~L.,  {Holtzman}
  J.~A.,  {Shetrone} M.,   {Beers} T.~C.,  2016, \mn@doi [\apj]
  {10.3847/0004-637X/823/1/30}, \href
  {http://adsabs.harvard.edu/abs/2016ApJ...823...30B} {823, 30}

\bibitem[\protect\citeauthoryear{{Buck}}{{Buck}}{2020}]{Buck2020}
{Buck} T.,  2020, \mn@doi [\mnras] {10.1093/mnras/stz3289}, \href
  {https://ui.adsabs.harvard.edu/abs/2020MNRAS.491.5435B} {491, 5435}

\bibitem[\protect\citeauthoryear{{Buck}, {Rybizki}, {Buder}, {Obreja},
  {Macci{\`o}}, {Pfrommer}, {Steinmetz}  \& {Ness}}{{Buck}
  et~al.}{2021}]{Buck2021}
{Buck} T.,  {Rybizki} J.,  {Buder} S.,  {Obreja} A.,  {Macci{\`o}} A.~V.,
  {Pfrommer} C.,  {Steinmetz} M.,   {Ness} M.,  2021, \mn@doi [\mnras]
  {10.1093/mnras/stab2736}, \href
  {https://ui.adsabs.harvard.edu/abs/2021MNRAS.508.3365B} {508, 3365}

\bibitem[\protect\citeauthoryear{{Buder} et~al.,}{{Buder}
  et~al.}{2018}]{Buder2018}
{Buder} S.,  et~al., 2018, \mn@doi [\mnras] {10.1093/mnras/sty1281}, \href
  {http://adsabs.harvard.edu/abs/2018MNRAS.478.4513B} {478, 4513}

\bibitem[\protect\citeauthoryear{{Buder} et~al.,}{{Buder}
  et~al.}{2019}]{Buder2019}
{Buder} S.,  et~al., 2019, \mn@doi [\aap] {10.1051/0004-6361/201833218}, \href
  {http://adsabs.harvard.edu/abs/2019A%26A...624A..19B} {624, A19}

\bibitem[\protect\citeauthoryear{{Buder} et~al.,}{{Buder}
  et~al.}{2021}]{Buder2021}
{Buder} S.,  et~al., 2021, \mn@doi [\mnras] {10.1093/mnras/stab1242}, \href
  {https://ui.adsabs.harvard.edu/abs/2021MNRAS.506..150B} {506, 150}

\bibitem[\protect\citeauthoryear{{Carollo} \& {Chiba}}{{Carollo} \&
  {Chiba}}{2021}]{Carollo2021}
{Carollo} D.,  {Chiba} M.,  2021, \mn@doi [\apj] {10.3847/1538-4357/abd7a4},
  \href {https://ui.adsabs.harvard.edu/abs/2021ApJ...908..191C} {908, 191}

\bibitem[\protect\citeauthoryear{{Carollo} et~al.,}{{Carollo}
  et~al.}{2007}]{Carollo2007}
{Carollo} D.,  et~al., 2007, \mn@doi [\nat] {10.1038/nature06460}, \href
  {http://adsabs.harvard.edu/abs/2007Natur.450.1020C} {450, 1020}

\bibitem[\protect\citeauthoryear{{Carollo} et~al.,}{{Carollo}
  et~al.}{2010}]{Carollo2010}
{Carollo} D.,  et~al., 2010, \mn@doi [\apj] {10.1088/0004-637X/712/1/692},
  \href {http://adsabs.harvard.edu/abs/2010ApJ...712..692C} {712, 692}

\bibitem[\protect\citeauthoryear{{Carretta} et~al.,}{{Carretta}
  et~al.}{2009}]{Carretta2009}
{Carretta} E.,  et~al., 2009, \mn@doi [\aap] {10.1051/0004-6361/200912096},
  \href {http://adsabs.harvard.edu/abs/2009A%26A...505..117C} {505, 117}

\bibitem[\protect\citeauthoryear{{Chaplin} et~al.,}{{Chaplin}
  et~al.}{2020}]{Chaplin2020}
{Chaplin} W.~J.,  et~al., 2020, \mn@doi [Nature Astronomy]
  {10.1038/s41550-019-0975-9}, \href
  {https://ui.adsabs.harvard.edu/abs/2020NatAs...4..382C} {4, 382}

\bibitem[\protect\citeauthoryear{{Chiappini} et~al.,}{{Chiappini}
  et~al.}{2015}]{Chiappini2015}
{Chiappini} C.,  et~al., 2015, \mn@doi [\aap] {10.1051/0004-6361/201525865},
  \href {http://adsabs.harvard.edu/abs/2015A%26A...576L..12C} {576, L12}

\bibitem[\protect\citeauthoryear{{Collet}, {Asplund}  \& {Trampedach}}{{Collet}
  et~al.}{2007}]{Collet2007}
{Collet} R.,  {Asplund} M.,   {Trampedach} R.,  2007, \mn@doi [\aap]
  {10.1051/0004-6361:20066321}, \href
  {https://ui.adsabs.harvard.edu/abs/2007A&A...469..687C} {469, 687}

\bibitem[\protect\citeauthoryear{{Conroy}, {Naidu}, {Zaritsky}, {Bonaca},
  {Cargile}, {Johnson}  \& {Caldwell}}{{Conroy} et~al.}{2019}]{Conroy2019}
{Conroy} C.,  {Naidu} R.~P.,  {Zaritsky} D.,  {Bonaca} A.,  {Cargile} P.,
  {Johnson} B.~D.,   {Caldwell} N.,  2019, \mn@doi [\apj]
  {10.3847/1538-4357/ab5710}, \href
  {https://ui.adsabs.harvard.edu/abs/2019ApJ...887..237C} {887, 237}

\bibitem[\protect\citeauthoryear{{Cordoni} et~al.,}{{Cordoni}
  et~al.}{2021}]{Cordoni2021}
{Cordoni} G.,  et~al., 2021, \mn@doi [\mnras] {10.1093/mnras/staa3417}, \href
  {https://ui.adsabs.harvard.edu/abs/2021MNRAS.503.2539C} {503, 2539}

\bibitem[\protect\citeauthoryear{{Dalton} et~al.,}{{Dalton}
  et~al.}{2018}]{WEAVE2018}
{Dalton} G.,  et~al., 2018, in J. E.~C.,  L. S.,   H. T.,  eds,  SPIE
  Conference Series Vol. 10702, \procspie. SPIE, p. 107021B,
  \mn@doi{10.1117/12.2312031}

\bibitem[\protect\citeauthoryear{{Das}, {Hawkins}  \& {Jofr{\'e}}}{{Das}
  et~al.}{2020}]{Das2020}
{Das} P.,  {Hawkins} K.,   {Jofr{\'e}} P.,  2020, \mn@doi [\mnras]
  {10.1093/mnras/stz3537}, \href
  {https://ui.adsabs.harvard.edu/abs/2020MNRAS.493.5195D} {493, 5195}

\bibitem[\protect\citeauthoryear{{De Silva} et~al.,}{{De Silva}
  et~al.}{2015}]{DeSilva2015}
{De Silva} G.~M.,  et~al., 2015, \mn@doi [\mnras] {10.1093/mnras/stv327}, \href
  {http://adsabs.harvard.edu/abs/2015MNRAS.449.2604D} {449, 2604}

\bibitem[\protect\citeauthoryear{Dempster, Laird  \& Rubin}{Dempster
  et~al.}{1977}]{Dempster1977}
Dempster A.~P.,  Laird N.~M.,   Rubin D.~B.,  1977, \mn@doi [Journal of the
  Royal Statistical Society: Series B (Methodological)]
  {https://doi.org/10.1111/j.2517-6161.1977.tb01600.x}, 39, 1

\bibitem[\protect\citeauthoryear{{Di Matteo}, {Haywood}, {Lehnert}, {Katz},
  {Khoperskov}, {Snaith}, {G{\'o}mez}  \& {Robichon}}{{Di Matteo}
  et~al.}{2019}]{DiMatteo2019}
{Di Matteo} P.,  {Haywood} M.,  {Lehnert} M.~D.,  {Katz} D.,  {Khoperskov} S.,
  {Snaith} O.~N.,  {G{\'o}mez} A.,   {Robichon} N.,  2019, \mn@doi [\aap]
  {10.1051/0004-6361/201834929}, \href
  {https://ui.adsabs.harvard.edu/abs/2019A&A...632A...4D} {632, A4}

\bibitem[\protect\citeauthoryear{{Di Matteo}, {Spite}, {Haywood}, {Bonifacio},
  {G{\'o}mez}, {Spite}  \& {Caffau}}{{Di Matteo} et~al.}{2020}]{DiMatteo2020}
{Di Matteo} P.,  {Spite} M.,  {Haywood} M.,  {Bonifacio} P.,  {G{\'o}mez} A.,
  {Spite} F.,   {Caffau} E.,  2020, \mn@doi [\aap]
  {10.1051/0004-6361/201937016}, \href
  {https://ui.adsabs.harvard.edu/abs/2020A&A...636A.115D} {636, A115}

\bibitem[\protect\citeauthoryear{{Eitner}, {Bergemann}, {Hansen}, {Cescutti},
  {Seitenzahl}, {Larsen}  \& {Plez}}{{Eitner} et~al.}{2020}]{Eitner2020}
{Eitner} P.,  {Bergemann} M.,  {Hansen} C.~J.,  {Cescutti} G.,  {Seitenzahl}
  I.~R.,  {Larsen} S.,   {Plez} B.,  2020, \mn@doi [\aap]
  {10.1051/0004-6361/201936603}, \href
  {https://ui.adsabs.harvard.edu/abs/2020A&A...635A..38E} {635, A38}

\bibitem[\protect\citeauthoryear{{Fabbian}, {Asplund}, {Barklem}, {Carlsson}
  \& {Kiselman}}{{Fabbian} et~al.}{2009}]{Fabbian2009}
{Fabbian} D.,  {Asplund} M.,  {Barklem} P.~S.,  {Carlsson} M.,   {Kiselman} D.,
   2009, \mn@doi [\aap] {10.1051/0004-6361/200809640}, \href
  {https://ui.adsabs.harvard.edu/abs/2009A&A...500.1221F} {500, 1221}

\bibitem[\protect\citeauthoryear{{Fensch} et~al.,}{{Fensch}
  et~al.}{2017}]{Fensch2017}
{Fensch} J.,  et~al., 2017, \mn@doi [\mnras] {10.1093/mnras/stw2920}, \href
  {https://ui.adsabs.harvard.edu/abs/2017MNRAS.465.1934F} {465, 1934}

\bibitem[\protect\citeauthoryear{{Fern{\'a}ndez-Alvar}
  et~al.,}{{Fern{\'a}ndez-Alvar} et~al.}{2017}]{FernandezAlvar2017}
{Fern{\'a}ndez-Alvar} E.,  et~al., 2017, \mn@doi [\mnras]
  {10.1093/mnras/stw2861}, \href
  {https://ui.adsabs.harvard.edu/abs/2017MNRAS.465.1586F} {465, 1586}

\bibitem[\protect\citeauthoryear{{Fern{\'a}ndez-Alvar}
  et~al.,}{{Fern{\'a}ndez-Alvar} et~al.}{2018}]{FernandezAlvar2018b}
{Fern{\'a}ndez-Alvar} E.,  et~al., 2018, \mn@doi [\apj]
  {10.3847/1538-4357/aa9ced}, \href
  {https://ui.adsabs.harvard.edu/abs/2018ApJ...852...50F} {852, 50}

\bibitem[\protect\citeauthoryear{{Feuillet}, {Feltzing}, {Sahlholdt}  \&
  {Casagrande}}{{Feuillet} et~al.}{2020}]{Feuillet2020}
{Feuillet} D.~K.,  {Feltzing} S.,  {Sahlholdt} C.~L.,   {Casagrande} L.,  2020,
  \mn@doi [\mnras] {10.1093/mnras/staa1888}, \href
  {https://ui.adsabs.harvard.edu/abs/2020MNRAS.497..109F} {497, 109}

\bibitem[\protect\citeauthoryear{{Feuillet}, {Sahlholdt}, {Feltzing}  \&
  {Casagrande}}{{Feuillet} et~al.}{2021}]{Feuillet2021}
{Feuillet} D.~K.,  {Sahlholdt} C.~L.,  {Feltzing} S.,   {Casagrande} L.,  2021,
  \mn@doi [\mnras] {10.1093/mnras/stab2614}, \href
  {https://ui.adsabs.harvard.edu/abs/2021MNRAS.508.1489F} {508, 1489}

\bibitem[\protect\citeauthoryear{{Fishlock}, {Yong}, {Karakas}, {Alves-Brito},
  {Mel{\'e}ndez}, {Nissen}, {Kobayashi}  \& {Casey}}{{Fishlock}
  et~al.}{2017}]{Fishlock2017}
{Fishlock} C.~K.,  {Yong} D.,  {Karakas} A.~I.,  {Alves-Brito} A.,
  {Mel{\'e}ndez} J.,  {Nissen} P.~E.,  {Kobayashi} C.,   {Casey} A.~R.,  2017,
  \mn@doi [\mnras] {10.1093/mnras/stx047}, \href
  {https://ui.adsabs.harvard.edu/abs/2017MNRAS.466.4672F} {466, 4672}

\bibitem[\protect\citeauthoryear{Foreman-Mackey}{Foreman-Mackey}{2016}]{corner}
Foreman-Mackey D.,  2016, \mn@doi [The Journal of Open Source Software]
  {10.21105/joss.00024}, 1, 24

\bibitem[\protect\citeauthoryear{{Freeman} \& {Bland-Hawthorn}}{{Freeman} \&
  {Bland-Hawthorn}}{2002}]{FreemanBlandHawthorn2002}
{Freeman} K.,  {Bland-Hawthorn} J.,  2002, \mn@doi [\araa]
  {10.1146/annurev.astro.40.060401.093840}, \href
  {http://adsabs.harvard.edu/abs/2002ARA%26A..40..487F} {40, 487}

\bibitem[\protect\citeauthoryear{{Gaia Collaboration} et~al.,}{{Gaia
  Collaboration} et~al.}{2016a}]{Gaia-Collaboration2016}
{Gaia Collaboration} et~al., 2016a, \mn@doi [\aap]
  {10.1051/0004-6361/201629272}, \href
  {http://adsabs.harvard.edu/abs/2016A%26A...595A...1G} {595, A1}

\bibitem[\protect\citeauthoryear{{Gaia Collaboration} et~al.,}{{Gaia
  Collaboration} et~al.}{2016b}]{Brown2016}
{Gaia Collaboration} et~al., 2016b, \mn@doi [\aap]
  {10.1051/0004-6361/201629512}, \href
  {http://adsabs.harvard.edu/abs/2016A%26A...595A...2G} {595, A2}

\bibitem[\protect\citeauthoryear{{Gaia Collaboration} et~al.,}{{Gaia
  Collaboration} et~al.}{2018a}]{Brown2018}
{Gaia Collaboration} et~al., 2018a, \mn@doi [\aap]
  {10.1051/0004-6361/201833051}, \href
  {http://adsabs.harvard.edu/abs/2018A%26A...616A...1G} {616, A1}

\bibitem[\protect\citeauthoryear{{Gaia Collaboration} et~al.,}{{Gaia
  Collaboration} et~al.}{2018b}]{Babusiaux2018}
{Gaia Collaboration} et~al., 2018b, \mn@doi [\aap]
  {10.1051/0004-6361/201832843}, \href
  {https://ui.adsabs.harvard.edu/abs/2018A&A...616A..10G} {616, A10}

\bibitem[\protect\citeauthoryear{{Gaia Collaboration} et~al.,}{{Gaia
  Collaboration} et~al.}{2018c}]{Helmi2018b}
{Gaia Collaboration} et~al., 2018c, \mn@doi [\aap]
  {10.1051/0004-6361/201832698}, \href
  {https://ui.adsabs.harvard.edu/abs/2018A&A...616A..12G} {616, A12}

\bibitem[\protect\citeauthoryear{{Gaia Collaboration} et~al.,}{{Gaia
  Collaboration} et~al.}{2021}]{Brown2021}
{Gaia Collaboration} et~al., 2021, \mn@doi [\aap]
  {10.1051/0004-6361/202039657}, \href
  {https://ui.adsabs.harvard.edu/abs/2021A&A...649A...1G} {649, A1}

\bibitem[\protect\citeauthoryear{{Gallart}, {Bernard}, {Brook}, {Ruiz-Lara},
  {Cassisi}, {Hill}  \& {Monelli}}{{Gallart} et~al.}{2019}]{Gallart2019}
{Gallart} C.,  {Bernard} E.~J.,  {Brook} C.~B.,  {Ruiz-Lara} T.,  {Cassisi} S.,
   {Hill} V.,   {Monelli} M.,  2019, \mn@doi [Nature Astronomy]
  {10.1038/s41550-019-0829-5}, \href
  {https://ui.adsabs.harvard.edu/abs/2019NatAs...3..932G} {3, 932}

\bibitem[\protect\citeauthoryear{{Gao} et~al.,}{{Gao} et~al.}{2020}]{Gao2020}
{Gao} X.,  et~al., 2020, \mn@doi [\mnras] {10.1093/mnrasl/slaa109}, \href
  {https://ui.adsabs.harvard.edu/abs/2020MNRAS.497L..30G} {497, L30}

\bibitem[\protect\citeauthoryear{{Gilmore} \& {Wyse}}{{Gilmore} \&
  {Wyse}}{1991}]{GilmoreWyse1991}
{Gilmore} G.,  {Wyse} R. F.~G.,  1991, \mn@doi [\apjl] {10.1086/185930}, \href
  {https://ui.adsabs.harvard.edu/abs/1991ApJ...367L..55G} {367, L55}

\bibitem[\protect\citeauthoryear{{Grand} et~al.,}{{Grand}
  et~al.}{2020}]{Grand2020}
{Grand} R. J.~J.,  et~al., 2020, \mn@doi [\mnras] {10.1093/mnras/staa2057},
  \href {https://ui.adsabs.harvard.edu/abs/2020MNRAS.497.1603G} {497, 1603}

\bibitem[\protect\citeauthoryear{{Gratton}}{{Gratton}}{1989}]{Gratton1989}
{Gratton} R.~G.,  1989, \aap, \href
  {https://ui.adsabs.harvard.edu/abs/1989A&A...208..171G} {208, 171}

\bibitem[\protect\citeauthoryear{{Gratton}, {Carretta}, {Desidera},
  {Lucatello}, {Mazzei}  \& {Barbieri}}{{Gratton} et~al.}{2003}]{Gratton2003}
{Gratton} R.~G.,  {Carretta} E.,  {Desidera} S.,  {Lucatello} S.,  {Mazzei} P.,
    {Barbieri} M.,  2003, \mn@doi [\aap] {10.1051/0004-6361:20030754}, \href
  {https://ui.adsabs.harvard.edu/abs/2003A&A...406..131G} {406, 131}

\bibitem[\protect\citeauthoryear{{Gustafsson}, {Edvardsson}, {Eriksson},
  {J{\o}rgensen}, {Nordlund}  \& {Plez}}{{Gustafsson}
  et~al.}{2008}]{Gustafsson2008}
{Gustafsson} B.,  {Edvardsson} B.,  {Eriksson} K.,  {J{\o}rgensen} U.~G.,
  {Nordlund} {\AA}.,   {Plez} B.,  2008, \mn@doi [\aap]
  {10.1051/0004-6361:200809724}, \href
  {http://adsabs.harvard.edu/abs/2008A%26A...486..951G} {486, 951}

\bibitem[\protect\citeauthoryear{{Hasselquist} et~al.,}{{Hasselquist}
  et~al.}{2017}]{Hasselquist2017}
{Hasselquist} S.,  et~al., 2017, \mn@doi [\apj] {10.3847/1538-4357/aa7ddc},
  \href {https://ui.adsabs.harvard.edu/abs/2017ApJ...845..162H} {845, 162}

\bibitem[\protect\citeauthoryear{{Hasselquist} et~al.,}{{Hasselquist}
  et~al.}{2019}]{Hasselquist2019}
{Hasselquist} S.,  et~al., 2019, \mn@doi [\apj] {10.3847/1538-4357/aafdac},
  \href {https://ui.adsabs.harvard.edu/abs/2019ApJ...872...58H} {872, 58}

\bibitem[\protect\citeauthoryear{{Hawkins}, {Jofr{\'e}}, {Gilmore}  \&
  {Masseron}}{{Hawkins} et~al.}{2014}]{Hawkins2014}
{Hawkins} K.,  {Jofr{\'e}} P.,  {Gilmore} G.,   {Masseron} T.,  2014, \mn@doi
  [\mnras] {10.1093/mnras/stu1910}, \href
  {http://adsabs.harvard.edu/abs/2014MNRAS.445.2575H} {445, 2575}

\bibitem[\protect\citeauthoryear{{Hawkins}, {Jofr{\'e}}, {Masseron}  \&
  {Gilmore}}{{Hawkins} et~al.}{2015}]{Hawkins2015}
{Hawkins} K.,  {Jofr{\'e}} P.,  {Masseron} T.,   {Gilmore} G.,  2015, \mn@doi
  [\mnras] {10.1093/mnras/stv1586}, \href
  {http://adsabs.harvard.edu/abs/2015MNRAS.453..758H} {453, 758}

\bibitem[\protect\citeauthoryear{{Hawkins} et~al.,}{{Hawkins}
  et~al.}{2016}]{Hawkins2016}
{Hawkins} K.,  et~al., 2016, \mn@doi [\aap] {10.1051/0004-6361/201628268},
  \href {http://adsabs.harvard.edu/abs/2016A%26A...592A..70H} {592, A70}

\bibitem[\protect\citeauthoryear{{Hayek}, {Asplund}, {Collet}  \&
  {Nordlund}}{{Hayek} et~al.}{2011}]{Hayek2011}
{Hayek} W.,  {Asplund} M.,  {Collet} R.,   {Nordlund} {\r{A}}.,  2011, \mn@doi
  [\aap] {10.1051/0004-6361/201015782}, \href
  {https://ui.adsabs.harvard.edu/abs/2011A&A...529A.158H} {529, A158}

\bibitem[\protect\citeauthoryear{{Hayes} et~al.,}{{Hayes}
  et~al.}{2018}]{Hayes2018}
{Hayes} C.~R.,  et~al., 2018, \mn@doi [\apj] {10.3847/1538-4357/aa9cec}, \href
  {http://adsabs.harvard.edu/abs/2018ApJ...852...49H} {852, 49}

\bibitem[\protect\citeauthoryear{{Haywood}, {Di Matteo}, {Lehnert}, {Snaith},
  {Fragkoudi}  \& {Khoperskov}}{{Haywood} et~al.}{2018a}]{Haywood2018}
{Haywood} M.,  {Di Matteo} P.,  {Lehnert} M.,  {Snaith} O.,  {Fragkoudi} F.,
  {Khoperskov} S.,  2018a, \mn@doi [\aap] {10.1051/0004-6361/201731363}, \href
  {https://ui.adsabs.harvard.edu/abs/2018A&A...618A..78H} {618, A78}

\bibitem[\protect\citeauthoryear{{Haywood}, {Di Matteo}, {Lehnert}, {Snaith},
  {Khoperskov}  \& {G{\'o}mez}}{{Haywood} et~al.}{2018b}]{Haywood2018b}
{Haywood} M.,  {Di Matteo} P.,  {Lehnert} M.~D.,  {Snaith} O.,  {Khoperskov}
  S.,   {G{\'o}mez} A.,  2018b, \mn@doi [\apj] {10.3847/1538-4357/aad235},
  \href {http://adsabs.harvard.edu/abs/2018ApJ...863..113H} {863, 113}

\bibitem[\protect\citeauthoryear{{Helmi}}{{Helmi}}{2020}]{Helmi2020}
{Helmi} A.,  2020, \mn@doi [\araa] {10.1146/annurev-astro-032620-021917}, \href
  {https://ui.adsabs.harvard.edu/abs/2020ARA&A..58..205H} {58, 205}

\bibitem[\protect\citeauthoryear{{Helmi}, {White}, {de Zeeuw}  \&
  {Zhao}}{{Helmi} et~al.}{1999}]{Helmi1999}
{Helmi} A.,  {White} S.~D.~M.,  {de Zeeuw} P.~T.,   {Zhao} H.,  1999, \mn@doi
  [\nat] {10.1038/46980}, \href
  {http://adsabs.harvard.edu/abs/1999Natur.402...53H} {402, 53}

\bibitem[\protect\citeauthoryear{{Helmi}, {Veljanoski}, {Breddels}, {Tian}  \&
  {Sales}}{{Helmi} et~al.}{2017}]{Helmi2017}
{Helmi} A.,  {Veljanoski} J.,  {Breddels} M.~A.,  {Tian} H.,   {Sales} L.~V.,
  2017, \mn@doi [\aap] {10.1051/0004-6361/201629990}, \href
  {https://ui.adsabs.harvard.edu/abs/2017A&A...598A..58H} {598, A58}

\bibitem[\protect\citeauthoryear{{Helmi}, {Babusiaux}, {Koppelman}, {Massari},
  {Veljanoski}  \& {Brown}}{{Helmi} et~al.}{2018}]{Helmi2018}
{Helmi} A.,  {Babusiaux} C.,  {Koppelman} H.~H.,  {Massari} D.,  {Veljanoski}
  J.,   {Brown} A.~G.~A.,  2018, \mn@doi [\nat] {10.1038/s41586-018-0625-x},
  \href {http://adsabs.harvard.edu/abs/2018Natur.563...85H} {563, 85}

\bibitem[\protect\citeauthoryear{{Holoien}, {Marshall}  \&
  {Wechsler}}{{Holoien} et~al.}{2017}]{Holoien2017}
{Holoien} T. W.~S.,  {Marshall} P.~J.,   {Wechsler} R.~H.,  2017, \mn@doi [\aj]
  {10.3847/1538-3881/aa68a1}, \href
  {https://ui.adsabs.harvard.edu/abs/2017AJ....153..249H} {153, 249}

\bibitem[\protect\citeauthoryear{{Horta} et~al.,}{{Horta}
  et~al.}{2021}]{Horta2021}
{Horta} D.,  et~al., 2021, \mn@doi [\mnras] {10.1093/mnras/staa2987}, \href
  {https://ui.adsabs.harvard.edu/abs/2021MNRAS.500.1385H} {500, 1385}

\bibitem[\protect\citeauthoryear{Hunter}{Hunter}{2007}]{matplotlib}
Hunter J.~D.,  2007, \mn@doi [Comput Sci Eng] {10.1109/MCSE.2007.55}, 9, 90

\bibitem[\protect\citeauthoryear{{Ibata}, {Gilmore}  \& {Irwin}}{{Ibata}
  et~al.}{1994}]{Ibata1994}
{Ibata} R.~A.,  {Gilmore} G.,   {Irwin} M.~J.,  1994, \mn@doi [\nat]
  {10.1038/370194a0}, \href
  {https://ui.adsabs.harvard.edu/abs/1994Natur.370..194I} {370, 194}

\bibitem[\protect\citeauthoryear{{Ishigaki}, {Chiba}  \& {Aoki}}{{Ishigaki}
  et~al.}{2012}]{Ishigaki2012}
{Ishigaki} M.~N.,  {Chiba} M.,   {Aoki} W.,  2012, \mn@doi [\apj]
  {10.1088/0004-637X/753/1/64}, \href
  {https://ui.adsabs.harvard.edu/abs/2012ApJ...753...64I} {753, 64}

\bibitem[\protect\citeauthoryear{{Ishigaki}, {Aoki}  \& {Chiba}}{{Ishigaki}
  et~al.}{2013}]{Ishigaki2013}
{Ishigaki} M.~N.,  {Aoki} W.,   {Chiba} M.,  2013, \mn@doi [\apj]
  {10.1088/0004-637X/771/1/67}, \href
  {https://ui.adsabs.harvard.edu/abs/2013ApJ...771...67I} {771, 67}

\bibitem[\protect\citeauthoryear{{Ishigaki} et~al.,}{{Ishigaki}
  et~al.}{2021}]{Ishigaki2021}
{Ishigaki} M.~N.,  et~al., 2021, \mn@doi [\mnras] {10.1093/mnras/stab1982},
  \href {https://ui.adsabs.harvard.edu/abs/2021MNRAS.506.5410I} {506, 5410}

\bibitem[\protect\citeauthoryear{{Ivezi{\'c}} et~al.,}{{Ivezi{\'c}}
  et~al.}{2008}]{Ivezic2008}
{Ivezi{\'c}} {\v Z}.,  et~al., 2008, \mn@doi [\apj] {10.1086/589678}, \href
  {http://adsabs.harvard.edu/abs/2008ApJ...684..287I} {684, 287}

\bibitem[\protect\citeauthoryear{{Jean-Baptiste}, {Di Matteo}, {Haywood},
  {G{\'o}mez}, {Montuori}, {Combes}  \& {Semelin}}{{Jean-Baptiste}
  et~al.}{2017}]{JeanBptiste2017}
{Jean-Baptiste} I.,  {Di Matteo} P.,  {Haywood} M.,  {G{\'o}mez} A.,
  {Montuori} M.,  {Combes} F.,   {Semelin} B.,  2017, \mn@doi [\aap]
  {10.1051/0004-6361/201629691}, \href
  {https://ui.adsabs.harvard.edu/abs/2017A&A...604A.106J} {604, A106}

\bibitem[\protect\citeauthoryear{{Jofr{\'e}} \& {Weiss}}{{Jofr{\'e}} \&
  {Weiss}}{2011}]{Jofre2011}
{Jofr{\'e}} P.,  {Weiss} A.,  2011, \mn@doi [\aap]
  {10.1051/0004-6361/201117131}, \href
  {http://adsabs.harvard.edu/abs/2011A%26A...533A..59J} {533, A59}

\bibitem[\protect\citeauthoryear{{Jofr{\'e}}, {Panter}, {Hansen}  \&
  {Weiss}}{{Jofr{\'e}} et~al.}{2010}]{Jofre2010}
{Jofr{\'e}} P.,  {Panter} B.,  {Hansen} C.~J.,   {Weiss} A.,  2010, \mn@doi
  [\aap] {10.1051/0004-6361/201014013}, \href
  {http://adsabs.harvard.edu/abs/2010A%26A...517A..57J} {517, A57}

\bibitem[\protect\citeauthoryear{{Jofr{\'e}}, {Heiter}  \&
  {Soubiran}}{{Jofr{\'e}} et~al.}{2019}]{Jofre2019}
{Jofr{\'e}} P.,  {Heiter} U.,   {Soubiran} C.,  2019, \mn@doi [\araa]
  {10.1146/annurev-astro-091918-104509}, \href
  {https://ui.adsabs.harvard.edu/abs/2019ARA&A..57..571J} {57, 571}

\bibitem[\protect\citeauthoryear{{Juri{\'c}} et~al.,}{{Juri{\'c}}
  et~al.}{2008}]{Juric2008}
{Juri{\'c}} M.,  et~al., 2008, \mn@doi [\apj] {10.1086/523619}, \href
  {http://adsabs.harvard.edu/abs/2008ApJ...673..864J} {673, 864}

\bibitem[\protect\citeauthoryear{{Karovicova}, {White}, {Nordlander},
  {Casagrand e}, {Ireland}, {Huber}  \& {Jofr{\'e}}}{{Karovicova}
  et~al.}{2020}]{Karovicova2020}
{Karovicova} I.,  {White} T.~R.,  {Nordlander} T.,  {Casagrand e} L.,
  {Ireland} M.,  {Huber} D.,   {Jofr{\'e}} P.,  2020, \mn@doi [\aap]
  {10.1051/0004-6361/202037590}, \href
  {https://ui.adsabs.harvard.edu/abs/2020A&A...640A..25K} {640, A25}

\bibitem[\protect\citeauthoryear{{Katz} et~al.,}{{Katz}
  et~al.}{2019}]{Katz2019}
{Katz} D.,  et~al., 2019, \mn@doi [\aap] {10.1051/0004-6361/201833273}, \href
  {http://adsabs.harvard.edu/abs/2019A%26A...622A.205K} {622, A205}

\bibitem[\protect\citeauthoryear{{Kobayashi} \& {Nakasato}}{{Kobayashi} \&
  {Nakasato}}{2011}]{Kobayashi2011}
{Kobayashi} C.,  {Nakasato} N.,  2011, \mn@doi [\apj]
  {10.1088/0004-637X/729/1/16}, \href
  {http://adsabs.harvard.edu/abs/2011ApJ...729...16K} {729, 16}

\bibitem[\protect\citeauthoryear{{Kobayashi}, {Umeda}, {Nomoto}, {Tominaga}  \&
  {Ohkubo}}{{Kobayashi} et~al.}{2006}]{Kobayashi2006}
{Kobayashi} C.,  {Umeda} H.,  {Nomoto} K.,  {Tominaga} N.,   {Ohkubo} T.,
  2006, \mn@doi [\apj] {10.1086/508914}, \href
  {http://adsabs.harvard.edu/abs/2006ApJ...653.1145K} {653, 1145}

\bibitem[\protect\citeauthoryear{{Kobayashi}, {Leung}  \& {Nomoto}}{{Kobayashi}
  et~al.}{2020a}]{Kobayashi2020b}
{Kobayashi} C.,  {Leung} S.-C.,   {Nomoto} K.,  2020a, \mn@doi [\apj]
  {10.3847/1538-4357/ab8e44}, \href
  {https://ui.adsabs.harvard.edu/abs/2020ApJ...895..138K} {895, 138}

\bibitem[\protect\citeauthoryear{{Kobayashi}, {Karakas}  \&
  {Lugaro}}{{Kobayashi} et~al.}{2020b}]{Kobayashi2020}
{Kobayashi} C.,  {Karakas} A.~I.,   {Lugaro} M.,  2020b, \mn@doi [\apj]
  {10.3847/1538-4357/abae65}, \href
  {https://ui.adsabs.harvard.edu/abs/2020ApJ...900..179K} {900, 179}

\bibitem[\protect\citeauthoryear{{Koch-Hansen}, {Hansen}  \&
  {McWilliam}}{{Koch-Hansen} et~al.}{2021}]{KochHansen2021}
{Koch-Hansen} A.~J.,  {Hansen} C.~J.,   {McWilliam} A.,  2021, \mn@doi [\aap]
  {10.1051/0004-6361/202141130}, \href
  {https://ui.adsabs.harvard.edu/abs/2021A&A...653A...2K} {653, A2}

\bibitem[\protect\citeauthoryear{{Kollmeier} et~al.,}{{Kollmeier}
  et~al.}{2017}]{Kollmeier2017}
{Kollmeier} J.~A.,  et~al., 2017, arXiv e-prints, \href
  {https://ui.adsabs.harvard.edu/abs/2017arXiv171103234K} {p. arXiv:1711.03234}

\bibitem[\protect\citeauthoryear{{Koppelman}, {Helmi}  \&
  {Veljanoski}}{{Koppelman} et~al.}{2018}]{Koppelman2018}
{Koppelman} H.,  {Helmi} A.,   {Veljanoski} J.,  2018, \mn@doi [\apjl]
  {10.3847/2041-8213/aac882}, \href
  {https://ui.adsabs.harvard.edu/abs/2018ApJ...860L..11K} {860, L11}

\bibitem[\protect\citeauthoryear{{Koppelman}, {Helmi}, {Massari},
  {Price-Whelan}  \& {Starkenburg}}{{Koppelman} et~al.}{2019}]{Koppelman2019}
{Koppelman} H.~H.,  {Helmi} A.,  {Massari} D.,  {Price-Whelan} A.~M.,
  {Starkenburg} T.~K.,  2019, \mn@doi [\aap] {10.1051/0004-6361/201936738},
  \href {https://ui.adsabs.harvard.edu/abs/2019A&A...631L...9K} {631, L9}

\bibitem[\protect\citeauthoryear{{Koppelman}, {Bos}  \& {Helmi}}{{Koppelman}
  et~al.}{2020}]{Koppelman2020b}
{Koppelman} H.~H.,  {Bos} R. O.~Y.,   {Helmi} A.,  2020, \mn@doi [\aap]
  {10.1051/0004-6361/202038652}, \href
  {https://ui.adsabs.harvard.edu/abs/2020A&A...642L..18K} {642, L18}

\bibitem[\protect\citeauthoryear{{Koppelman}, {Hagen}  \& {Helmi}}{{Koppelman}
  et~al.}{2021}]{Koppelman2021}
{Koppelman} H.~H.,  {Hagen} J. H.~J.,   {Helmi} A.,  2021, \mn@doi [\aap]
  {10.1051/0004-6361/202039390}, \href
  {https://ui.adsabs.harvard.edu/abs/2021A&A...647A..37K} {647, A37}

\bibitem[\protect\citeauthoryear{{Lane}, {Bovy}  \& {Mackereth}}{{Lane}
  et~al.}{2021}]{Lane2021}
{Lane} J. M.~M.,  {Bovy} J.,   {Mackereth} J.~T.,  2021, arXiv e-prints, \href
  {https://ui.adsabs.harvard.edu/abs/2021arXiv210609699L} {p. arXiv:2106.09699}

\bibitem[\protect\citeauthoryear{{Leaman} et~al.,}{{Leaman}
  et~al.}{2017}]{Leaman2017}
{Leaman} R.,  et~al., 2017, \mn@doi [\mnras] {10.1093/mnras/stx2014}, \href
  {https://ui.adsabs.harvard.edu/abs/2017MNRAS.472.1879L} {472, 1879}

\bibitem[\protect\citeauthoryear{{Limberg} et~al.,}{{Limberg}
  et~al.}{2021}]{Limberg2021}
{Limberg} G.,  et~al., 2021, \mn@doi [\apjl] {10.3847/2041-8213/ac0056}, \href
  {https://ui.adsabs.harvard.edu/abs/2021ApJ...913L..28L} {913, L28}

\bibitem[\protect\citeauthoryear{{Lindegren} \& {Feltzing}}{{Lindegren} \&
  {Feltzing}}{2013}]{Lindegren2013}
{Lindegren} L.,  {Feltzing} S.,  2013, \mn@doi [\aap]
  {10.1051/0004-6361/201321057}, \href
  {http://adsabs.harvard.edu/abs/2013A%26A...553A..94L} {553, A94}

\bibitem[\protect\citeauthoryear{{Lindegren} et~al.,}{{Lindegren}
  et~al.}{2021a}]{Lindegren2021a}
{Lindegren} L.,  et~al., 2021a, \mn@doi [\aap] {10.1051/0004-6361/202039709},
  \href {https://ui.adsabs.harvard.edu/abs/2021A&A...649A...2L} {649, A2}

\bibitem[\protect\citeauthoryear{{Lindegren} et~al.,}{{Lindegren}
  et~al.}{2021b}]{Lindegren2021b}
{Lindegren} L.,  et~al., 2021b, \mn@doi [\aap] {10.1051/0004-6361/202039653},
  \href {https://ui.adsabs.harvard.edu/abs/2021A&A...649A...4L} {649, A4}

\bibitem[\protect\citeauthoryear{{Lu}, {Ness}, {Buck}  \& {Zinn}}{{Lu}
  et~al.}{2021}]{Lu2021}
{Lu} Y.,  {Ness} M.,  {Buck} T.,   {Zinn} J.,  2021, arXiv e-prints, \href
  {https://ui.adsabs.harvard.edu/abs/2021arXiv210212003Y} {p. arXiv:2102.12003}

\bibitem[\protect\citeauthoryear{{Mackereth} \& {Bovy}}{{Mackereth} \&
  {Bovy}}{2018}]{Mackereth2018}
{Mackereth} J.~T.,  {Bovy} J.,  2018, \mn@doi [\pasp]
  {10.1088/1538-3873/aadcdd}, \href
  {https://ui.adsabs.harvard.edu/abs/2018PASP..130k4501M} {130, 114501}

\bibitem[\protect\citeauthoryear{{Mackereth} et~al.,}{{Mackereth}
  et~al.}{2019}]{Mackereth2019}
{Mackereth} J.~T.,  et~al., 2019, \mn@doi [\mnras] {10.1093/mnras/sty2955},
  \href {http://adsabs.harvard.edu/abs/2019MNRAS.482.3426M} {482, 3426}

\bibitem[\protect\citeauthoryear{{Malhan} \& {Ibata}}{{Malhan} \&
  {Ibata}}{2018}]{Malhan2018}
{Malhan} K.,  {Ibata} R.~A.,  2018, \mn@doi [\mnras] {10.1093/mnras/sty912},
  \href {https://ui.adsabs.harvard.edu/abs/2018MNRAS.477.4063M} {477, 4063}

\bibitem[\protect\citeauthoryear{{Massari}, {Koppelman}  \& {Helmi}}{{Massari}
  et~al.}{2019}]{Massari2019}
{Massari} D.,  {Koppelman} H.~H.,   {Helmi} A.,  2019, \mn@doi [\aap]
  {10.1051/0004-6361/201936135}, \href
  {https://ui.adsabs.harvard.edu/abs/2019A&A...630L...4M} {630, L4}

\bibitem[\protect\citeauthoryear{{Matsuno}, {Aoki}  \& {Suda}}{{Matsuno}
  et~al.}{2019}]{Matsuno2019}
{Matsuno} T.,  {Aoki} W.,   {Suda} T.,  2019, \mn@doi [\apjl]
  {10.3847/2041-8213/ab0ec0}, \href
  {https://ui.adsabs.harvard.edu/abs/2019ApJ...874L..35M} {874, L35}

\bibitem[\protect\citeauthoryear{{Matsuno}, {Hirai}, {Tarumi}, {Hotokezaka},
  {Tanaka}  \& {Helmi}}{{Matsuno} et~al.}{2021}]{Matsuno2021}
{Matsuno} T.,  {Hirai} Y.,  {Tarumi} Y.,  {Hotokezaka} K.,  {Tanaka} M.,
  {Helmi} A.,  2021, \mn@doi [\aap] {10.1051/0004-6361/202040227}, \href
  {https://ui.adsabs.harvard.edu/abs/2021A&A...650A.110M} {650, A110}

\bibitem[\protect\citeauthoryear{{Matteucci}}{{Matteucci}}{2021}]{Matteucci2021}
{Matteucci} F.,  2021, \mn@doi [\aapr] {10.1007/s00159-021-00133-8}, \href
  {https://ui.adsabs.harvard.edu/abs/2021A&ARv..29....5M} {29, 5}

\bibitem[\protect\citeauthoryear{{McMillan}}{{McMillan}}{2017}]{McMillan2017}
{McMillan} P.~J.,  2017, \mn@doi [\mnras] {10.1093/mnras/stw2759}, \href
  {https://ui.adsabs.harvard.edu/abs/2017MNRAS.465...76M} {465, 76}

\bibitem[\protect\citeauthoryear{{McWilliam}}{{McWilliam}}{1997}]{McWilliam1997}
{McWilliam} A.,  1997, \mn@doi [\araa] {10.1146/annurev.astro.35.1.503}, \href
  {http://adsabs.harvard.edu/abs/1997ARA%26A..35..503M} {35, 503}

\bibitem[\protect\citeauthoryear{{Miglio} et~al.,}{{Miglio}
  et~al.}{2021}]{Miglio2021}
{Miglio} A.,  et~al., 2021, \mn@doi [\aap] {10.1051/0004-6361/202038307}, \href
  {https://ui.adsabs.harvard.edu/abs/2021A&A...645A..85M} {645, A85}

\bibitem[\protect\citeauthoryear{{Minchev}, {Steinmetz}, {Chiappini}, {Martig},
  {Anders}, {Matijevic}  \& {de Jong}}{{Minchev} et~al.}{2017}]{Minchev2017}
{Minchev} I.,  {Steinmetz} M.,  {Chiappini} C.,  {Martig} M.,  {Anders} F.,
  {Matijevic} G.,   {de Jong} R.~S.,  2017, \mn@doi [\apj]
  {10.3847/1538-4357/834/1/27}, \href
  {http://adsabs.harvard.edu/abs/2017ApJ...834...27M} {834, 27}

\bibitem[\protect\citeauthoryear{{Molaro}, {Cescutti}  \& {Fu}}{{Molaro}
  et~al.}{2020}]{Molaro2020}
{Molaro} P.,  {Cescutti} G.,   {Fu} X.,  2020, \mn@doi [\mnras]
  {10.1093/mnras/staa1653}, \href
  {https://ui.adsabs.harvard.edu/abs/2020MNRAS.496.2902M} {496, 2902}

\bibitem[\protect\citeauthoryear{{Monachesi} et~al.,}{{Monachesi}
  et~al.}{2019}]{Monachesi2019}
{Monachesi} A.,  et~al., 2019, \mn@doi [\mnras] {10.1093/mnras/stz538}, \href
  {https://ui.adsabs.harvard.edu/abs/2019MNRAS.485.2589M} {485, 2589}

\bibitem[\protect\citeauthoryear{{Montalb{\'a}n} et~al.,}{{Montalb{\'a}n}
  et~al.}{2021}]{Montalban2021}
{Montalb{\'a}n} J.,  et~al., 2021, \mn@doi [Nature Astronomy]
  {10.1038/s41550-021-01347-7}, \href
  {https://ui.adsabs.harvard.edu/abs/2021NatAs...5..640M} {5, 640}

\bibitem[\protect\citeauthoryear{{Monty}, {Venn}, {Lane}, {Lokhorst}  \&
  {Yong}}{{Monty} et~al.}{2020}]{Monty2020}
{Monty} S.,  {Venn} K.~A.,  {Lane} J. M.~M.,  {Lokhorst} D.,   {Yong} D.,
  2020, \mn@doi [\mnras] {10.1093/mnras/staa1995}, \href
  {https://ui.adsabs.harvard.edu/abs/2020MNRAS.497.1236M} {497, 1236}

\bibitem[\protect\citeauthoryear{{Myeong}, {Evans}, {Belokurov}, {Amorisco}  \&
  {Koposov}}{{Myeong} et~al.}{2018a}]{Myeong2018c}
{Myeong} G.~C.,  {Evans} N.~W.,  {Belokurov} V.,  {Amorisco} N.~C.,   {Koposov}
  S.~E.,  2018a, \mn@doi [\mnras] {10.1093/mnras/stx3262}, \href
  {https://ui.adsabs.harvard.edu/abs/2018MNRAS.475.1537M} {475, 1537}

\bibitem[\protect\citeauthoryear{{Myeong}, {Evans}, {Belokurov}, {Sanders}  \&
  {Koposov}}{{Myeong} et~al.}{2018b}]{Myeong2018b}
{Myeong} G.~C.,  {Evans} N.~W.,  {Belokurov} V.,  {Sanders} J.~L.,   {Koposov}
  S.~E.,  2018b, \mn@doi [\mnras] {10.1093/mnras/sty1403}, \href
  {http://adsabs.harvard.edu/abs/2018MNRAS.478.5449M} {478, 5449}

\bibitem[\protect\citeauthoryear{{Myeong}, {Evans}, {Belokurov}, {Sanders}  \&
  {Koposov}}{{Myeong} et~al.}{2018c}]{Myeong2018a}
{Myeong} G.~C.,  {Evans} N.~W.,  {Belokurov} V.,  {Sanders} J.~L.,   {Koposov}
  S.~E.,  2018c, \mn@doi [\apjl] {10.3847/2041-8213/aab613}, \href
  {http://adsabs.harvard.edu/abs/2018ApJ...856L..26M} {856, L26}

\bibitem[\protect\citeauthoryear{{Myeong}, {Vasiliev}, {Iorio}, {Evans}  \&
  {Belokurov}}{{Myeong} et~al.}{2019}]{Myeong2019}
{Myeong} G.~C.,  {Vasiliev} E.,  {Iorio} G.,  {Evans} N.~W.,   {Belokurov} V.,
  2019, \mn@doi [\mnras] {10.1093/mnras/stz1770}, \href
  {https://ui.adsabs.harvard.edu/abs/2019MNRAS.488.1235M} {488, 1235}

\bibitem[\protect\citeauthoryear{{Naidu}, {Conroy}, {Bonaca}, {Johnson},
  {Ting}, {Caldwell}, {Zaritsky}  \& {Cargile}}{{Naidu}
  et~al.}{2020}]{Naidu2020}
{Naidu} R.~P.,  {Conroy} C.,  {Bonaca} A.,  {Johnson} B.~D.,  {Ting} Y.-S.,
  {Caldwell} N.,  {Zaritsky} D.,   {Cargile} P.~A.,  2020, \mn@doi [\apj]
  {10.3847/1538-4357/abaef4}, \href
  {https://ui.adsabs.harvard.edu/abs/2020ApJ...901...48N} {901, 48}

\bibitem[\protect\citeauthoryear{{Naidu} et~al.,}{{Naidu}
  et~al.}{2021}]{Naidu2021}
{Naidu} R.~P.,  et~al., 2021, arXiv preprints, \href
  {https://ui.adsabs.harvard.edu/abs/2021arXiv210303251N} {p. 2103.03251}

\bibitem[\protect\citeauthoryear{{Navarro}, {Abadi}, {Venn}, {Freeman}  \&
  {Anguiano}}{{Navarro} et~al.}{2011}]{Navarro2011}
{Navarro} J.~F.,  {Abadi} M.~G.,  {Venn} K.~A.,  {Freeman} K.~C.,   {Anguiano}
  B.,  2011, \mn@doi [\mnras] {10.1111/j.1365-2966.2010.17975.x}, \href
  {https://ui.adsabs.harvard.edu/abs/2011MNRAS.412.1203N} {412, 1203}

\bibitem[\protect\citeauthoryear{{Necib} et~al.,}{{Necib}
  et~al.}{2020}]{Necib2020}
{Necib} L.,  et~al., 2020, \mn@doi [Nature Astronomy]
  {10.1038/s41550-020-1131-2}, \href
  {https://ui.adsabs.harvard.edu/abs/2020NatAs...4.1078N} {4, 1078}

\bibitem[\protect\citeauthoryear{{Ness} et~al.,}{{Ness}
  et~al.}{2018}]{Ness2018}
{Ness} M.,  et~al., 2018, \mn@doi [\apj] {10.3847/1538-4357/aa9d8e}, \href
  {http://adsabs.harvard.edu/abs/2018ApJ...853..198N} {853, 198}

\bibitem[\protect\citeauthoryear{{Ness}, {Johnston}, {Blancato}, {Rix},
  {Beane}, {Bird}  \& {Hawkins}}{{Ness} et~al.}{2019}]{Ness2019b}
{Ness} M.~K.,  {Johnston} K.~V.,  {Blancato} K.,  {Rix} H.~W.,  {Beane} A.,
  {Bird} J.~C.,   {Hawkins} K.,  2019, \mn@doi [\apj]
  {10.3847/1538-4357/ab3e3c}, \href
  {https://ui.adsabs.harvard.edu/abs/2019ApJ...883..177N} {883, 177}

\bibitem[\protect\citeauthoryear{{Nikakhtar} et~al.,}{{Nikakhtar}
  et~al.}{2021}]{Nikakhtar2021}
{Nikakhtar} F.,  et~al., 2021, \mn@doi [\apj] {10.3847/1538-4357/ac1a10}, \href
  {https://ui.adsabs.harvard.edu/abs/2021ApJ...921..106N} {921, 106}

\bibitem[\protect\citeauthoryear{{Nissen} \& {Gustafsson}}{{Nissen} \&
  {Gustafsson}}{2018}]{Nissen2018}
{Nissen} P.~E.,  {Gustafsson} B.,  2018, \mn@doi [\aapr]
  {10.1007/s00159-018-0111-3}, \href
  {http://adsabs.harvard.edu/abs/2018A%26ARv..26....6N} {26, 6}

\bibitem[\protect\citeauthoryear{{Nissen} \& {Schuster}}{{Nissen} \&
  {Schuster}}{1997}]{Nissen1997b}
{Nissen} P.~E.,  {Schuster} W.~J.,  1997, \aap, \href
  {https://ui.adsabs.harvard.edu/abs/1997A&A...326..751N} {326, 751}

\bibitem[\protect\citeauthoryear{{Nissen} \& {Schuster}}{{Nissen} \&
  {Schuster}}{2010}]{Nissen2010}
{Nissen} P.~E.,  {Schuster} W.~J.,  2010, \mn@doi [\aap]
  {10.1051/0004-6361/200913877}, \href
  {http://adsabs.harvard.edu/abs/2010A%26A...511L..10N} {511, L10}

\bibitem[\protect\citeauthoryear{{Nissen} \& {Schuster}}{{Nissen} \&
  {Schuster}}{2011}]{Nissen2011}
{Nissen} P.~E.,  {Schuster} W.~J.,  2011, \mn@doi [\aap]
  {10.1051/0004-6361/201116619}, \href
  {http://adsabs.harvard.edu/abs/2011A%26A...530A..15N} {530, A15}

\bibitem[\protect\citeauthoryear{{Nissen} \& {Schuster}}{{Nissen} \&
  {Schuster}}{2012}]{Nissen2012}
{Nissen} P.~E.,  {Schuster} W.~J.,  2012, \mn@doi [\aap]
  {10.1051/0004-6361/201219342}, \href
  {http://adsabs.harvard.edu/abs/2012A%26A...543A..28N} {543, A28}

\bibitem[\protect\citeauthoryear{{Nissen}, {Hoeg}  \& {Schuster}}{{Nissen}
  et~al.}{1997}]{Nissen1997}
{Nissen} P.~E.,  {Hoeg} E.,   {Schuster} W.~J.,  1997, in {Bonnet} R.~M.,
  et~al., eds,  ESA Special Publication Vol. 402, Hipparcos - Venice '97. pp
  225--230

\bibitem[\protect\citeauthoryear{{Nissen}, {Chen}, {Carigi}, {Schuster}  \&
  {Zhao}}{{Nissen} et~al.}{2014}]{Nissen2014}
{Nissen} P.~E.,  {Chen} Y.~Q.,  {Carigi} L.,  {Schuster} W.~J.,   {Zhao} G.,
  2014, \mn@doi [\aap] {10.1051/0004-6361/201424184}, \href
  {http://adsabs.harvard.edu/abs/2014A%26A...568A..25N} {568, A25}

\bibitem[\protect\citeauthoryear{{Nomoto}, {Kobayashi}  \& {Tominaga}}{{Nomoto}
  et~al.}{2013}]{Nomoto2013}
{Nomoto} K.,  {Kobayashi} C.,   {Tominaga} N.,  2013, \mn@doi [\araa]
  {10.1146/annurev-astro-082812-140956}, \href
  {https://ui.adsabs.harvard.edu/abs/2013ARA&A..51..457N} {51, 457}

\bibitem[\protect\citeauthoryear{{Obreja} et~al.,}{{Obreja}
  et~al.}{2019}]{Obreja2019}
{Obreja} A.,  et~al., 2019, \mn@doi [\mnras] {10.1093/mnras/stz1563}, \href
  {https://ui.adsabs.harvard.edu/abs/2019MNRAS.487.4424O} {487, 4424}

\bibitem[\protect\citeauthoryear{{Ochsenbein}, {Bauer}  \&
  {Marcout}}{{Ochsenbein} et~al.}{2000}]{Vizier2000}
{Ochsenbein} F.,  {Bauer} P.,   {Marcout} J.,  2000, \mn@doi [\aaps]
  {10.1051/aas:2000169}, \href
  {https://ui.adsabs.harvard.edu/abs/2000A&AS..143...23O} {143, 23}

\bibitem[\protect\citeauthoryear{{Ostdiek} et~al.,}{{Ostdiek}
  et~al.}{2020}]{Ostdiek2020}
{Ostdiek} B.,  et~al., 2020, \mn@doi [\aap] {10.1051/0004-6361/201936866},
  \href {https://ui.adsabs.harvard.edu/abs/2020A&A...636A..75O} {636, A75}

\bibitem[\protect\citeauthoryear{Pedregosa et~al.,}{Pedregosa
  et~al.}{2011}]{scikit-learn}
Pedregosa F.,  et~al., 2011, J Mach Learn Res, 12, 2825

\bibitem[\protect\citeauthoryear{P\'erez \& Granger}{P\'erez \&
  Granger}{2007}]{ipython}
P\'erez F.,  Granger B.~E.,  2007, \mn@doi [Comput Sci Eng]
  {10.1109/MCSE.2007.53}, 9, 21

\bibitem[\protect\citeauthoryear{{Pillepich}, {Madau}  \& {Mayer}}{{Pillepich}
  et~al.}{2015}]{Pillepich2015}
{Pillepich} A.,  {Madau} P.,   {Mayer} L.,  2015, \mn@doi [\apj]
  {10.1088/0004-637X/799/2/184}, \href
  {http://adsabs.harvard.edu/abs/2015ApJ...799..184P} {799, 184}

\bibitem[\protect\citeauthoryear{{Piskunov} \& {Valenti}}{{Piskunov} \&
  {Valenti}}{2017}]{Piskunov2017}
{Piskunov} N.,  {Valenti} J.~A.,  2017, \mn@doi [\aap]
  {10.1051/0004-6361/201629124}, \href
  {http://adsabs.harvard.edu/abs/2017A%26A...597A..16P} {597, A16}

\bibitem[\protect\citeauthoryear{{Ram{\'\i}rez}, {Mel{\'e}ndez}  \&
  {Chanam{\'e}}}{{Ram{\'\i}rez} et~al.}{2012}]{Ramirez2012b}
{Ram{\'\i}rez} I.,  {Mel{\'e}ndez} J.,   {Chanam{\'e}} J.,  2012, \mn@doi
  [\apj] {10.1088/0004-637X/757/2/164}, \href
  {https://ui.adsabs.harvard.edu/abs/2012ApJ...757..164R} {757, 164}

\bibitem[\protect\citeauthoryear{{Re Fiorentin}, {Lattanzi}, {Spagna}  \&
  {Curir}}{{Re Fiorentin} et~al.}{2015}]{ReFiorentin2015}
{Re Fiorentin} P.,  {Lattanzi} M.~G.,  {Spagna} A.,   {Curir} A.,  2015,
  \mn@doi [\aj] {10.1088/0004-6256/150/4/128}, \href
  {https://ui.adsabs.harvard.edu/abs/2015AJ....150..128R} {150, 128}

\bibitem[\protect\citeauthoryear{{Recio-Blanco}, {Fern{\'a}ndez-Alvar}, {de
  Laverny}, {Antoja}, {Helmi}  \& {Crida}}{{Recio-Blanco}
  et~al.}{2021}]{Recio-Blanco2021}
{Recio-Blanco} A.,  {Fern{\'a}ndez-Alvar} E.,  {de Laverny} P.,  {Antoja} T.,
  {Helmi} A.,   {Crida} A.,  2021, \mn@doi [\aap]
  {10.1051/0004-6361/202038943}, \href
  {https://ui.adsabs.harvard.edu/abs/2021A&A...648A.108R} {648, A108}

\bibitem[\protect\citeauthoryear{{Reid} \& {Brunthaler}}{{Reid} \&
  {Brunthaler}}{2004}]{Reid2004}
{Reid} M.~J.,  {Brunthaler} A.,  2004, \mn@doi [\apj] {10.1086/424960}, \href
  {https://ui.adsabs.harvard.edu/abs/2004ApJ...616..872R} {616, 872}

\bibitem[\protect\citeauthoryear{{Renaud}, {Agertz}, {Read}, {Ryde},
  {Andersson}, {Bensby}, {Rey}  \& {Feuillet}}{{Renaud}
  et~al.}{2021a}]{Renaud2021}
{Renaud} F.,  {Agertz} O.,  {Read} J.~I.,  {Ryde} N.,  {Andersson} E.~P.,
  {Bensby} T.,  {Rey} M.~P.,   {Feuillet} D.~K.,  2021a, \mn@doi [\mnras]
  {10.1093/mnras/stab250}, \href
  {https://ui.adsabs.harvard.edu/abs/2021MNRAS.503.5846R} {503, 5846}

\bibitem[\protect\citeauthoryear{{Renaud}, {Agertz}, {Andersson}, {Read},
  {Ryde}, {Bensby}, {Rey}  \& {Feuillet}}{{Renaud} et~al.}{2021b}]{Renaud2021b}
{Renaud} F.,  {Agertz} O.,  {Andersson} E.~P.,  {Read} J.~I.,  {Ryde} N.,
  {Bensby} T.,  {Rey} M.~P.,   {Feuillet} D.~K.,  2021b, \mn@doi [\mnras]
  {10.1093/mnras/stab543}, \href
  {https://ui.adsabs.harvard.edu/abs/2021MNRAS.503.5868R} {503, 5868}

\bibitem[\protect\citeauthoryear{{Rix} \& {Bovy}}{{Rix} \&
  {Bovy}}{2013}]{Rix2013}
{Rix} H.-W.,  {Bovy} J.,  2013, \mn@doi [\aapr] {10.1007/s00159-013-0061-8},
  \href {http://adsabs.harvard.edu/abs/2013A%26ARv..21...61R} {21, 61}

\bibitem[\protect\citeauthoryear{{Sanders}, {Belokurov}  \& {Man}}{{Sanders}
  et~al.}{2021}]{Sanders2021}
{Sanders} J.~L.,  {Belokurov} V.,   {Man} K. T.~F.,  2021, \mn@doi [\mnras]
  {10.1093/mnras/stab1951}, \href
  {https://ui.adsabs.harvard.edu/abs/2021MNRAS.506.4321S} {506, 4321}

\bibitem[\protect\citeauthoryear{{Sch{\"o}nrich}, {Binney}  \&
  {Dehnen}}{{Sch{\"o}nrich} et~al.}{2010}]{Schoenrich2010}
{Sch{\"o}nrich} R.,  {Binney} J.,   {Dehnen} W.,  2010, \mn@doi [\mnras]
  {10.1111/j.1365-2966.2010.16253.x}, \href
  {http://adsabs.harvard.edu/abs/2010MNRAS.403.1829S} {403, 1829}

\bibitem[\protect\citeauthoryear{{Schuler}, {Andrews}, {Clanzy}, {Mourabit},
  {Chanam{\'e}}  \& {Ag{\"u}eros}}{{Schuler} et~al.}{2021}]{Schuler2021}
{Schuler} S.~C.,  {Andrews} J.~J.,  {Clanzy} V.~R.,  {Mourabit} M.,
  {Chanam{\'e}} J.,   {Ag{\"u}eros} M.~A.,  2021, \mn@doi [\aj]
  {10.3847/1538-3881/ac10c6}, \href
  {https://ui.adsabs.harvard.edu/abs/2021AJ....162..109S} {162, 109}

\bibitem[\protect\citeauthoryear{{Schuster}, {Moreno}, {Nissen}  \&
  {Pichardo}}{{Schuster} et~al.}{2012}]{Schuster2012}
{Schuster} W.~J.,  {Moreno} E.,  {Nissen} P.~E.,   {Pichardo} B.,  2012,
  \mn@doi [\aap] {10.1051/0004-6361/201118035}, \href
  {http://adsabs.harvard.edu/abs/2012A%26A...538A..21S} {538, A21}

\bibitem[\protect\citeauthoryear{{Schwarz}}{{Schwarz}}{1978}]{Schwarz1978}
{Schwarz} G.,  1978, Annals of Statistics, \href
  {https://ui.adsabs.harvard.edu/abs/1978AnSta...6..461S} {6, 461}

\bibitem[\protect\citeauthoryear{{Searle} \& {Zinn}}{{Searle} \&
  {Zinn}}{1978}]{Searle1978}
{Searle} L.,  {Zinn} R.,  1978, \mn@doi [\apj] {10.1086/156499}, \href
  {http://adsabs.harvard.edu/abs/1978ApJ...225..357S} {225, 357}

\bibitem[\protect\citeauthoryear{{Sellwood}}{{Sellwood}}{2014}]{Sellwood2014}
{Sellwood} J.~A.,  2014, \mn@doi [Reviews of Modern Physics]
  {10.1103/RevModPhys.86.1}, \href
  {https://ui.adsabs.harvard.edu/abs/2014RvMP...86....1S} {86, 1}

\bibitem[\protect\citeauthoryear{{Sestito} et~al.,}{{Sestito}
  et~al.}{2021}]{Sestito2021}
{Sestito} F.,  et~al., 2021, \mn@doi [\mnras] {10.1093/mnras/staa3479}, \href
  {https://ui.adsabs.harvard.edu/abs/2021MNRAS.500.3750S} {500, 3750}

\bibitem[\protect\citeauthoryear{{Sharma} et~al.,}{{Sharma}
  et~al.}{2018}]{Sharma2018}
{Sharma} S.,  et~al., 2018, \mn@doi [\mnras] {10.1093/mnras/stx2582}, \href
  {http://adsabs.harvard.edu/abs/2018MNRAS.473.2004S} {473, 2004}

\bibitem[\protect\citeauthoryear{{Sheinis} et~al.,}{{Sheinis}
  et~al.}{2015}]{Sheinis2015}
{Sheinis} A.,  et~al., 2015, \mn@doi [J. Astron. Telesc. Instrum. Syst.]
  {10.1117/1.JATIS.1.3.035002}, \href
  {http://adsabs.harvard.edu/abs/2015JATIS...1c5002S} {1, 035002}

\bibitem[\protect\citeauthoryear{{Simpson} et~al.,}{{Simpson}
  et~al.}{2021}]{Simpson2021}
{Simpson} J.~D.,  et~al., 2021, \mn@doi [\mnras] {10.1093/mnras/stab2012},
  \href {https://ui.adsabs.harvard.edu/abs/2021MNRAS.507...43S} {507, 43}

\bibitem[\protect\citeauthoryear{Skrutskie et~al.,}{Skrutskie
  et~al.}{2006}]{Skrutskie2006}
Skrutskie M.~F.,  et~al., 2006, \mn@doi [\aj] {10.1086/498708}, 131, 1163

\bibitem[\protect\citeauthoryear{{Soderblom}}{{Soderblom}}{2010}]{Soderblom2010}
{Soderblom} D.~R.,  2010, \mn@doi [\araa]
  {10.1146/annurev-astro-081309-130806}, \href
  {http://adsabs.harvard.edu/abs/2010ARA%26A..48..581S} {48, 581}

\bibitem[\protect\citeauthoryear{{Spite} \& {Spite}}{{Spite} \&
  {Spite}}{1982}]{Spite1982}
{Spite} F.,  {Spite} M.,  1982, \aap, \href
  {http://adsabs.harvard.edu/abs/1982A%26A...115..357S} {115, 357}

\bibitem[\protect\citeauthoryear{{Taylor}}{{Taylor}}{2005}]{Taylor2005}
{Taylor} M.~B.,  2005, in {Shopbell} P.,  {Britton} M.,   {Ebert} R.,  eds,
  ~ASPC Vol. 347, Astronomical Data Analysis Software and Systems XIV. p.~29

\bibitem[\protect\citeauthoryear{{Ting} \& {Weinberg}}{{Ting} \&
  {Weinberg}}{2021}]{Ting2021}
{Ting} Y.-S.,  {Weinberg} D.~H.,  2021, arXiv e-prints, \href
  {https://ui.adsabs.harvard.edu/abs/2021arXiv210204992T} {p. arXiv:2102.04992}

\bibitem[\protect\citeauthoryear{{Ting}, {Freeman}, {Kobayashi}, {De Silva}  \&
  {Bland-Hawthorn}}{{Ting} et~al.}{2012}]{Ting2012}
{Ting} Y.-S.,  {Freeman} K.~C.,  {Kobayashi} C.,  {De Silva} G.~M.,
  {Bland-Hawthorn} J.,  2012, \mn@doi [\mnras]
  {10.1111/j.1365-2966.2011.20387.x}, \href
  {http://adsabs.harvard.edu/abs/2012MNRAS.421.1231T} {421, 1231}

\bibitem[\protect\citeauthoryear{{Ting}, {Rix}, {Bovy}  \& {van de Ven}}{{Ting}
  et~al.}{2013}]{Ting2013}
{Ting} Y.-S.,  {Rix} H.-W.,  {Bovy} J.,   {van de Ven} G.,  2013, \mn@doi
  [\mnras] {10.1093/mnras/stt1053}, \href
  {https://ui.adsabs.harvard.edu/abs/2013MNRAS.434..652T} {434, 652}

\bibitem[\protect\citeauthoryear{{Ting}, {Conroy}  \& {Goodman}}{{Ting}
  et~al.}{2015}]{Ting2015}
{Ting} Y.-S.,  {Conroy} C.,   {Goodman} A.,  2015, \mn@doi [\apj]
  {10.1088/0004-637X/807/1/104}, \href
  {http://adsabs.harvard.edu/abs/2015ApJ...807..104T} {807, 104}

\bibitem[\protect\citeauthoryear{{Tinsley}}{{Tinsley}}{1979}]{Tinsley1979}
{Tinsley} B.~M.,  1979, \mn@doi [\apj] {10.1086/157039}, \href
  {http://adsabs.harvard.edu/abs/1979ApJ...229.1046T} {229, 1046}

\bibitem[\protect\citeauthoryear{{Tsujimoto}, {Nomoto}, {Yoshii}, {Hashimoto},
  {Yanagida}  \& {Thielemann}}{{Tsujimoto} et~al.}{1995}]{Tsujimoto1995}
{Tsujimoto} T.,  {Nomoto} K.,  {Yoshii} Y.,  {Hashimoto} M.,  {Yanagida} S.,
  {Thielemann} F.-K.,  1995, \mn@doi [\mnras] {10.1093/mnras/277.3.945}, \href
  {http://adsabs.harvard.edu/abs/1995MNRAS.277..945T} {277, 945}

\bibitem[\protect\citeauthoryear{{Valenti} \& {Piskunov}}{{Valenti} \&
  {Piskunov}}{1996}]{Valenti1996}
{Valenti} J.~A.,  {Piskunov} N.,  1996, \aaps, \href
  {http://adsabs.harvard.edu/abs/1996A%26AS..118..595V} {118, 595}

\bibitem[\protect\citeauthoryear{VanderPlas}{VanderPlas}{2016}]{VanderPlas2016}
VanderPlas J.,  2016, Python Data Science Handbook: Essential Tools for Working
  with Data, 1st edn.
O'Reilly Media, Inc.

\bibitem[\protect\citeauthoryear{{Vanderplas}, {Connolly}, {Ivezi{\'c}}  \&
  {Gray}}{{Vanderplas} et~al.}{2012}]{astroml}
{Vanderplas} J.,  {Connolly} A.,  {Ivezi{\'c}} {\v Z}.,   {Gray} A.,  2012, in
  Conference on Intelligent Data Understanding (CIDU). pp 47 --54

\bibitem[\protect\citeauthoryear{{Vasiliev}}{{Vasiliev}}{2019}]{Vasiliev2019}
{Vasiliev} E.,  2019, \mn@doi [\mnras] {10.1093/mnras/stz171}, \href
  {https://ui.adsabs.harvard.edu/abs/2019MNRAS.484.2832V} {484, 2832}

\bibitem[\protect\citeauthoryear{{Venn}, {Irwin}, {Shetrone}, {Tout}, {Hill}
  \& {Tolstoy}}{{Venn} et~al.}{2004}]{Venn2004}
{Venn} K.~A.,  {Irwin} M.,  {Shetrone} M.~D.,  {Tout} C.~A.,  {Hill} V.,
  {Tolstoy} E.,  2004, \mn@doi [\aj] {10.1086/422734}, \href
  {http://adsabs.harvard.edu/abs/2004AJ....128.1177V} {128, 1177}

\bibitem[\protect\citeauthoryear{{Villalobos} \& {Helmi}}{{Villalobos} \&
  {Helmi}}{2008}]{Villalobos2008}
{Villalobos} {\'A}.,  {Helmi} A.,  2008, \mn@doi [\mnras]
  {10.1111/j.1365-2966.2008.13979.x}, \href
  {http://adsabs.harvard.edu/abs/2008MNRAS.391.1806V} {391, 1806}

\bibitem[\protect\citeauthoryear{{Vincenzo}, {Spitoni}, {Calura}, {Matteucci},
  {Silva Aguirre}, {Miglio}  \& {Cescutti}}{{Vincenzo}
  et~al.}{2019}]{Vincenzo2019}
{Vincenzo} F.,  {Spitoni} E.,  {Calura} F.,  {Matteucci} F.,  {Silva Aguirre}
  V.,  {Miglio} A.,   {Cescutti} G.,  2019, \mn@doi [\mnras]
  {10.1093/mnrasl/slz070}, \href
  {https://ui.adsabs.harvard.edu/abs/2019MNRAS.487L..47V} {487, L47}

\bibitem[\protect\citeauthoryear{Virtanen et~al.,}{Virtanen
  et~al.}{2020}]{scipy}
Virtanen P.,  et~al., 2020, \mn@doi [Nature Methods]
  {10.1038/s41592-019-0686-2}, \href {https://rdcu.be/b08Wh} {17, 261}

\bibitem[\protect\citeauthoryear{Walt, Colbert  \& Varoquaux}{Walt
  et~al.}{2011}]{numpy}
Walt S. v.~d.,  Colbert S.~C.,   Varoquaux G.,  2011, \mn@doi [Comput Sci Eng]
  {10.1109/MCSE.2011.37}, 13, 22

\bibitem[\protect\citeauthoryear{{Weinberg} et~al.,}{{Weinberg}
  et~al.}{2021}]{Weinberg2021}
{Weinberg} D.~H.,  et~al., 2021, arXiv e-prints, \href
  {https://ui.adsabs.harvard.edu/abs/2021arXiv210808860W} {p. arXiv:2108.08860}

\bibitem[\protect\citeauthoryear{{Wisnioski} et~al.,}{{Wisnioski}
  et~al.}{2015}]{Wisnioski2015}
{Wisnioski} E.,  et~al., 2015, \mn@doi [\apj] {10.1088/0004-637X/799/2/209},
  \href {https://ui.adsabs.harvard.edu/abs/2015ApJ...799..209W} {799, 209}

\bibitem[\protect\citeauthoryear{{Woosley}, {Heger}  \& {Weaver}}{{Woosley}
  et~al.}{2002}]{Woosley2002}
{Woosley} S.~E.,  {Heger} A.,   {Weaver} T.~A.,  2002, \mn@doi [Reviews of
  Modern Physics] {10.1103/RevModPhys.74.1015}, \href
  {https://ui.adsabs.harvard.edu/abs/2002RvMP...74.1015W} {74, 1015}

\bibitem[\protect\citeauthoryear{{Wu}, {Valluri}, {Panithanpaisal},
  {Sanderson}, {Freese}, {Wetzel}  \& {Sharma}}{{Wu} et~al.}{2021}]{Wu2021}
{Wu} Y.,  {Valluri} M.,  {Panithanpaisal} N.,  {Sanderson} R.~E.,  {Freese} K.,
   {Wetzel} A.,   {Sharma} S.,  2021, \mn@doi [\mnras]
  {10.1093/mnras/stab3306}, \href
  {https://ui.adsabs.harvard.edu/abs/2021MNRAS.tmp.3021W} {}

\bibitem[\protect\citeauthoryear{{Wyse}}{{Wyse}}{2001}]{Wyse2001}
{Wyse} R.~F.~G.,  2001, in {Funes} J.~G.,  {Corsini} E.~M.,  eds, ~ASPC Vol.
  230, Galaxy Disks and Disk Galaxies. pp 71--80

\bibitem[\protect\citeauthoryear{{Yuan}, {Chang}, {Banerjee}, {Han}, {Kang}  \&
  {Smith}}{{Yuan} et~al.}{2018}]{Yuan2018}
{Yuan} Z.,  {Chang} J.,  {Banerjee} P.,  {Han} J.,  {Kang} X.,   {Smith} M.~C.,
   2018, \mn@doi [\apj] {10.3847/1538-4357/aacd0d}, \href
  {https://ui.adsabs.harvard.edu/abs/2018ApJ...863...26Y} {863, 26}

\bibitem[\protect\citeauthoryear{{Yuan} et~al.,}{{Yuan}
  et~al.}{2020a}]{Yuan2020}
{Yuan} Z.,  et~al., 2020a, \mn@doi [\apj] {10.3847/1538-4357/ab6ef7}, \href
  {https://ui.adsabs.harvard.edu/abs/2020ApJ...891...39Y} {891, 39}

\bibitem[\protect\citeauthoryear{{Yuan}, {Chang}, {Beers}  \& {Huang}}{{Yuan}
  et~al.}{2020b}]{Yuan2020a}
{Yuan} Z.,  {Chang} J.,  {Beers} T.~C.,   {Huang} Y.,  2020b, \mn@doi [\apjl]
  {10.3847/2041-8213/aba49f}, \href
  {https://ui.adsabs.harvard.edu/abs/2020ApJ...898L..37Y} {898, L37}

\bibitem[\protect\citeauthoryear{{Zhang} et~al.,}{{Zhang}
  et~al.}{2021}]{Zhang2021}
{Zhang} M.,  et~al., 2021, arXiv e-prints, \href
  {https://ui.adsabs.harvard.edu/abs/2021arXiv210900746Z} {p. arXiv:2109.00746}

\bibitem[\protect\citeauthoryear{{Zolotov}, {Willman}, {Brooks}, {Governato},
  {Hogg}, {Shen}  \& {Wadsley}}{{Zolotov} et~al.}{2010}]{Zolotov2010}
{Zolotov} A.,  {Willman} B.,  {Brooks} A.~M.,  {Governato} F.,  {Hogg} D.~W.,
  {Shen} S.,   {Wadsley} J.,  2010, \mn@doi [\apj]
  {10.1088/0004-637X/721/1/738}, \href
  {https://ui.adsabs.harvard.edu/abs/2010ApJ...721..738Z} {721, 738}

\bibitem[\protect\citeauthoryear{{de Jong} et~al.,}{{de Jong}
  et~al.}{2019}]{deJong2019}
{de Jong} R.~S.,  et~al., 2019, \mn@doi [The Messenger]
  {10.18727/0722-6691/5117}, \href
  {https://ui.adsabs.harvard.edu/abs/2019Msngr.175....3D} {175, 3}

\bibitem[\protect\citeauthoryear{{de los Reyes}, {Kirby}, {Seitenzahl}  \&
  {Shen}}{{de los Reyes} et~al.}{2020}]{delosReyes2020}
{de los Reyes} M. A.~C.,  {Kirby} E.~N.,  {Seitenzahl} I.~R.,   {Shen} K.~J.,
  2020, \mn@doi [\apj] {10.3847/1538-4357/ab736f}, \href
  {https://ui.adsabs.harvard.edu/abs/2020ApJ...891...85D} {891, 85}

\makeatother
\end{thebibliography}
